\documentclass[aps,prd,twocolumn,preprintnumbers,showpacs,nofootinbib,amssymb]{revtex4}
\usepackage{graphicx}
\usepackage{amsmath}
\usepackage{amssymb}
\usepackage{amsfonts}
\usepackage{bm}
\usepackage{color}

\def\be{\begin{equation}}
\def\ee{\end{equation}}
\def\bea{\begin{eqnarray}}
\def\eea{\end{eqnarray}}

\begin{document}

\title{Predictive model of fermionic dark matter halos with a quantum core \\
and an isothermal atmosphere}
\author{Pierre-Henri Chavanis}
\affiliation{Laboratoire de Physique Th\'eorique, Universit\'e de Toulouse,
CNRS, UPS, France}

\begin{abstract}

We develop a thermodynamical model of fermionic dark matter halos at
finite temperature. Statistical
equilibrium states may be justified by a process of violent collisionless
relaxation in the sense of 
Lynden-Bell or from a collisional relaxation of
nongravitational origin if the fermions
are self-interacting. 
The most probable state (maximum entropy state) generically has a
``core-halo'' structure with a quantum core (fermion ball) surrounded by an
isothermal atmosphere. The quantum core is equivalent to a polytrope of index
$n=3/2$.  The Pauli exclusion principle creates a quantum pressure that
prevents gravitational collapse and solves the core-cusp problem of the cold
dark
matter model. The isothermal atmosphere (which is similar to the NFW profile of
cold dark matter) accounts for the  flat rotation curves of the galaxies at
large distances. We numerically solve the equation of hydrostatic equilibrium with the Fermi-Dirac equation of state and determine the density profiles and rotation curves of fermionic
dark matter halos. We
impose that the surface density of the dark matter halos has the universal value
$\Sigma_0=\rho_0r_h=141\, M_{\odot}/{\rm pc}^2$ obtained from the observations.
For a fermion mass $m=165 \, {\rm eV}/c^2$, the  ``minimum
halo'' has a mass  $(M_h)_{\rm min}=10^8\, M_{\odot}$ and a radius
$(r_h)_{\rm min}=597\, {\rm pc}$ similar to dwarf spheroidals like Fornax. This
ultracompact halo
corresponds to a completely degenerate fermion ball at $T=0$. This is the ground state of the
self-gravitating Fermi gas. For ultracompact dark matter
halos with a mass $(M_h)_{\rm
min}< M_h<(M_h)_{\rm CCP}=6.73\times 10^{8}\, M_{\odot}$ (canonical critical point), the quantum core is
surrounded by a tenuous classical isothermal atmosphere. Dark matter halos with
a mass
$M_h>(M_h)_{\rm CCP}$ are dominated by
the classical isothermal atmosphere.
 They may
be purely gaseous (similar to the Burkert profile) or harbor a fermion ball. The
gaseous solution is stable in all statistical ensembles. The core-halo solution
is canonically unstable (having a negative specific
heat) but, for small dark matter halos with
a mass $(M_h)_{\rm CCP}<M_h<(M_h)_{\rm MCP}=1.08\times 10^{10}\,
M_{\odot}$ (microcanonical critical point), it is microcanonically stable. By
maximizing the entropy at fixed mass
and energy we find that the mass of the quantum core scales with the halo
mass as
$M_c/(M_h)_{\rm min}=1.47 \, [M_h/(M_h)_{\rm min}]^{3/8}$. This relation is
equivalent to the ``velocity dispersion tracing'' relation according to which
the velocity dispersion in the core $v_c^2\sim GM_c/R_c$ is of the same order as
the velocity dispersion in the halo $v_h^2\sim GM_h/r_h$. We provide therefore a
justification of this relation from thermodynamical arguments. The fermion
ball represents a large quantum bulge   which is either present
now or may have, in the past, triggered the
collapse of the surrounding gas, leading to a supermassive black hole
and a quasar. When $M_h>(M_h)_{\rm MCP}$, the quantum core-halo solution is
microcanonically unstable. Large dark matter halos
may undergo a
gravothermal
catastrophe leading ultimately to the formation of  a small out-of-equilibrium
condensed core or, in the case of very large dark matter halos, to a
supermassive black
hole when the core mass overcomes the Oppenheimer-Volkoff limit.
The isothermal halo is left undisturbed and is in agreement
with the Burkert
profile. Our
model has no free parameter (the mass  $m=165\, {\rm eV}/c^2$ of the
fermionic particle is determined by the minimum halo) so it is completely
predictive. It predicts that the Milky Way should harbor a fermionic dark
matter bulge of
mass $M_c=9.45\times 10^{9}\, M_{\odot}$ and  radius $R_c=240\,
{\rm pc}$ in possible agreement with the observations.
 We also consider another model
involving a  larger fermion mass $m=54.6\, {\rm keV/c^2}$. In this
model, a fermion ball of mass $M_c=4.2\times 10^{6}\, M_{\odot}$ and  radius
$R_c=6\times 10^{-4}\,
{\rm pc}$ could mimic the
effect of a supermassive black hole at the
center of the Milky Way (Sagittarius A$^*$). In bigger
galaxies,  the fermion ball should be replaced by a supermassive black hole of
mass $M_{\rm BH}=2.10\times 10^8\, M_{\odot}$ which could account for active
galactic nuclei. For an even larger fermion mass
$m=386\,
{\rm keV/c^2}$, a supermassive black hole of
mass $M_{\rm BH}=4.2\times 10^{6}\, M_{\odot}$ should be formed in the Milky
Way instead of a fermion ball. However, models with a fermion mass $m=
54.6\, {\rm keV/c^2}$ predict that
ultracompact dark matter
halos of mass $\sim 10^8\, M_{\odot}$ should contain a fermionic core of mass
$M_c\sim 10^4\, M_{\odot}$ and radius $R_c\sim 5\, {\rm mpc}$ similar to
intermediate mass black holes, a prediction which may be
challenged by observations.

\end{abstract}


\maketitle

\section{Introduction}
\label{sec_intro}

The cold dark matter (CDM) model of cosmology is remarkably successful in
explaining the large scale structure of the universe  but it experiences several
difficulties at small scales: (i) classical CDM simulations lead to a universal
cuspy density profile -- the Navarro-Frenk-White (NFW) profile \cite{nfw} which
decreases  as $r^{-3}$ at large distances and diverges as $r^{-1}$ at the center
-- while observations rather favor core-like centers -- the Burkert
\cite{observations} profile which  also decreases as $r^{-3}$ at large distances
but tends to a constant at the center; (ii) the number of sub-halos obtained in
CDM simulations is much larger than the number of satellites observed in the
Galaxy \cite{satellites1,satellites2,satellites3}; (iii) dissipationless CDM
simulations predict that the majority of the most massive subhaloes of the Milky
Way are too dense to host any of its bright satellites;  (iv) the stellar
velocity dispersions measured in CDM simulations are larger than those observed
in the satellites of the Galaxy \cite{svd}. These problems are called the
core-cusp problem \cite{moore}, the missing satellites 
problem \cite{satellites1,satellites2,satellites3} and the ``too big to fail''
problem \cite{boylan}. They are responsible for the small-scale crisis of CDM
\cite{crisis}. To solve these problem one solution might be to take into account
baryonic feedback that can
transform cusps into cores \cite{romano1,romano2,romano3}. A possible
alternative is  to take into account the quantum (or wave) nature of the
particles. Indeed, quantum mechanics 
creates an effective pressure which can
balance gravitational attraction and lead to cores instead of cusps. Moreover,
the quantum Jeans length is finite even at $T=0$ (contrary to the classical
Jeans length) and this may solve the missing satellites problem and other
small-scale problems experienced by the classical CDM model.

If the DM particle is a boson,\footnote{See the Introduction of Ref.
\cite{tunnel} for a short review and an exhaustive list of references on bosonic
DM.} like an ultra-light axion (ULA) \cite{marshrevue}, the quantum pressure is
due to the Heisenberg uncertainty principle which is equivalent to an
anisotropic pressure or to a quantum potential. If the bosons are
self-interacting, there is an additional (isotropic) pressure arising from the
self-interaction. At $T=0$, bosons form Bose-Einstein condensates (BECs).
Newtonian self-gravitating BECs are described by the Schr\"odinger-Poisson (SP)
if they are noninteracting or by the Gross-Pitaevskii-Poisson (GPP) equations
if they are self-interacting \cite{prd1}. Numerical simulations
of the SP equations
\cite{ch2,ch3,schwabe,mocz,moczSV,veltmaat,moczprl,moczmnras,veltmaat2} show
that BECDM halos typically have a core-halo structure with a quantum core
(soliton) surrounded by a halo made of quantum interferences. The halo typically
has a NFW or Burkert profile. It can be approximated 
in certain cases by an isothermal profile with an effective temperature (see
Sec. III.C
of \cite{modeldm}). The halo leads to approximately flat rotation curves and the
quantum core solves the core-cusp problem. This core-halo structure results
from
a process of gravitational cooling \cite{seidel94,gul0,gul} or
from a collisional relaxation  of nongravitational origin if
the bosons are self-interacting.

\begin{figure}[!h]
\begin{center}
\includegraphics[clip,scale=0.3]{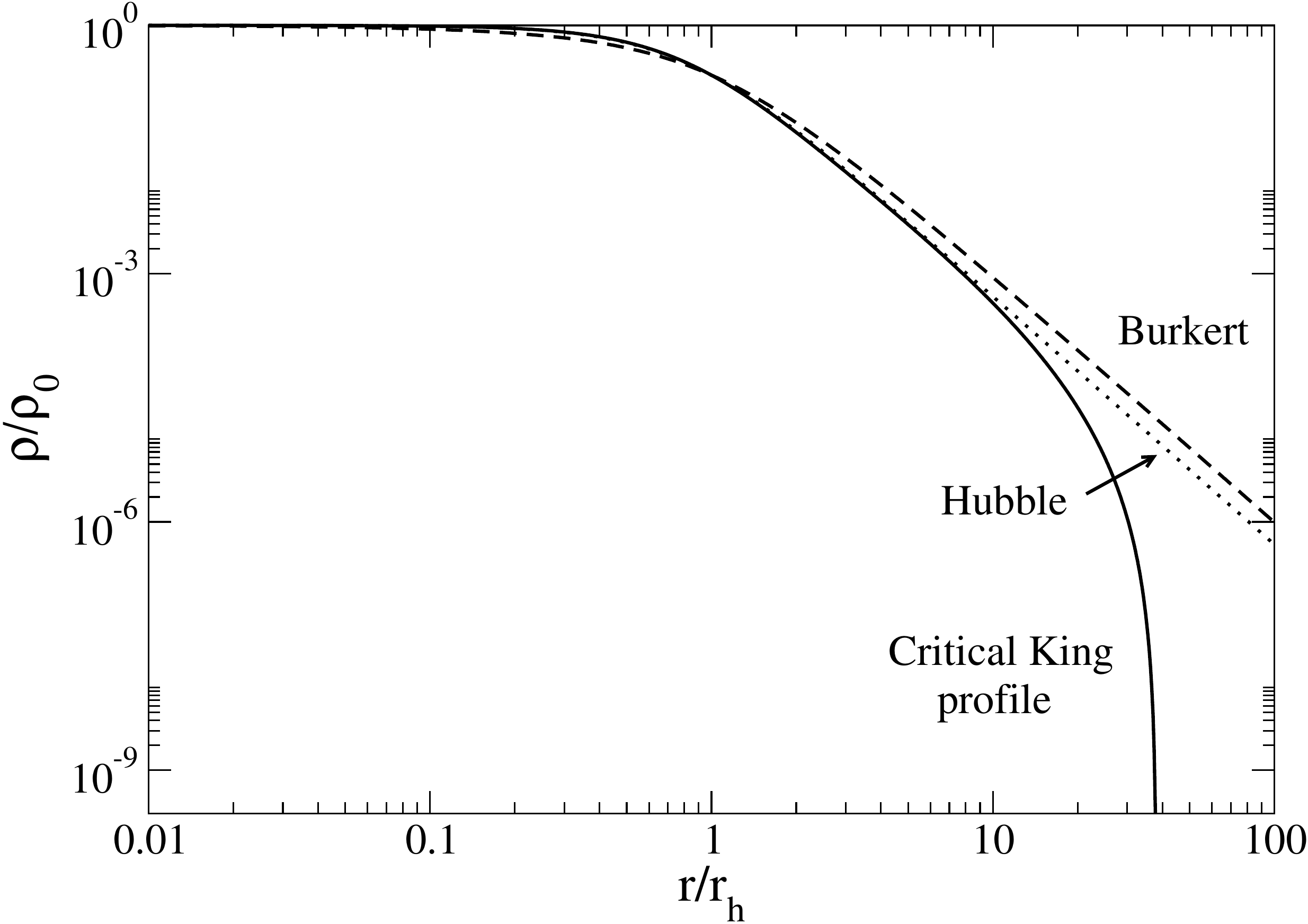}
\caption{Normalized density profiles in logarithmic scales
(zoom of Fig. 18 in \cite{clm1}). Solid line: Critical (marginal) King profile;
Dotted line: Modified Hubble profile; Dashed line: Burkert profile.}
\label{densityLOG}
\end{center}
\end{figure}

If the DM particle is a fermion,\footnote{See the Introduction of Ref.
\cite{gr1} for a short review and an exhaustive list of references on fermionic
DM.} like a sterile neutrino, the quantum pressure is due to the Pauli exclusion
principle which creates an effective isotropic pressure. Fermionic DM halos are
described by the Fermi-Dirac distribution at finite (effective) temperature
which may be
justified by Lynden-Bell's theory of violent relaxation
\cite{lb,csmnras} as argued in \cite{clm1,clm2}. They
typically have a core-halo structure with a quantum core (fermion ball)
surrounded by an isothermal halo \cite{csmnras}. An
isothermal halo is not very different from the NFW or Burkert profile and may
provide a good approximation of it in certain
cases.\footnote{It is shown in Figs. 5 and 6 of \cite{modeldm} that the
isothermal profile is almost indistinguishable from the empirical
(observational) Burkert profile up to a few halo radii. This is even more true
if we account for tidal effects by using the fermionic King model
\cite{clm1,clm2}. It is shown in \cite{clm1} that the critical (marginal) King
profile
triggering the gravothermal catastrophe at
the turning point of energy is
relatively
close to the Burkert profile (see Fig. \ref{densityLOG}).} The
isothermal halo leads to flat rotation curves and the
quantum core solves
the core-cusp problem.  This core-halo structure results from a process of
violent collisionless relaxation \cite{lb,csmnras} or from a collisional
relaxation of nongravitational origin if the fermions are
self-interacting (see Sec. \ref{sec_kt}).\footnote{Fermions and
bosons behave antisymmetrically regarding their collisional relaxation.
The Pauli blocking $f(\eta_0-f)$ for fermions has the tendency to slow down the
relaxation and the Bose enhancement $f(\eta_0+f)$ for bosons, leading to the
formation of ``granules'' or ``quasiparticles'', has the tendency to
accelerate the relaxation.
Gravitational encounters (``collisions'') are completely negligible in fermionic
DM halos. In bosonic DM halos, they manifest themselves on a
(secular) timescale of the order of the age of the universe (see
\cite{kinquant} and references therein).}

The analogy between bosonic and fermionic DM halos suggests that (i) 
the process of gravitational cooling is similar to the process of violent
relaxation (they may even correspond to the same phenomenon); (ii) the NFW
profile (excluding the cusp) and the Burkert profile, which are similar to the
isothermal profile, may be physically justified by Lynden-Bell's theory of
violent relaxation; (iii) the fermion ball in the fermionic model is the
counterpart of the soliton in the BEC model. 

A problem of considerable interest in the physics of quantum (fermionic and
bosonic) DM halos is to construct core-halo profiles of DM halos and predict
the quantum core mass -- halo mass relation $M_c(M_h)$. One can then compare
theoretical predictions with direct numerical simulations and observations.

In a previous paper \cite{modeldm}, we 
have developed a predictive model of BECDM halos in the case where the bosons
have a strong repulsive self-interaction so that the Thomas-Fermi (TF)
approximation can be implemented. We considered a generalized GPP
equation\footnote{This equation was introduced heuristically
in \cite{chavtotal} and justified with more precise arguments in
\cite{wignerPH} from a coarse-graining of the Wigner-Poisson equations.} which
provides a parametrization of the complicated processes of violent relaxation
and gravitational cooling. With respect to the ordinary GPP equation, this
new wave equation includes an effective thermal term and a source of
dissipation. We
determined the equilibrium states of this equation and obtained core-halo
profiles with a quantum core and an isothermal halo similar to those observed in
direct numerical simulations of BECDM.  We obtained the core -- halo mass
relation $M_c(M_h)$
from an effective thermodynamical approach by maximizing the entropy at fixed
mass and energy. We showed that this relation is equivalent to the velocity
dispersion tracing relation stating that the velocity dispersion in the core
$v_c^2\sim GM_c/R_c$ is of the same order as the velocity dispersion in the halo
$v_h^2\sim GM_h/r_h$. We could therefore provide a justification of this
relation from thermodynamical arguments (maximum entropy principle)
\cite{modeldm}.  In subsequent papers \cite{mcmh,mcmhbh,mrjeans}, we derived
the core -- halo mass relation $M_c(M_h)$  from a simple analytical model again
based on a maximum entropy principle and we obtained a general formula for
$M_c(M_h)$ valid for noninteracting bosons, for bosons with a
repulsive or attractive self-interaction, and for fermions. We showed that the
velocity dispersion tracing relation is fulfilled in all cases. In the present
paper, we adapt the bosonic model developed
in \cite{modeldm} to the case of fermions.\footnote{The fact that the
results obtained for bosons could be transposed to fermions was
mentioned in \cite{modeldm}. Conversely, some of the
results obtained here for fermions can be exported to bosons and can complete
the
discussion given in \cite{modeldm}.}

In the recent years, three types of studies have been conducted in the context of fermionic DM: 

(i) Chavanis and coworkers studied phase
transitions in the self-gravitating Fermi gas in Newtonian gravity 
\cite{csmnras,pt,ptdimd,ijmpb,clm2,clmh} and general relativity
\cite{acf,caf,rcf}.
They showed that three possibilities can arise in the caloric curve depending on
the size of the system (measured by the so-called ``degeneracy parameter''
$\mu$). For small systems $\mu<\mu_{\rm CCP}$, no phase transition occur. For
intermediate size systems $\mu_{\rm CCP}<\mu<\mu_{\rm MCP}$, phase transitions
can occur in the canonical ensemble but not in the microcanonical ensemble. For
large systems $\mu>\mu_{\rm MCP}$, both canonical and microcanonical phase
transitions can occur. They also showed that above the
Oppenheimer-Volkoff limit $N>N_{\rm OV}$, a new turning point of energy appears
in the caloric curve and triggers a general relativistic instability
leading to the formation of a black hole by gravitational collapse.

(ii) de Vega and coworkers \cite{dvs1,dvs2,vss,vsedd,vs2,vsbh} constructed
models of DM halos in Newtonian gravity adopting 
a fermion mass of the order of $1\, {\rm keV}/c^2$. This mass determines a
minimum halo of mass $(M_h)_{\rm min}=0.39\times 10^6\, M_{\odot}$ and size
$(r_h)_{\rm min}=33\, {\rm pc}$ corresponding to Willman I assumed to be
completely degenerate. They argued that larger halos are nondegenerate (without
quantum core) so they coincide with the well-known classical isothermal sphere
\cite{chandrabook}. 

(iii) Arg\"uelles and coworkers \cite{ar,arf,sar,rar,krut,krut2,rarnew}
constructed
general relativistic models of DM 
halos adopting a fermion mass of the order of $48\, {\rm keV}/c^2$ and applied
these models to the Milky Way. The system has a core-halo
structure made of a quantum
core
(fermion ball) surrounded by an isothermal atmosphere. Reviving the original
idea of Bilic and coworkers \cite{bvn,bmv,bvr,bmtv,btv}, they argued that the
fermion ball could mimic a supermassive black hole (SMBH) that is purported to
exist at
the center of the Galaxy.

In the present paper, we develop a general model valid for fermions of
arbitrary mass $m$. Then, we consider specifically the case of a ``small'' mass
$m=165\, {\rm eV}/c^2$ or $m\sim 1\, {\rm keV}/c^2$ and the case of a ``large''
mass $m=54.6\, {\rm keV/c^2}$ or $m=386\, {\rm keV/c^2}$ and discuss the
connection with previous works. A brief and
synthetic presentation of our results is given in Ref.
\cite{mg16F} that the readers may
consult in a first reading. The present paper provides a justification and a
detailed description of these results.

The paper is organized as 
follows. In Sec. \ref{sec_kt}, we explain why fermionic DM halos may be in a
maximum entropy state described by the Fermi-Dirac distribution at finite
(effective) temperature. In
Sec. \ref{sec_sgfgb}, we recall basic results concerning the thermodynamics of a
self-gravitating Fermi gas in a box. In Sec.
\ref{sec_nd}, we consider the nondegenerate limit of the self-gravitating Fermi
gas which explains the external  structure of large DM halos. In Sec.
\ref{sec_cdl}, we consider the completely degenerate limit of the
self-gravitating Fermi gas which explains the structure of ultracompact DM
halos and the cores of large DM halos. In Sec. \ref{sec_app}, we consider
partially degenerate DM halos. We show
that they have a core-halo structure  with a quantum core (fermion
ball)
surrounded by an isothermal halo. The isothermal halo leads to flat rotation
curves and the quantum core solves the core-cusp problem. We determine the  core
-- halo mass
relation $M_c(M_h)$ from thermodynamical arguments. In Sec.
\ref{sec_astapp}, we
consider astrophysical applications of our model for a fermion mass $m=165\,
{\rm eV}/c^2$.  We evidence a bifurcation above a 
canonical critical point 
$\mu_{\rm CCP}$ between purely gaseous states and core-halo
states containing a
fermion ball. For the core-halo states, we also evidence a transition at a
microcanonical critical point $\mu_{\rm MCP}$. We argue that small DM
halos with $M_h<(M_h)_{\rm MCP}$ should contain a large quantum bulge while
large DM halos
with  $M_h>(M_h)_{\rm MCP}$  should
rather contain a small out-of-equilibrium quantum core resulting from a
gravothermal catastrophe
arrested by quantum effects. For very large DM
halos, when the core mass $M_c$ passes
above the Oppenheimer-Volkoff limit $M_{\rm OV}$, quantum mechanics cannot
prevent gravitational collapse and the gravothermal catastrophe 
leads to the formation of a SMBH. The
isothermal halo is left undisturbed and is in agreement with the Burkert
profile. We argue that the Milky
Way should contain a large fermionic bulge of
mass $M_c=9.45\times 10^{9}\, M_{\odot}$ and  radius $R_c=240\,
{\rm pc}$ in possible agreement with the observations. In Sec.
\ref{sec_smbh}, we
consider another model  with a fermion mass $m=54.6\, {\rm keV}/c^2$ in which
the fermion ball could mimic a SMBH at the center of
the Milky Way. We argue that, for larger DM halos or for a larger fermion mass,
the fermion ball should be replaced by a real SMBH.  In Sec \ref{sec_paradox},
we propose possible
solutions to an
apparent paradox related to the universal  surface density of DM halos. In
Sec. \ref{sec_tqc} we discuss the difference between isothermal and quantum
cores and their formation process. In Sec.
\ref{sec_conclusion}, we summarize our main results and conclude.

\section{Formation and evolution of DM halos}
\label{sec_kt}

In this section, we recall basic 
elements of kinetic theory related to the formation and the evolution of
fermionic DM halos in order to justify our thermodynamical approach (we refer to
\cite{kinquant} for a more detailed discussion).

It is well-known that self-gravitating systems experience two successive types of relaxation:

(i) In a first regime, gravitational encounters
can be neglected and the evolution of the system is described by the Vlasov (or
collisionless Boltzmann) equation. The Vlasov-Poisson equations experience a
complicated process of collisionless violent relaxation as described by
Lynden-Bell \cite{lb} in the context of stellar
systems.
Through violent relaxation, the system reaches a quasistationary state
(virialized state) on the coarse-grained scale which is a stable stationary
solution of the Vlasov equation. The Vlasov-Poisson equations admit an infinite
number of stationary solutions. In addition, all spherical DFs of the form
$f=f(\epsilon)$ with $f'(\epsilon)<0$, where $\epsilon=v^2/2+\Phi({\bf r})$ is
the energy of the particles, are
dynamically (Vlasov) stable in Newtonian gravity
\cite{bt}.\footnote{This is no more true in general relativity (see the
discussion in \cite{gr1}).} Lynden-Bell
proposed to determine the
``most probable state'' of the system resulting from a collisionless
relaxation by using arguments of statistical mechanics and
thermodynamics.  This equilibrium state is obtained by maximizing a mixing
entropy while taking into account all the constraints of the Vlasov equation. In
the single level approximation, the Lynden-Bell entropy is similar to the
Fermi-Dirac entropy and the constraints of the Vlasov equation reduce to the
conservation of mass and energy. In addition, if the particles are fermions, the
Lynden-Bell exclusion principle $\overline{f}\le \eta_0$, where $\eta_0$ is the
initial DF, coincides with the Pauli exclusion principle  ${f}\le 2m^4/h^3$ up
to a numerical factor of order unity (see footnote 34 of \cite{clm1}). In order
to select the most probable structure arising from a violent collisionless
relaxation, one has therefore to maximize the Lynden-Bell (or Fermi-Dirac)
entropy at fixed mass and energy. The
extremization problem leads to the Lynden-Bell (or Fermi-Dirac) DF. Then, we
have to make sure that the equilibrium state is an entropy maximum (most
probable state) not an entropy minimum or a saddle point.\footnote{If several
stable equilibrium states (entropy maxima) are found for the same values of mass
and energy, they may all be equally relevant. Indeed, metastable states (local
entropy maxima) have a very long lifetime and are as much
relevant as fully
stable states (global entropy maxima). Their selection is related to a notion of
basin of attraction and cannot be decided simply by comparing their entropies.
Finally, we note that the predictive power of Lynden-Bell's theory is limited by
the problem of incomplete relaxation: the system may reach a dynamically stable
quasistationary state that is not a maximum entropy state.} 

(ii) In a second regime, collisions must be taken into account.\footnote{We
consider collisions of all sort. They may correspond to weak gravitational
encounters or strong (hard core-like) collisions if the particles have a
self-interaction, leading to the notion of
self-interacting dark matter (SIDM) halos. DM
halos may also
experience a stochastic forcing due to the
presence of baryons or other external  sources that can induce a secular
relaxation of the system (see Appendix B of \cite{modeldm}).} The collisional
relaxation of the system is described by a kinetic equation such as the
gravitational Boltzmann, Landau 
or Lenard-Balescu equation.\footnote{See the introduction of Ref. \cite{aakin}
for a short review and an exhaustive list of references on the kinetic theory of
self-gravitating systems.} If the particles are fermions, the collisional
relaxation leads to a statistical equilibrium state described by the ordinary
Fermi-Dirac distribution which maximizes the Fermi-Dirac entropy at fixed mass
and energy. This corresponds to the ``most probable state'' of the system
resulting from a collisional relaxation. This is also a 
stable stationary solution of
the kinetic equation.\footnote{Actually, because of
evaporation and because of its interaction with nearby galaxies, the system is
tidally truncated and the Fermi-Dirac distribution must be
replaced by the fermionic King distribution \cite{stella,mnras}. This DF
can be derived from
a kinetic theory based on the fermionic Landau equation 
\cite{mnras}. The fermionic King model is studied in \cite{clm2} in Newtonian
gravity and in \cite{krut,krut2,rarnew} in general relativity.}

In the two situations described 
above we are led to maximizing the Lynden-Bell or Fermi-Dirac entropy at fixed
mass and energy.  We stress that the justification of this maximization problem 
is different in the collisionless (Lynden-Bell) and collisional (Fermi-Dirac)
regimes. If the system is
collisionless, the temperature is effective. We also
note that the proper thermodynamical ensemble to consider is
the microcanonical ensemble. Indeed, the system is assumed to be isolated so
that the energy and the mass are conserved. This remark is important since
statistical ensembles may be inequivalent for systems with long-range
interactions such as self-gravitating systems \cite{paddy,campa,ijmpb}.

If the particles interact only via (weak) two-body gravitational encounters, the
collisional relaxation time is extremely long, scaling as $t_R\sim (N/\ln
N)t_D$ \cite{bt}, where $N$ is the number of particles and $t_D\sim R/v\sim
1/\sqrt{G\rho}\sim 0.1\, {\rm Gyrs}$ is the dynamical time (we have taken
$R\sim 20.1\, {\rm kpc}$, $\rho\sim
7.02\times 10^{-3}\, M_{\odot}/{\rm pc}^3$ and $v\sim 146\, {\rm km/s}$  in a
galaxy of mass $M\sim 10^{12}\, M_{\odot}$
like the Milky Way).
For fermionic DM halos,  
$N$ is huge ($N\sim 10^{75}$ for keV
fermions) so the relaxation time is
much larger than the age of the universe $t_{\rm U}\sim 13.8\, {\rm Gyrs}$ by
many orders of magnitude (the Pauli blocking even increases
this relaxation time). In that
case, the system is essentially collisionless and only the Lynden-Bell type of
relaxation is relevant. However, in order to be more general, we consider the
possibility that the particles have a (strong) self-interaction that can cause
a faster collisional evolution of nongravitational origin. This
allows us to consider the possibility of a
collisional relaxation (especially in the core of the system where the density
is high and the relaxation time short) towards a Fermi-Dirac equilibrium state
on a timescale smaller than the age of the universe. For
example, if they have a cross section per unit mass
$\sigma_m\equiv \sigma/m=1.25\, {\rm cm^2/g}$ consistent with the Bullet Cluster
constraint \cite{bullet} we get $t_{\rm self}\sim 1/(\rho
\sigma_m v)\sim 3.66\, {\rm Gyrs}<t_{\rm U}$. In that case, 
fermionic DM
halos
behave
similarly to globular clusters with additional quantum
effects. Because of
collisions and evaporation, they follow a series of equilibria towards
configurations of higher and higher central density. If the equilibrium state
becomes
thermodynamically unstable, a fermionic DM halo may experience a phase
transition from a gaseous phase to a condensed phase (with a quantum
core and an isothermal envelope) associated with a form of
gravothermal catastrophe \cite{lbw} stopped by quantum degeneracy
\cite{csmnras}. This may be followed by a dynamical
instability of general relativity origin leading to the formation of a SMBH
\cite{caf,acf,rcf}.

\section{Self-gravitating Fermi gas in a box}
\label{sec_sgfgb}

In this section, we consider the statistical mechanics of a
self-gravitating Fermi gas in a box. We  summarize the main results obtained in
our previous papers \cite{csmnras,pt,ptdimd,ijmpb} and detail the theoretical
framework that will be needed in the
present study.

\subsection{Theoretical framework}
\label{sec_tf}

We consider a gas of  nonrelativistic fermions interacting via Newtonian
gravity. Let $f({\bf
r},{\bf v},t)$ denote its distribution function (DF) in phase space giving the
mass
density of
fermions with position ${\bf r}$ and velocity ${\bf v}$ at time $t$. The mass
density in configuration space is $\rho=\int f\, d{\bf v}$. The total mass of
the system is
\begin{eqnarray}
\label{tf1}
M=\int f\, d{\bf r}d{\bf v}
\end{eqnarray}
and its total energy is
\begin{eqnarray}
\label{tf2}
E=\int f\frac{v^2}{2}\, d{\bf r}d{\bf v}+\frac{1}{2}\int\rho\Phi\, d{\bf r},
\end{eqnarray}
where the first term is the kinetic energy and the second term is the
gravitational energy ($E=E_{\rm
kin}+W$). We introduce the Fermi-Dirac entropy
\begin{equation}
\label{tf3}
\frac{S}{k_B}=-\frac{\eta_0}{m}\int\Biggl\lbrace \frac{f}{\eta_0}\ln
\frac{f}{\eta_0}
+\left (1-\frac{f}{\eta_0}\right )\ln\left
(1-\frac{f}{\eta_0}\right )\Biggr\rbrace\, d{\bf r}d{\bf v},
\end{equation}
where
\begin{equation}
\label{tf3b}
\eta_0=\frac{2m^4}{h^3}
\end{equation}
is the maximum value of the DF fixed by the Pauli exclusion principle
(the factor $2$ accounts for
the multiplicity $2s+1$ of quantum states for particles of spin $s=1/2$). The
Fermi-Dirac entropy is equal to the logarithm of the number of microstates,
specified by the precise position and velocity $\lbrace {\bf r}_i,{\bf
v}_i\rbrace$ of all the fermions, corresponding to a given macrostate specified
by the DF $f({\bf r},{\bf v})$ giving the density
of fermions around the point $({\bf r},{\bf v})$ in phase space.

In the microcanonical ensemble, the statistical equilibrium state of a
self-gravitating gas of fermions is
obtained by maximizing the Fermi-Dirac entropy $S$ at fixed energy $E$ and
mass $M$. One has therefore to solve the optimization problem
\begin{eqnarray}
\max\ \lbrace {S}\, |\,  E, M \,\, {\rm fixed} \rbrace.
\label{tf4}
\end{eqnarray}
This thermodynamic approach is justified in a mean field approximation which
is exact in a proper thermodynamic limit $N\rightarrow +\infty$ (see
Sec. 7.1 of \cite{ijmpb} and Appendix B of \cite{acf}).

An extremum of entropy at fixed energy and mass is determined by the variational
principle
\begin{equation}
\label{tf5}
\frac{\delta S}{k_B}-\beta\delta E+\frac{\alpha}{m}\delta M=0,
\end{equation}
where $\beta=1/(k_B T)$ and $\alpha=\mu/(k_B T)$ are
Lagrange multipliers ($T$ is the temperature and $\mu$ is the global chemical
potential). This leads to the Fermi-Dirac distribution
\begin{equation}
\label{tf6}
f=\frac{\eta_0}{1+e^{\left \lbrack m {v^2}/{2}+m\Phi({\bf r})-\mu\right
\rbrack/k_B T}}.
\end{equation}

The density of particles $\rho=\int f\, d{\bf v}$   and the pressure
$P=\frac{1}{3}\int f v^2\, d{\bf v}$  are related
to the gravitational potential $\Phi({\bf r})$ by
\begin{equation}
\rho({\bf r})=\frac{4\pi \sqrt{2}\eta_0 }{(\beta m)^{3/2}}I_{1/2}\left
\lbrack \lambda e^{\beta m\Phi({\bf r})}\right \rbrack,
\label{tf7}
\end{equation}
\begin{equation}
P({\bf r})=\frac{8\pi\sqrt{2}\eta_0}{3(\beta m)^{5/2}}I_{3/2}\left
\lbrack \lambda e^{\beta m\Phi({\bf r})}\right \rbrack,
\label{tf8}
\end{equation}
where  $\lambda=e^{-\beta\mu}$  and $I_{n}(t)$ denotes the Fermi
integrals
\begin{equation}
I_{n}(t)=\int_{0}^{+\infty}\frac{x^{n}}{1+t e^{x}}\, dx.
\label{tf9}
\end{equation}
We recall the identity
\begin{equation}
I'_{n}(t)=-\frac{n}{t}I_{n-1}(t) \qquad (n>0),
\label{tf10}
\end{equation}
which can be established from Eq. (\ref{tf9}) by an integration by parts.
Eliminating $\lambda e^{\beta m\Phi({\bf r})}$ between Eqs. (\ref{tf7}) and
(\ref{tf8}), we obtain the equation of state $P(\rho)$ of the
nonrelativistic Fermi gas at
finite temperature in parametric form.

Combining the condition of hydrostatic
equilibrium
\begin{equation}
\nabla P+\rho\nabla\Phi={\bf 0}
\label{tf11}
\end{equation}
with the Poisson equation
\begin{equation}
\Delta\Phi=4\pi G\rho,
\label{tf12}
\end{equation}
we obtain the fundamental
differential equation of hydrostatic equilibrium
\begin{equation}
\nabla\cdot \left (\frac{\nabla P}{\rho}\right )=-4\pi G \rho.
\label{tf13}
\end{equation}
Together with the barotropic equation of state $P(\rho)$ specified by
Eqs. (\ref{tf7}) and (\ref{tf8}) this equation determines the density profile of
the self-gravitating Fermi gas at statistical equilibrium.

Alternatively, substituting the density-potential relation from Eq.
(\ref{tf7}) into the Poisson equation (\ref{tf12}), we obtain a
differential equation determining the gravitational potential
\begin{equation}
\Delta\Phi=\frac{16\pi^2\sqrt{2}G\eta_0 }{(\beta m)^{3/2}}I_{1/2}\left
\lbrack \lambda e^{\beta m\Phi({\bf r})}\right \rbrack,
\label{tf14}
\end{equation}
which is called the Fermi-Poisson equation or the finite temperature Thomas-Fermi
equation. The density is then obtained from Eq. (\ref{tf7}). The two equations
(\ref{tf13}) and (\ref{tf14}) are equivalent.

\begin{figure}
\begin{center}
\includegraphics[clip,scale=0.3]{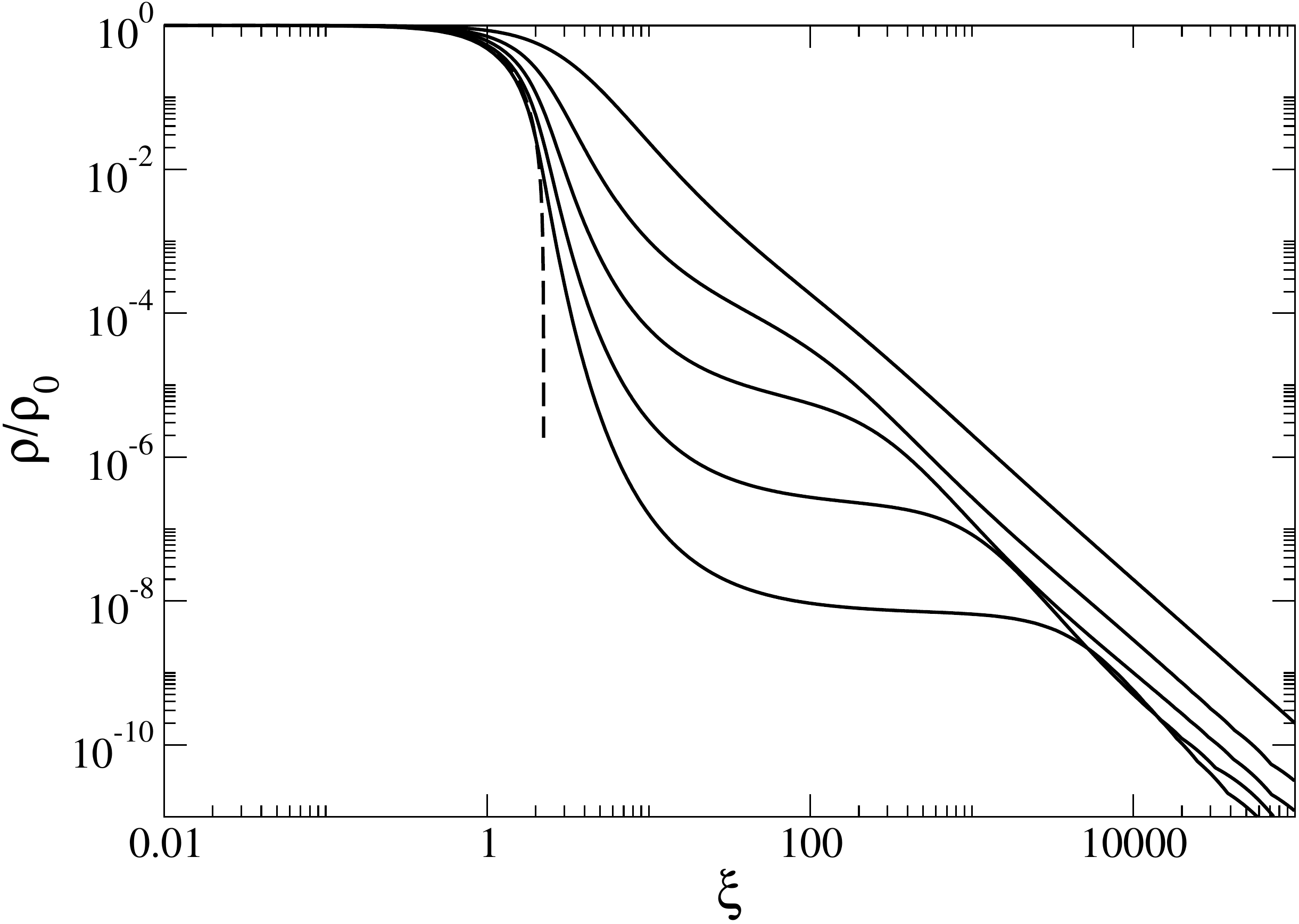}
\caption{Normalized density
profiles of
fermionic DM halos for different values of $k$ ($k=10^{-8}, 10^{-6}, 10^{-4},
10^{-2}$ from bottom to top). The dashed line corresponds to a completely
degenerate fermion ball. The upper curve corresponds to a nondegenerate
isothermal DM halo.}
\label{multirho}
\end{center}
\end{figure}

We now assume that the system is spherically symmetric and introduce the
dimensionless variables
\begin{equation}
\psi=\beta m(\Phi-\Phi_{0}),\qquad k=\lambda e^{\beta m\Phi_{0}},
\label{tf15}
\end{equation}
and
\begin{equation}
\xi=\left \lbrack \frac{16\pi^2\sqrt{2}G\eta_0}{(\beta
m)^{1/2}}\right\rbrack^{1/2}r,
\label{tf17}
\end{equation}
where $\Phi_0$ is the central value of the gravitational potential. We can then
rewrite the density and the
pressure under the form
\begin{equation}
\rho(r)=\frac{4\pi \sqrt{2}\eta_0 }{(\beta m)^{3/2}}I_{1/2}\left
\lbrack k e^{\psi(\xi)}\right \rbrack,
\label{tf18}
\end{equation}
\begin{equation}
P(r)=\frac{8\pi\sqrt{2}\eta_0}{3(\beta m)^{5/2}}I_{3/2}\left
\lbrack k e^{\psi(\xi)}\right \rbrack.
\label{tf19}
\end{equation}
On the other hand, Eqs. (\ref{tf13}) and (\ref{tf14}) lead to the fermionic
Emden equation
\begin{equation}
\frac{1}{\xi^2}\frac{d}{d\xi}\left (\xi^2\frac{d\psi}{d\xi}\right
)=I_{1/2}(ke^{\psi})
\label{tf20}
\end{equation}
with the boundary conditions
\begin{equation}
\psi(0)=\psi'(0)=0.
\label{tf21}
\end{equation}
This equation determines the structure of the system as a
function of the parameter $k$. For $k\rightarrow +\infty$, the system is
nondegenerate: this corresponds to the gaseous phase (see Sec. \ref{sec_nd}).
For $k\rightarrow 0$, the system is completely degenerate: this corresponds to
the condensed phase (see Sec. \ref{sec_cdl}). For intermediate values of $k$,
the system is partially degenerate: it typically has a core-halo
structure with a quantum core (fermion ball) surrounded by a classical
isothermal halo (see Secs. \ref{sec_app}).\footnote{See Ref.
\cite{csmnras} for a
description of this core-halo configuration and some analytical approximations.
See also Sec. V of Ref. \cite{modeldm} for similar results obtained in the case
of bosonic DM halos which can be directly exported to the case of fermionic DM
halos.} Some density profiles are plotted in Fig. \ref{multirho} for different
values of $k$.

As is well-known, self-gravitating systems at nonzero temperature
have the tendency to evaporate. Therefore, there is no equilibrium
state in a strict sense and the statistical mechanics of
self-gravitating systems is essentially an out-of-equilibrium
problem. However, the evaporation rate is small in general and the
system can be found in a quasi-equilibrium state for a relatively long
time.  In order to describe the thermodynamics of the self-gravitating
Fermi gas rigorously, we shall use an artifice and enclose the system
within a spherical box of radius $R$.\footnote{A more rigorous approach would be
to use a truncated model (fermionic King model) like in \cite{clm1,clm2}.} The
box typically represents the
size of the cluster under consideration (see Sec.  \ref{sec_app}). In that case,
the solution
of Eq. (\ref{tf20}) is terminated by the box at the normalized radius
\begin{equation}
\alpha=\left \lbrack \frac{16\pi^2\sqrt{2}G\eta_0}{(\beta
m)^{1/2}}\right\rbrack^{1/2}R.
\label{tf22}
\end{equation}
Since $\alpha$ is the value of $\xi$ at the box radius $R$ we can write
\begin{equation}
\xi=\alpha\frac{r}{R}.
\label{tf23}
\end{equation}

Let us first calculate the normalized inverse temperature
\begin{equation}
\eta=\frac{\beta GMm}{R}.
\label{tf24}
\end{equation}
For a spherically symmetric distribution of matter, the Poisson
equation (\ref{tf12}) is equivalent to Newton's law
\begin{equation}
\frac{d\Phi}{dr}=\frac{GM(r)}{r^2},
\label{tf25}
\end{equation}
where $M(r)=\int_0^r \rho(r')4\pi {r'}^2\, dr'$ is the mass contained within
the sphere of radius $r$. Applying Newton's law at $r=R$ and using Eqs.
(\ref{tf15}), (\ref{tf23}) and (\ref{tf24}), we get
\begin{equation}
\eta=\alpha\psi'_k(\alpha).
\label{tf26}
\end{equation}
This equation relates the dimensionless box radius $\alpha$ and the
concentration variable $k$ to the dimensionless inverse temperature $\eta$.

On the other hand, according to Eqs. (\ref{tf22}) and (\ref{tf24}), $\alpha$ and
$k$ are linked to each other by the
relation
\begin{equation}
\alpha^{2}\sqrt{\eta}=\mu
\label{tf27}
\end{equation}
or, more explicitly [using Eq. (\ref{tf26})]
\begin{equation}
\alpha^{5}\psi'_{k}(\alpha)=\mu^{2},
\label{tf28}
\end{equation}
where
\begin{equation}
\label{tf29}
\mu=\eta_0\sqrt{512\pi^4G^3MR^3}
\end{equation}
is the so-called degeneracy parameter \cite{ijmpb}.\footnote{It should not be
confused
with
the chemical potential which is denoted by the same symbol. In principle, no
confusion should arise.} We
shall give a
physical interpretation of this parameter in Sec. \ref{sec_app}.

The calculation of the energy is a little more involved.
The kinetic energy of a nonrelativistic gas can be written as
\begin{equation}
E_{\rm kin}=\frac{3}{2}\int P\, d{\bf r}.
\label{tf30}
\end{equation}
Using Eq. (\ref{tf19}), we obtain
\begin{equation}
\frac{E_{\rm kin}R}{GM^2}=\frac{\alpha^7}{\mu^4}\int_0^{\alpha}
I_{3/2}\left\lbrack ke^{\psi_k(\xi)}\right\rbrack\xi^2\, d\xi.
\label{tf31}
\end{equation}
In order to determine the potential energy, we can use the
Virial theorem (see, e.g., \cite{gr1})
\begin{equation}
2E_{\rm kin}+W=3P_{\rm b}V,
\label{tf32}
\end{equation}
where $P_b=P(R)$ is the pressure on the boundary of the box and
$V=\frac{4}{3}\pi R^3$ is the volume of the spherical box.
Using the expression of the pressure from Eq. (\ref{tf19}) at the box radius
$R$, we get
\begin{equation}
\frac{W R}{GM^{2}}=\frac{2\alpha^{10}}{3\mu^{4}}I_{3/2}\left\lbrack k
e^{\psi_k(\alpha)}\right\rbrack-\frac{2E_{\rm kin}R}{GM^{2}}.
\label{tf33}
\end{equation}
Introducing the normalized energy
\begin{equation}
\Lambda=-\frac{ER}{GM^2}
\label{tf34}
\end{equation}
and combining Eqs. (\ref{tf31}) and (\ref{tf33}), we finally obtain
\begin{equation}
\Lambda=\frac{\alpha^7}{\mu^4}\int_0^{\alpha}
I_{3/2}\left\lbrack ke^{\psi_k(\xi)}\right\rbrack\xi^2\,
d\xi-\frac{2\alpha^{10}}{3\mu^4}
I_{3/2}\left\lbrack ke^{\psi_k(\alpha)}\right\rbrack.
\label{tf35}
\end{equation}
The expression of the entropy is derived in Appendix \ref{sec_ent}.

Finally, using Eqs. (\ref{tf15}), (\ref{tf18}), (\ref{tf24}), (\ref{tf25}),
(\ref{tf29}) and (\ref{hr4})  the normalized density and velocity profiles can
be written as
\begin{equation}
\frac{\rho(r)}{M/R^3}=\frac{\mu}{4\pi\eta^{3/2}}I_{1/2}[k e^{\psi(\xi)}],
\label{tf35b}
\end{equation}
\begin{equation}
\frac{v^2(r)}{GM/R}=\frac{1}{\eta}\xi \psi'(\xi).
\label{tf35c}
\end{equation}

\subsection{Caloric curves and ensembles inequivalence}
\label{sec_ineq}

\begin{figure}
\begin{center}
\includegraphics[clip,scale=0.3]{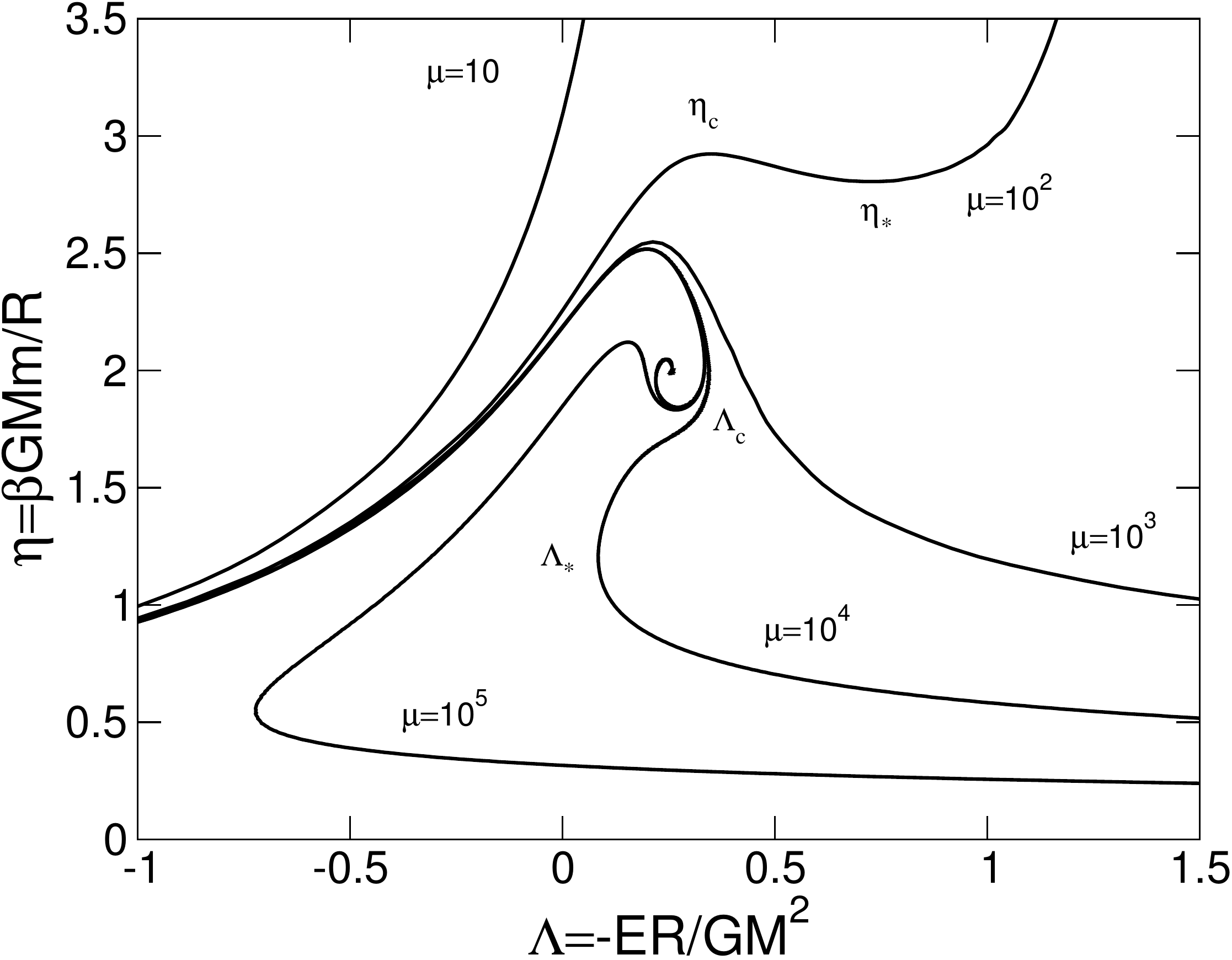}
\caption{Caloric curves (series of equilibria) of the  self-gravitating Fermi
gas for
different values of $\mu$.}
\label{multimu2}
\end{center}
\end{figure}

Using the foregoing formulae, we can obtain the caloric curve $\eta(\Lambda)$
of the self-gravitating Fermi gas for a specified value of $\mu$ as follows. For
a given value of $k$, we can solve the ordinary differential equation
(\ref{tf20}) with the initial conditions (\ref{tf21})  until the value of
$\alpha$ at which the relation (\ref{tf28}) is satisfied. Then,
Eqs. (\ref{tf26}) and (\ref{tf35}) determine the normalized inverse
temperature $\eta$ and the
normalized energy $\Lambda$ of the configuration. By varying the parameter $k$
from $0$ to $+\infty$, we can determine the full
caloric curve  $\eta(\Lambda)$ for the specified value of the degeneracy
parameter $\mu$ (see Fig. \ref{multimu2}). We can then study the occurrence of
phase
transitions as a function of $\mu$. This study has
been made in detail in \cite{ijmpb}.\footnote{Very similar results
apply to the case of tidally truncated
systems described by the fermionic King model \cite{clm1,clm2}.} Below, we
summarize the main results of
this study that will be useful in the following.

We have to be careful that, for self-gravitating systems (which have a
long-range interaction), the statistical
ensembles are inequivalent. In the previous section, we have worked in the
microcanonical ensemble. This is
the statistical ensemble associated with isolated systems where the energy $E$
is fixed. By contrast, systems in contact with a heat bath fixing the
temperature $T$ are described by the canonical ensemble. In the canonical
ensemble, the statistical equilibrium state of a
self-gravitating gas of fermions is
obtained by minimizing the Fermi-Dirac free energy $F=E-TS$ at fixed
mass $M$. One has therefore to solve the optimization problem
\begin{eqnarray}
\min\ \lbrace {F}\, |\,   M \,\, {\rm fixed} \rbrace.
\label{tf4b}
\end{eqnarray}

One can easily show that the equilibrium states in the microcanonical and in the
canonical ensembles
are the same: an extremum of free energy at fixed mass is also an extremum of
entropy at fixed mass and energy \cite{cc}. However, their stability may be
different in the two ensembles. An equilibrium state that is canonically
stable is always microcanonically stable (a minimum of free energy at fixed mass
is always a  maximum of entropy at fixed mass and
energy), but the converse is wrong:  a maximum of entropy at fixed mass and
energy is not necessarily a
minimum of free energy at fixed mass \cite{cc}. For example, equilibrium
states with a negative specific heat are always unstable in the canonical
ensemble while they may be stable in the microcanonical ensemble. This
corresponds to the concept of ensembles inequivalence for systems with
long-range interaction \cite{paddy,ijmpb,campa}. As a result, we may miss
important solutions if we use the canonical ensemble instead of the
microcanonical one.\footnote{In particular, as shown in
\cite{mcmh} and further discussed in Sec. \ref{sec_astapp}, the core-halo
solution that could correspond to real DM halos
is unstable in the canonical ensemble while it is stable in the microcanonical
ensemble. This suggests that the microcanoncal ensemble is more adapted to DM
halos than the canonical ensemble. Since DM halos are
isolated instead of being in contact with a thermal bath, the use of the
microcanonical ensemble is physically
justified \cite{clm1,clm2}.}

The self-gravitating Fermi gas presents two critical points, one in each
ensemble: the function $\Lambda(\eta)$ becomes multivalued at the canonical
critical point $\mu_{\rm CCP}$ and the function $\eta(\Lambda)$ becomes
multivalued at the microcanonical critical point $\mu_{\rm MCP}$ whose values
are \cite{ijmpb}
\begin{equation}
\label{cpoi}
\mu_{\rm CCP}=83,\qquad \mu_{\rm MCP}=2670.
\end{equation}

For $\mu<\mu_{\rm CCP}$ the series of equilibria $\eta(\Lambda)$ is
monotonic (see the curve $\mu=10$ in Fig. \ref{multimu2}; see also
Fig. \ref{multimu2PH} below). The equilibrium states are stable in
both canonical and
microcanonical ensembles.

For $\mu>\mu_{\rm CCP}$ the series of equilibria $\eta(\Lambda)$ presents
turning points of temperature (see the curves $\mu=10^2-10^5$ in
Fig. \ref{multimu2}; see also Figs. \ref{fel} and \ref{le5} below).
Following the series of equilibria towards
higher and higher density contrasts, and using the Poincar\'e-Katz
\cite{poincare,katzpoincare1} criterion,
one can show that the equilibrium states are canonically stable before the first
turning point of temperature $\eta_c$ (gaseous phase) and after the last turning
point of temperature $\eta_*$ (condensed phase). They are canonically unstable
in between. In the canonical ensemble, the system undergoes an isothermal
collapse at $\eta_c$ from the gaseous phase to the condensed phase and
an explosion at $\eta_*$ from the condensed
phase to the gaseous phase. The first order phase transition that is expected
at $\eta_t$ (at which the free energy of the two phases coincides) does not take
place in practice because of the very long lifetime of metastable states for
systems with long-range interactions scaling as $e^N$ \cite{lifetime}.

For $\mu>\mu_{\rm MCP}$ the series of equilibria $\eta(\Lambda)$ presents
turning points of energy (see the curves $\mu=10^4$ and $10^5$ in
Fig. \ref{multimu2}; see also Fig. \ref{le5} below). Following the
series of equilibria towards
higher and higher density contrasts, and using the Poincar\'e-Katz
\cite{poincare,katzpoincare1} criterion,
one can show that the equilibrium
states are microcanonically stable before the first turning point of energy
$\Lambda_c$ (gaseous phase) and after the last
turning
point of energy $\Lambda_*$ (condensed phase). They
are microcanonically
unstable in between. In the microcanonical ensemble, the
system undergoes a gravothermal catastrophe at $\Lambda_c$ from the gaseous
phase to the condensed phase and an explosion at $\Lambda_*$ from the condensed
phase to the gaseous phase. The first order phase transition that is expected
at $\Lambda_t$ (at which the entropy of the two phases coincides) does not take
place in practice because of the very long lifetime of metastable states for
systems with long-range interactions scaling as $e^N$ \cite{lifetime}.

For $\mu\rightarrow +\infty$ we recover the series of equilibria
$\eta(\Lambda)$ of a classical isothermal self-gravitating gas (see
Fig. \ref{etalambda} below). It has a snail-like structure (spiral). Using the
Poincar\'e-Katz
\cite{poincare,katzpoincare1} criterion,
one can show that the equilibrium
states become (and remain) unstable after the first turning point of temperature
in the canonical ensemble and after the first turning point of energy in
the microcanonical ensemble.

For $\mu_{\rm CCP}<\mu<\mu_{\rm MCP}$  (see the curves $\mu=10^2$ and $10^3$ in
Fig. \ref{multimu2}; see also Fig. \ref{fel} below) all the equilibrium
states are stable in the microcanoncal ensemble while the equilibrium states
between the first
turning point of temperature $\eta_c$ and the last turning point of temperature
$\eta_*$ (core-halo solution) are  unstable in the canonical ensemble. They
have a core-halo structure and a negative specific heat. The region of negative
specific heat that is allowed in the microcanonical ensemble is replaced by a
phase transition in the canonical ensemble. This corresponds to a region of
ensembles inequivalence.\footnote{The physical nature of the core-halo
solution is very different in the two ensembles. In the canonical ensemble, the
degenerate core
represents a ``germ'' or a ``critical droplet'' (in the language of phase
transitions and nucleation) that the system must form to pass from the gaseous
phase to the condensed phase. This is a saddle point of free energy at fixed mass.
The probability to form this configuration  is
very low, scaling with the number of particles as $e^{-N}$. This is a rare event.
This explains why metastable gaseous states have a very long lifetime scaling as
$e^N$ \cite{ijmpb,lifetime}. In the microcanonical ensemble, the core-halo
solution is fully stable
and corresponds to the most probable state of the system for the corresponding
energy. It is therefore expected to be physically selected by the system. }

We now apply these results to DM halos. We have explained that
metastable states (local entropy maxima) are as much relevant as fully stable
states (global entropy maxima). They are robust and long-lived. Therefore,
we shall not make a distinction between fully stable and metastable states in
our study.

\section{Nondegenerate limit: External structure of large DM halos}
\label{sec_nd}

Let us first consider the nondegenerate limit of the self-gravitating Fermi gas 
which describes the external structure (envelope) of large DM halos.

\subsection{Isothermal equation of state}

In the nondegenerate limit  $T\rightarrow +\infty$ (or  $T\gg T_{F}$ where
$T_F\sim \hbar^2\rho^{2/3}/m^{5/3}k_B$ is the Fermi temperature) the Fermi-Dirac
DF (\ref{tf6}) reduces to the Maxwell-Boltzmann distribution
\begin{equation}
\label{nd1}
f=\eta_0 e^{\beta\mu} e^{-\beta m\left \lbrack\frac{v^2}{2}+\Phi({\bf
r})\right \rbrack}.
\end{equation}
In that case, the density and the pressure are given by
\begin{equation}
\label{nd2}
\rho=\eta_0 e^{\beta\mu}\left (\frac{2\pi}{\beta m}\right )^{3/2}e^{-\beta
m\Phi({\bf r})},
\end{equation}
\begin{equation}
\label{nd3}
P=\eta_0 e^{\beta\mu}\left (\frac{2\pi}{\beta m}\right
)^{3/2}\frac{1}{\beta m}e^{-\beta
m\Phi({\bf r})},
\end{equation}
leading to the classical
isothermal equation of state
\begin{equation}
\label{nd4}
P({\bf r})=\rho({\bf r})\frac{k_B T}{m}.
\end{equation}
The fundamental differential equation (\ref{tf13}) of hydrostatic equilibrium 
takes the form
\begin{equation}
\label{nd5}
-\frac{k_B
T}{m}\Delta\ln\rho=4\pi
G\rho.
\end{equation}
It describes the balance between the gravitational attraction and the thermal
pressure. It is equivalent to the Boltzmann-Poisson equation
\begin{equation}
\label{nd6}
\Delta\Phi=4\pi G \eta_0 e^{\beta\mu}\left (\frac{2\pi}{\beta m}\right
)^{3/2}e^{-\beta
m\Phi({\bf r})}
\end{equation}
obtained by combining Eqs. (\ref{tf12}) and (\ref{nd2}). Equations (\ref{nd5})
and (\ref{nd6}) can be reduced to the Emden equation (\ref{i5})
\cite{chandrabook}.

\subsection{Flat rotation curves}

The  differential equation of hydrostatic equilibrium (\ref{nd5}) has
no simple analytical solution and must be solved
numerically. However, its asymptotic behavior is known
analytically \cite{chandrabook,bt}. The
density of a self-gravitating isothermal
halo decreases as
\begin{equation}
\label{nd7}
\rho(r)\sim \frac{k_B T}{2\pi G m r^{2}}
\end{equation}
for $r\rightarrow +\infty$, corresponding to an accumulated mass $M(r)\sim 2k_B
Tr/(Gm)$ increasing linearly with $r$. This leads to flat rotation curves
\begin{equation}
\label{nd8}
v^2(r)=\frac{GM(r)}{r}\quad \rightarrow \quad v_{\infty}^2=\frac{2k_B T}{m}
\end{equation}
in agreement with the
observations \cite{bt}.

\subsection{Thermal core radius}

The isothermal density profile has not a compact support so it
extends to infinity. Furthermore, its total mass is infinite \cite{bt}. As a result, self-gravitating systems have no statistical equilibrium state in an
unbounded domain. In practice, the isothermal equation of state is not valid
at arbitrarily large distances and the halo is confined by other effects, either
incomplete relaxation \cite{lb,incomplete} or tidal confinement
\cite{king,clm1,clm2}. From the scaling of Eq. (\ref{nd5}) we can define a
characteristic radius
\begin{equation}
\label{nd9}
r_0=\left (\frac{k_B T}{4\pi G\rho_0 m}\right )^{1/2}
\end{equation}
that we shall call the thermal core radius.\footnote{This is the scale
radius that is used in Eq. (\ref{i3b}) to obtain the Emden equation (\ref{i5}).
We add the word
``thermal'' to avoid confusion with the quantum core radius considered later
(see also Sec. \ref{sec_tqc}).}
It represents the typical radius of an isothermal halo of
central density $\rho_0$. The halo mass $M_h$, the halo radius $r_h$, the
temperature $T$ and the circular velocity $v_h$ at
the halo radius are defined in Appendix \ref{sec_hr}. For an isothermal profile
they are given by (see Appendix \ref{sec_i})
\begin{eqnarray}
\label{nd10}
\frac{r_h}{r_0}=3.63,\qquad \frac{M_h}{\rho_0 r_h^3}=1.76,
\end{eqnarray}
\begin{eqnarray}
\label{nd11}
 \frac{k_B T}{Gm\rho_0
r_h^2}=0.954,\qquad \frac{v_h^2}{4\pi G\rho_0r_h^2}=0.140.
\end{eqnarray}
We note that the dimensionless inverse temperature has the value
\begin{eqnarray}
\label{nd12}
\eta_v=\frac{\beta GM_hm}{r_h}=1.84.
\end{eqnarray}
This is essentially a consequence of the virial theorem. Equation (\ref{nd12}) will be called the  ``virial
condition''.

The density and  circular velocity
profiles of a purely isothermal halo are plotted in Figs. 2-6 of \cite{modeldm}.
The
isothermal profile is
relatively close
to the empirical (observational) Burkert profile  \cite{observations} up to
$r/r_h=6$.

\subsection{The constant surface density}

It is an
observational evidence that the surface density of DM halos is independent of
their mass and size and has a  universal value \cite{kormendy,spano,donato}:
\begin{equation}
\label{nd13}
\Sigma_0=\rho_0r_h=141_{-52}^{+83}\, M_{\odot}/{\rm
pc}^2.
\end{equation}
This result is valid for all the galaxies even if their sizes and masses vary by
several orders of magnitude (up to $14$ orders of magnitude in luminosity). The
reason for this universality is not known but it is crucial to take this result
into account in any modeling of DM halos. Therefore, we shall assume this
relation as an empirical fact.\footnote{In Refs.
\cite{epjp,lettre,jcap,pdu,graal} we have explained this universal value,
without adjustable parameter, from a logotropic
model of DM halos.}

\subsection{Halo mass-radius relation}

Substituting the constraint (\ref{nd13}) of a uniform surface density into
Eqs. (\ref{nd10})
and (\ref{nd11}), we obtain the relations
\begin{equation}
\label{nd14}
M_h=1.76\, \Sigma_0 r_h^2,\qquad
\frac{k_B T}{m}=0.954\, G\Sigma_0r_h,
\end{equation}
\begin{equation}
\label{nd15}
v_h^2=1.76\, G\Sigma_0r_h,\qquad \rho_0=\frac{\Sigma_0}{r_h}.
\end{equation}
They determine the halo mass, the halo temperature, the halo velocity and the
halo central density as a function of the halo radius. The halo mass scales with
the size as
$M_h\propto r_h^2$ and the temperature as $k_B T/m\propto r_h$
(basically, these scalings stem from the universality of the
surface density of DM halos $M_h/r_h^2\sim \Sigma_0$ and from the virial theorem
$k_B
T/m \sim v_h^2\sim GM_h/r_h\sim G\Sigma_0 r_h$).

\begin{figure}
\begin{center}
\includegraphics[clip,scale=0.3]{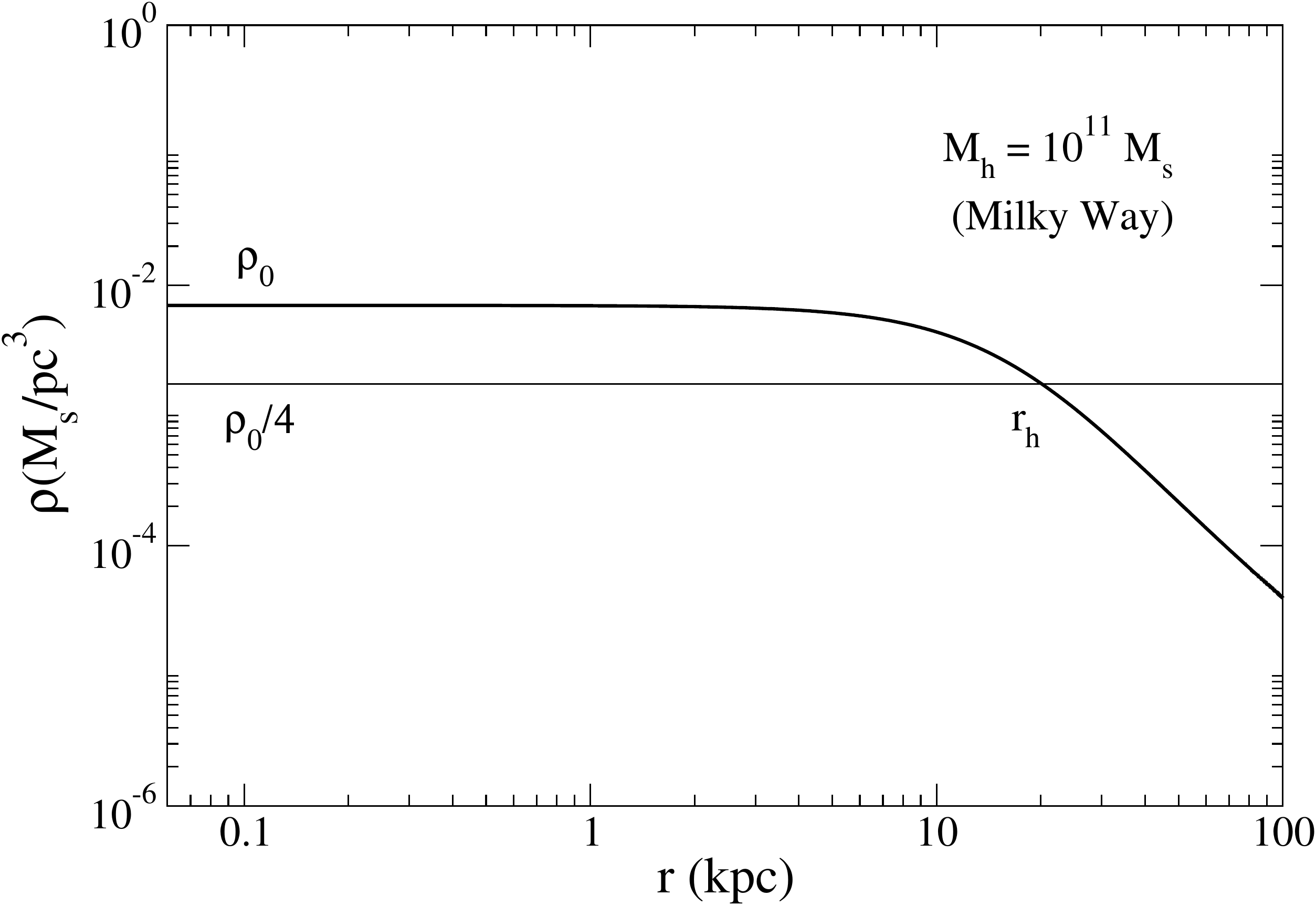}
\caption{Density profile
of a classical isothermal DM halo of mass $M_h=10^{11}\,
M_{\odot}$ (Milky Way).}
\label{profileisotherme}
\end{center}
\end{figure}

\begin{figure}
\begin{center}
\includegraphics[clip,scale=0.3]{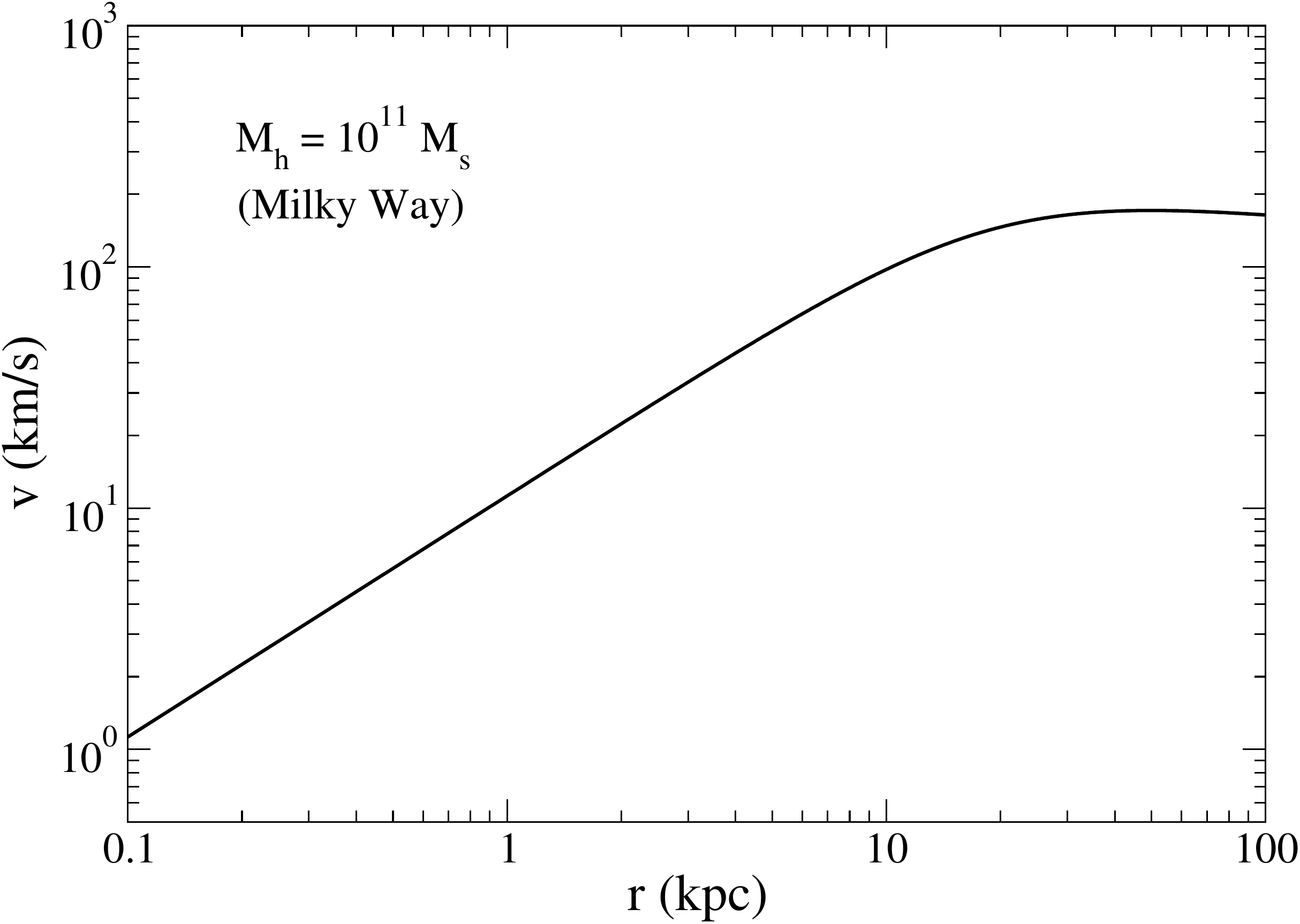}
\caption{Rotation curve of a classical
isothermal DM halo of mass $M_h=10^{11}\, M_{\odot}$ (Milky Way).}
\label{Viso}
\end{center}
\end{figure}

For
a halo of mass $M_h=10^{11}\, M_\odot$, similar to the halo
that surrounds our Galaxy, we find
$r_h=20.1\, {\rm kpc}$, $\rho_0=7.02\times 10^{-3}M_{\odot}/{\rm
pc}^3$, $(k_B T/m)^{1/2}=108\, {\rm km/s}$, and
$v_h=(GM_h/r_h)^{1/2}=146\,  {\rm km/s}$ (we also have $v_\infty=153\, {\rm
km/s}$).  We stress that these results
are independent of the characteristics of the DM particle. The
corresponding density and velocity profiles are plotted in Figs.
\ref{profileisotherme} and \ref{Viso}.

{\it Remark:} The (effective) temperature of the DM halos
depends on the fermion mass $m$ and on the halo mass $M_h$ through the
law $k_B T=0.719\,Gm\sqrt{\Sigma_0 M_h}$. Let us consider a halo mass
$M_h=10^{11}\,
M_\odot$ as before. For $m=165\, {\rm eV}/c^2$ (see Sec. \ref{sec_mh}), we get
$T=0.247\, {\rm K}$. For $m=1\, {\rm keV}/c^2$ (see Sec. \ref{sec_mh}), we get 
$T=1.50\, {\rm K}$. For $m=54.6\, {\rm keV}/c^2$ (see Sec. \ref{sec_smbh}), we
get
$T=81.9\, {\rm K}$. These values, which are of the order of the Kelvin scale,
are much more physical that those obtained in the case of bosonic DM, which are
of the order of $T\sim 10^{-25}\, {\rm K}$ \cite{modeldm}.

\subsection{Classical isothermal gas in a box}
\label{sec_cigb}

Let us finally derive the equations determining the caloric curve of a self-gravitating isothermal gas in a box.

\begin{figure}
\begin{center}
\includegraphics[clip,scale=0.3]{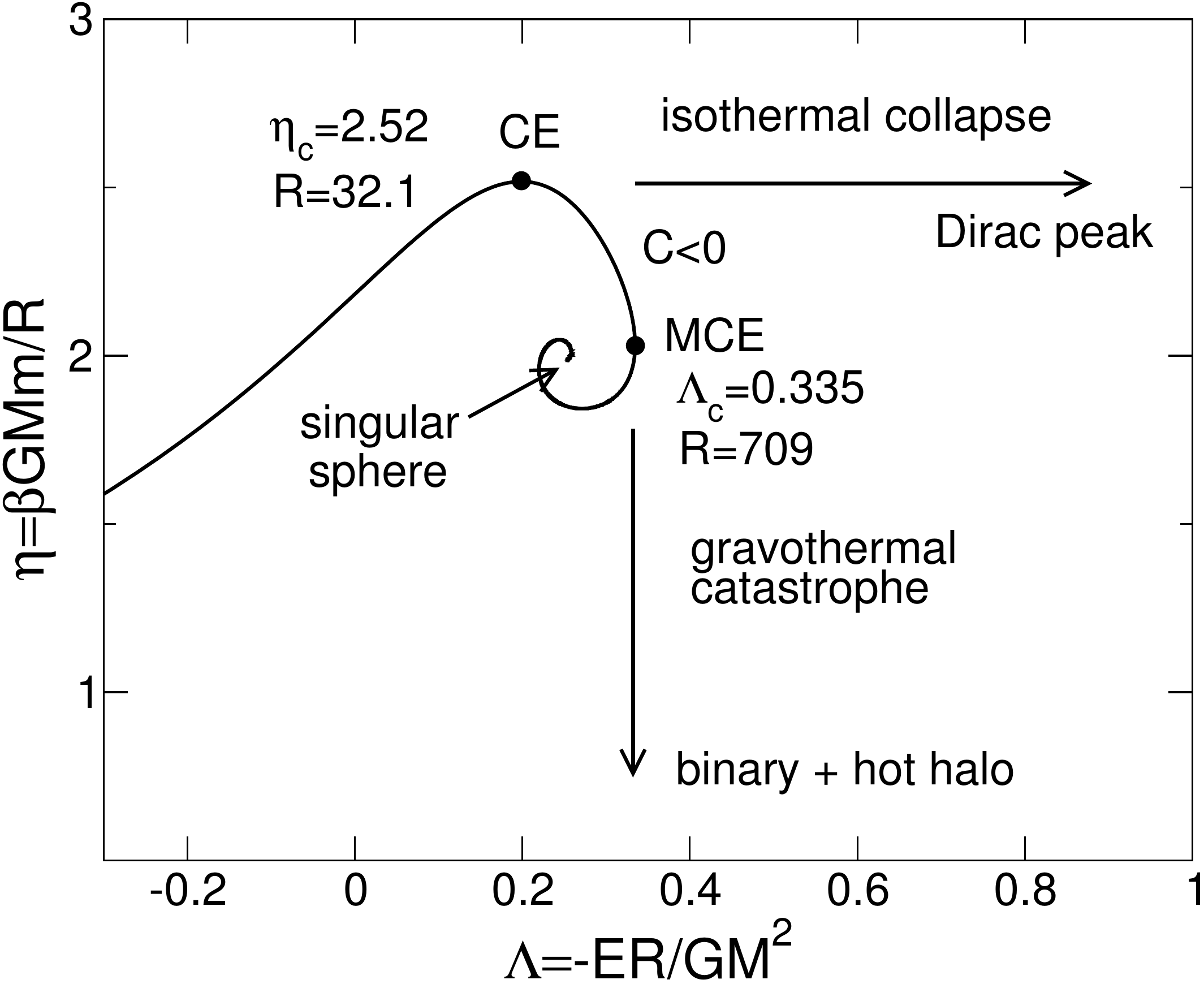}
\caption{Caloric curve of the classical self-gravitating
gas.}
\label{etalambda}
\end{center}
\end{figure}

The density profile (\ref{nd2}) can be written as
\begin{equation}
\rho(r)=\rho_{0}e^{-\beta m(\Phi(r)-\Phi_0)},
\label{a70}
\end{equation}
where $\rho_{0}$ is the central density and $\Phi_0$ is the central potential.
The Boltzmann-Poisson equation (\ref{nd6}) then becomes
\begin{equation}
\Delta\Phi=4\pi G \rho_{0}e^{-\beta m(\Phi-\Phi_{0})}.
\label{a71}
\end{equation}
If we assume that the system is spherically symmetric and introduce the
dimensionless variables
\begin{equation}
\rho=\rho_{0}e^{-\psi(\xi)},\qquad \psi=\beta m(\Phi-\Phi_{0}),
\label{a73}
\end{equation}
and
\begin{equation}
 \xi=(4\pi
G\beta m\rho_{0})^{1/2}r
\label{a73b}
\end{equation}
into Eq. (\ref{a71}), we obtain the Emden equation (\ref{i5}). If we denote by
$\alpha$ the value of $\xi$ at the edge of the box, we
have
\begin{equation}
\alpha=(4\pi G\beta m\rho_{0})^{1/2}R\qquad {\rm and}\qquad
\xi=\alpha\frac{r}{R}.
\label{a75}
\end{equation}
Introducing the inverse normalized temperature from Eq. (\ref{tf24}) and
applying Newton's law (\ref{tf25}) at $r=R$, we get
\begin{equation}
\eta=\alpha\psi'(\alpha).
\label{a77}
\end{equation}
To compute the energy, we proceed as follows. The kinetic energy of an
isothermal gas is
\begin{equation}
E_{\rm kin}=\frac{3}{2}Nk_B T.
\label{a77b}
\end{equation}
Using the virial theorem from Eq. (\ref{tf32}) we can  compute the gravitational
energy as
\begin{equation}
W=-2E_{\rm kin}+3P_{b}V=-3Nk_B T+\frac{4\pi
R^3\rho(R)k_B T}{m}.
\label{a79}
\end{equation}
The total energy is $E=E_{\rm kin}+W$. Introducing the normalized
energy from Eq. (\ref{tf34}) we obtain
\begin{equation}
\Lambda=\frac{3}{2\alpha\psi'(\alpha)}-\frac{e^{-\psi(\alpha)}}{\psi'(\alpha)^2}
.
\label{a82}
\end{equation}
The expression of the entropy is derived in Appendix \ref{sec_ent}. The caloric
curve $\eta(\Lambda)$ of the classical self-gravitating gas
\cite{antonov,lbw,katzpoincare1} is represented in Fig. \ref{etalambda}. It has
the form of a spiral (see Sec. \ref{sec_ineq}). It leads to an isothermal
collapse \cite{aaiso} in the canonical ensemble above $\eta_c=2.52$ and to a
gravothermal catastrophe \cite{lbw} in the microcanoncal ensemble above
$\Lambda_c=0.335$.

{\it Remark:} The equations of this section can be recovered from the general
equations of Sec. \ref{sec_sgfgb} by taking the
nondegenerate limit $k\rightarrow +\infty$ and replacing the Fermi integrals by
their asymptotic
expressions
\begin{equation}
I_{n}(t)\sim \frac{1}{t}\Gamma(n+1), \qquad (t\rightarrow +\infty).
\label{d9nd}
\end{equation}

\section{Completely degenerate limit: minimum halo (ground state) and quantum core of DM halos}
\label{sec_cdl}

We now consider the completely degenerate limit of the self-gravitating Fermi
gas which describes (i) ultracompact dwarf spheroidals (dSphs) 
like Fornax or Willman I and (ii) the quantum core of bigger DM
halos.

\subsection{Polytropic equation of state}

In the completely degenerate limit $T=0$ (or $T\ll T_F$), the Fermi-Dirac
DF (\ref{tf6})
reduces to the step function
\begin{eqnarray}
\label{cdl1}
f({\bf r},{\bf
v})=\eta_0\qquad v\le v_F({\bf r}),\nonumber\\
f({\bf r},{\bf
v})=0\qquad v\ge v_F({\bf r}),
\end{eqnarray}
where
\begin{equation}
\label{cdl2}
v_F({\bf r})=\sqrt{2\left \lbrack \frac{\mu}{m}-\Phi({\bf r})\right\rbrack}
\end{equation}
is the Fermi velocity.\footnote{The Fermi energy is $\epsilon_F({\bf
r})=\frac{1}{2}mv_F^{2}({\bf r})=\mu-m\Phi({\bf r})$. It is equal
to the local chemical potential $\mu({\bf r})=\mu-m\Phi({\bf r})$.} The
density and the pressure are explicitly given by
\begin{equation}
\rho=\int
f\, d{\bf v}=\int_{0}^{v_{F}}\eta_0 4\pi v^{2}\, dv=\frac{4\pi}{3}\eta_0
v_{F}^3({\bf r}),
\label{cdl3}
\end{equation}
\begin{equation}
P=\frac{1}{3}\int
f v^2\, d{\bf v}=\frac{1}{3}\int_{0}^{v_{F}}\eta_0 v^2 4\pi
v^{2}\,
dv=\frac{4\pi}{15}\eta_0 v_{F}^5({\bf r}).
\label{cdl4}
\end{equation}
Eliminating the Fermi velocity between these two expressions, we find that  the
equation of state of a cold Fermi gas  is
\cite{chandrabook}
\begin{eqnarray}
\label{cdl5}
P=\frac{1}{20}\left (\frac{3}{\pi}\right )^{2/3}\frac{h^2}{m^{8/3}}\rho^{5/3}.
\end{eqnarray}
This is  a polytropic equation of state $P=K_1 \rho^{5/3}$ of index $\gamma=5/3$
(i.e. $n=3/2$).
The fundamental differential equation of hydrostatic equilibrium
determining the density profile of a fermion ball at
$T=0$ with the equation of state (\ref{cdl5}) reads [see Eq.
(\ref{tf13})]
\begin{equation}
\label{cdl6}
\frac{1}{8}\left (\frac{3}{\pi}\right
)^{2/3}\frac{h^2}{m^{8/3}}\Delta\rho^{2/3}=-4\pi G\rho.
\end{equation}
It describes the balance between the gravitational attraction and the quantum
pressure. It is equivalent to the Thomas-Fermi equation
\begin{equation}
\label{cdl7}
\Delta\Phi=\frac{16}{3}\pi^2G\eta_0\left\lbrack
2\left(\frac{\mu}{m}-\Phi\right )\right\rbrack^{3/2}
\end{equation}
obtained by combining Eqs. (\ref{tf12}) and (\ref{cdl3}) with Eq. (\ref{cdl2}).
Equations (\ref{cdl6}) and (\ref{cdl7}) can be reduced to the Lane-Emden
equation (\ref{p5}) of index $n=3/2$.

\subsection{Mass-radius relation}

The density profile of a fermion
ball at $T=0$ (corresponding to a polytrope of index $n=3/2$)
has a compact support: the density vanishes at a finite distance
$r=R$ representing the radius of the object (see Fig. \ref{profilepolytrope}
below). The
radius, the mass and the central density of the object satisfy the relations
(see Appendix \ref{sec_p})
\begin{equation}
\label{cdl8}
R=0.359\, \frac{h}{m^{4/3}G^{1/2}\rho_0^{1/6}},
\end{equation}
\begin{equation}
\label{cdl9}
M=0.699\, \rho_0 R^3,
\end{equation}
\begin{equation}
\label{cdl10}
MR^3=0.00149\, \frac{h^6}{G^3 m^8}.
\end{equation}
Similarly, the halo mass, the halo radius and the central density satisfy the
relations (see
Appendix \ref{sec_p})
\begin{equation}
\label{cdl11}
r_h=  0.223   \, \frac{h}{m^{4/3}G^{1/2}\rho_0^{1/6}},
\end{equation}
\begin{equation}
\label{cdl12}
M_h=1.99\, \rho_0 r_h^3,
\end{equation}
\begin{equation}
\label{cdl13}
M_hr_h^3=2.45\times 10^{-4}\, \frac{h^6}{G^3 m^8}.
\end{equation}
Therefore
\begin{equation}
\frac{M}{M_h}=1.46,\qquad \frac{R}{r_h}=1.61,
\label{qwe}
\end{equation}
yielding $GM/R=0.907\, GM_h/r_h$.

\subsection{Minimum halo}
\label{sec_mh}

\begin{figure}
\begin{center}
\includegraphics[clip,scale=0.3]{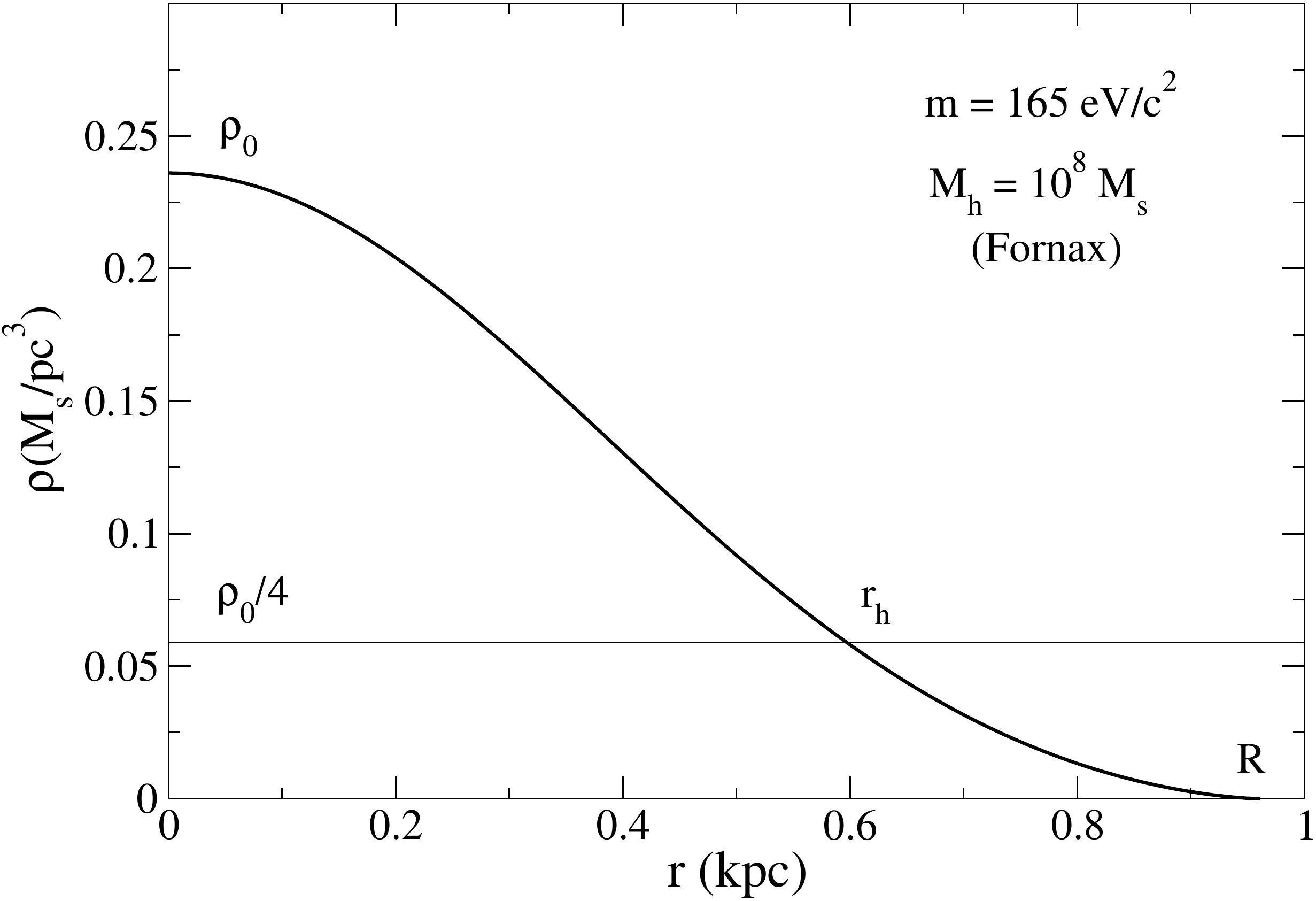}
\caption{Density profile
of a completely degenerate DM halo of mass $M_h=10^{8}\,
M_{\odot}$ (Fornax).}
\label{profilepolytrope}
\end{center}
\end{figure}

\begin{figure}
\begin{center}
\includegraphics[clip,scale=0.3]{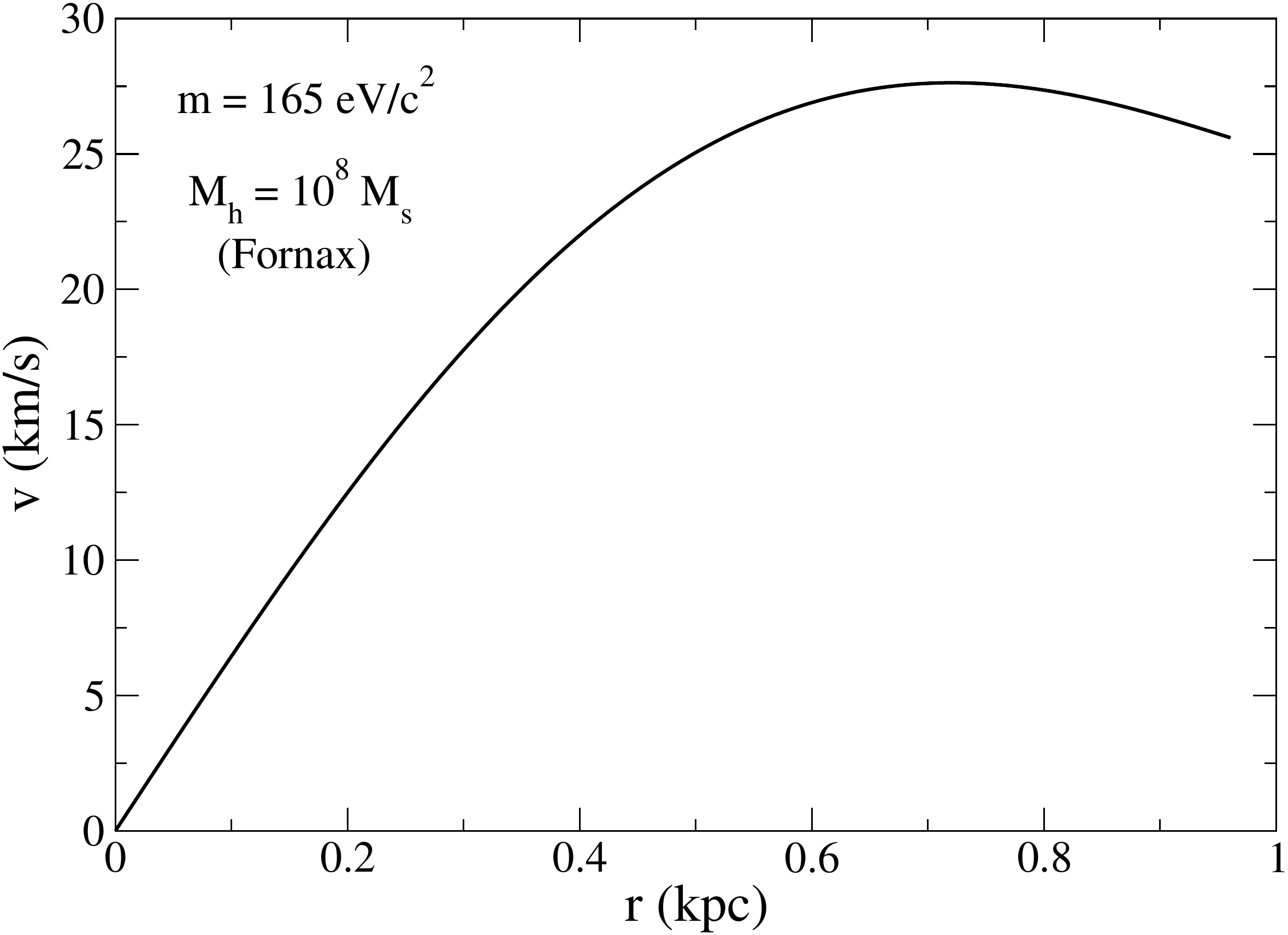}
\caption{Rotation curve of a
completely
degenerate DM halo of mass $M_h=10^{8}\, M_{\odot}$ (Fornax).}
\label{Vpoly}
\end{center}
\end{figure}

The foregoing equations determine the ground state  ($T=0$) of the
self-gravitating Fermi gas.  This fermion ball corresponds either to the
smallest and most
compact DM halo of the Universe (which has no isothermal atmosphere) that we
call the ``minimum
halo'', or to the
quantum cores of larger DM halos (which are surrounded by an isothermal
atmosphere) \cite{mcmh}. We consider here the first possibility (minimum halo).
Using Eqs.
(\ref{cdl11})-(\ref{cdl13}) and the constraint from Eq.
(\ref{nd13}), we
obtain \cite{mcmh}
\begin{equation}
\label{cdl14}
(r_{h})_{\rm min}=1.50\, \left (\frac{\hbar^6}{G^3m^8\Sigma_0}\right
)^{1/5},
\end{equation}
\begin{equation}
\label{cdl15}
(M_{h})_{\rm min}=4.47\, \left (\frac{\hbar^{12}\Sigma_0^3}{G^6m^{16}}\right
)^{1/5},
\end{equation}
\begin{equation}
\label{cdl16}
(\rho_{0})_{\rm max}=0.667\, \left (\frac{\Sigma_0
m^{4/3}G^{1/2}}{\hbar}\right
)^{6/5},
\end{equation}
\begin{equation}
\label{cdl16w}
(v_h^2)_{\rm min}=2.98\, \left (\frac{\Sigma_0^4
G^{2}\hbar^6}{m^8}\right
)^{1/5}.
\end{equation}
These
equations determine the radius, the mass, and the central density of the
minimum halo as a function of the fermion mass $m$ and the universal
surface density $\Sigma_0$. In practice, they are used the other way
round in order to determine the fermion mass $m$. Assuming that the mass
$(M_h)_{\rm min}$ of the minimum halo is known, we obtain the fermion mass under
the form 
\begin{equation}
\label{cdl17}
m=1.60\, \frac{\hbar^{3/4}\Sigma_0^{3/16}}{G^{3/8}(M_h)_{\rm min}^{5/16}}.
\end{equation}
If we take $(M_h)_{\rm min}=10^8\, M_{\odot}$ we find that $m=165 \, {\rm
eV}/c^2$. We then obtain $(r_{h})_{\rm min}=597\, {\rm pc}$, $(\rho_{0})_{\rm
max}=0.236\, M_{\odot}/{\rm pc}^3$, and $(v_h)_{\rm min}=26.8\, {\rm km/s}$.
Using Eq. (\ref{qwe}), we also have $M_{\rm
min}=1.46\times 10^8\,
M_{\odot}$ and $R_{\rm min}=961\, {\rm pc}$. The
corresponding density and velocity profiles are plotted in Figs.
\ref{profilepolytrope} and \ref{Vpoly}.

The Fermi temperature is defined by 
\begin{equation}
k_B T_{\rm F}=\frac{\hbar^2\rho^{2/3}}{m^{5/3}}.
\end{equation}
It can be obtained qualitatively by equating Eqs.
(\ref{nd4}) and (\ref{cdl5}). For the minimum halo, using Eq.
(\ref{cdl16}), it reads
\begin{equation}
k_B T_{\rm F}=\left (\frac{\hbar^6\Sigma_0^4G^2}{m^{3}}\right )^{1/5}.
\end{equation}
For $m=165 \, {\rm eV}/c^2$, we get $T_F=5.15\times 10^{-3}\, {\rm K}$. We note
that the minimum halo is determined by the condition
$T\sim T_F$, where $T$ is the temperature given by Eq. (\ref{nd14}).

{\it Remark:} The choice of the mass $(M_h)_{\rm min}=10^8\, M_{\odot}$
(Fornax) for the minimum halo is a little bit arbitrary and open to criticisms.
We shall adopt this value, however, in order to be consistent with our other
papers \cite{abrilphas,modeldm,mcmh,mrjeans}. Nevertheless, our model is
perfectible. If a more relevant minimum halo mass is considered, our analytical
results remain valid but the numerical applications must be reconsidered. For
example, if we take a minimum halo mass $(M_h)_{\rm min}=0.39\times 10^6\,
M_{\odot}$ corresponding to Willman I (as in Refs.
\cite{dvs1,dvs2,vss,vsedd,vs2,vsbh} and Refs. \cite{clm2,abrilph}) we
obtain $m\sim 1\, {\rm
keV}/c^2$. We then obtain $(r_{h})_{\rm min}\sim 33\, {\rm pc}$ (in very good
agreement with the measured value reported in
Ref. \cite{dvs1}),
$(\rho_{0})_{\rm
max}\sim 4.3\, M_{\odot}/{\rm pc}^3$, and $(v_h)_{\rm min}=6.35\, {\rm km/s}$.
It is also possible that the concept of a
``minimum halo'' which would be completely degenerate is wrong (see
Sec. \ref{sec_smbh}). In that case, our general results remain valid but the
fermion mass cannot be obtained from the considerations of this section.

\subsection{Quantum core of DM halos}

The relations from Eqs. (\ref{cdl8})-(\ref{cdl17}) apply to the minimum halo
which is a pure fermion ball without isothermal atmosphere. The relations from
Eqs. (\ref{cdl8})-(\ref{cdl10}) also apply to the quantum core of larger DM
halos. If we normalize the core mass $M_c$ by the minimum halo mass $(M_h)_{\rm min}$ and the core
radius
$R_c$ by the minimum halo radius $(r_h)_{\rm min}$, we get
\begin{equation}
\label{cdl18}
\frac{M_c}{(M_h)_{\rm min}}\left \lbrack \frac{R_c}{(r_{h})_{\rm min}}\right
\rbrack^3=6.09.
\end{equation}
We can check that Eq. (\ref{cdl18}) is verified for  the minimum
halo for which $M_c=1.46\, (M_h)_{\rm min}$ and $R_c=1.61\,
(r_{h})_{\rm min}$ [see Eq. (\ref{qwe})].

\subsection{Maximum mass due to general relativity}
\label{sec_ov}

The maximum mass and the minimum stable radius of a fermion ball at $T=0$ set by
general
relativity  are
\begin{equation}
M_{\rm OV}=0.384\, \left (\frac{\hbar
c}{G}\right )^{3/2}\frac{1}{m^2},\quad R_{\rm OV}=8.73 \,
\frac{GM_{\rm max}}{c^2}.
\label{cdl19}
\end{equation}
 They were first  determined by Oppenheimer and Volkoff
\cite{ov} in the context of neutron stars.
For a fermion of mass $m=165\, {\rm eV}/c^2$, we
obtain $M_{\rm OV}=2.30\times
10^{13}\,
M_{\odot}$
and $R_{\rm OV}=9.61\, {\rm pc}$. For $m=1\, {\rm keV}/c^2$, we get $M_{\rm
OV}=6.26\times
10^{11}\,
M_{\odot}$
and $R_{\rm OV}=0.262\, {\rm pc}$.  
The maximum mass is much larger than the
typical core mass of any DM halo. Assuming that a fermion ball at $T=0$
describes the quantum core of a DM halo, we conclude that such cores are
nonrelativistic since $M_c\ll M_{\rm max}$ in general. Since the maximum mass
is much larger than the core mass, gravity can
be treated within a Newtonian framework.\footnote{This statement is valid for
the fermion mass $m=165\, {\rm eV}/c^2$ that we consider here. It is less
valid for larger masses, of the order of $m=48\, {\rm keV}/c^2$, such as those
considered in Sec. \ref{sec_smbh}.}

\subsection{Energy of a fermion ball}

Let us finally derive the energy of a completely degenerate fermion ball (ground state)

A nonrelativistic fermion ball at $T=0$ is equivalent to a
polytrope of index $n=3/2$ [see Eq. (\ref{cdl5})]. Its kinetic
energy is given by [see Eq. (\ref{tf30})]
\begin{equation}
\label{bx6}
E_{\rm kin}=\frac{3}{2}\int P\, d{\bf r}=\frac{3}{2}K_1\int
\rho^{5/3}\, d{\bf r}.
\end{equation}
It gravitational energy is given by the
Betti-Ritter formula \cite{chandrabook}
\begin{equation}
\label{pn12b}
W=-\frac{6}{7}\frac{GM^2}{R}
\end{equation}
with the mass-radius relation from Eq. (\ref{cdl10}).
The virial theorem reduces to [see Eq. (\ref{tf32})]
\begin{equation}
\label{vth}
2E_{\rm kin}+W=0
\end{equation}
since there is no pressure on the boundary of the fermion ball.
Therefore, the total energy $E=E_{\rm kin}+W$ of the fermion ball is
\begin{equation}
\label{vtha}
E=-E_{\rm kin}=\frac{W}{2}=-\frac{3}{7}\frac{GM^2}{R}.
\end{equation}
This is the minimum energy $E_{\rm min}$ of the self-gravitating Fermi gas
(ground state). Using the mass-radius relation (\ref{cdl10}), we obtain
\begin{equation}
\label{a11}
E_{\rm min}=-3.750\frac{G^2m^{8/3}}{h^2}M^{7/3}.
\end{equation}
Considering the box model of Sec. \ref{sec_sgfgb} and introducing the normalized
energy from Eq. (\ref{tf34}) we get
\begin{equation}
\label{a11b}
\Lambda_{\rm max}=0.0642\, \mu^{2/3}.
\end{equation}

{\it Remark:} The equations of this section can be recovered from the general
equations of Sec. \ref{sec_sgfgb} by taking the completely degenerate limit
$k\rightarrow 0$ and replacing the Fermi integrals by their asymptotic
expressions
\begin{equation}
I_{n}(t)\sim {(-\ln t)^{n+1}\over n+1}, \qquad (t\rightarrow 0).
\label{d9}
\end{equation}

\section{Partially degenerate DM halos: core-halo structure}
\label{sec_app}

To study partially degenerate DM halos with a  core-halo structure, we shall
use the box model of Sec. \ref{sec_sgfgb}. In order to apply this model to
real DM halos,
we identify the mass $M$ with the halo mass $M_h$ and the box radius $R$ with
the halo radius $r_h$:
\begin{eqnarray}
\label{conn1}
M=M_h,\qquad R=r_h.
\end{eqnarray}
We first have to determine the relation between the degeneracy parameter $\mu$
and the halo mass $M_h$.

\subsection{Relation between the degeneracy parameter $\mu$ and the halo mass
$M_h$}
\label{sec_mu}

Sufficiently large DM halos are dominated by their classical isothermal
envelope (the quantum core of mass $M_c\ll M_h$ does not affect
their external structure). As a result, the halo mass is
related to the halo radius by [see Eq. (\ref{nd14})]
\begin{eqnarray}
\label{mu1}
M_h=1.76\, \Sigma_0 r_h^2.
\end{eqnarray}
Using this relation, the degeneracy parameter defined by Eq. (\ref{tf29}) can be
written
as
\begin{equation}
\mu=1.18\, \frac{G^{3/2}m^4M_h^{5/4}}{\hbar^3\Sigma_0^{3/4}}.
\label{mu2}
\end{equation}
Normalizing the halo mass $M_h$ by the minimum halo mass $(M_{h})_{\rm min}$
given by Eq. (\ref{cdl15}) we obtain
\begin{equation}
\mu=7.66\, \left \lbrack\frac{M_h}{(M_{h})_{\rm min}}\right \rbrack^{5/4}.
\label{mu3}
\end{equation}
This equation relates the degeneracy parameter $\mu$ to the halo mass $M_h$. As
a result, the canonical and microcanonical critical points given by Eq.
(\ref{cpoi}) may be expressed in terms of the halo mass by using $(M_h)_{\rm
CCP}=(\mu_{\rm CCP}/7.66)^{4/5}(M_h)_{\rm min}$ and $(M_h)_{\rm MCP}=(\mu_{\rm
MCP}/7.66)^{4/5}(M_h)_{\rm min}$ yielding
\begin{equation}
(M_{h})_{\rm CCP}=6.73\, (M_h)_{\rm min}=30.1\, \left
(\frac{\hbar^{12}\Sigma_0^3}{G^6m^{16}}\right
)^{1/5},
\label{mu4exp}
\end{equation}
\begin{equation}
(M_{h})_{\rm MCP}=108\, (M_h)_{\rm min}=483\, \left
(\frac{\hbar^{12}\Sigma_0^3}{G^6m^{16}}\right
)^{1/5}.
\label{mu4expb}
\end{equation}
Taking $(M_{h})_{\rm min}=10^8\, M_{\odot}$
(Fornax) corresponding to a fermion mass $m=165\, {\rm eV}/c^2$ we get
\begin{equation}
(M_{h})_{\rm CCP}=6.73\times 10^8\, M_{\odot},\quad (M_{h})_{\rm
MCP}=1.08\times 10^{10}\, M_{\odot}.
\label{mu4}
\end{equation}
If we take $(M_h)_{\rm
min}=0.39\times 10^6\, M_{\odot}$ (Willman I) instead,
corresponding to a fermion mass $m\sim 1\, {\rm keV}/c^2$, we
get $(M_{h})_{\rm CCP}=2.62\times 10^6\, M_{\odot}$ and $(M_{h})_{\rm
MCP}=4.21\times 10^{7}\, M_{\odot}$. We note that $(M_h)_{\rm CCP}$ and
$(M_h)_{\rm MCP}$ are very sensitive to the value
of
$m$ since it occurs in their expressions [see Eqs. (\ref{mu4exp}) and
(\ref{mu4expb})] with the power $16/5$.

{\it Remark:} We
can also write the degeneracy parameter $\mu$ under the form \cite{ijmpb}
\begin{equation}
\mu=17.3\, \left \lbrack\frac{R}{R_F(M)}\right \rbrack^{3/2},
\label{mu5}
\end{equation}
where $R_F(M)$ is the (Fermi) radius of a completely degenerate fermion
ball of mass $M$ given by Eq.
(\ref{cdl10}). The condition that $R>R_c$ imposes
$\mu>\mu_{\rm min}=17.3$. Applying this inequality to real DM halos, using Eq.
(\ref{mu3}), we find that  $M_h>1.92\, (M_{h})_{\rm min}$. Up to a
factor of order unity, we recover the fact that the ground state of the
self-gravitating
Fermi gas ($T=0$) determines the existence of a minimum halo mass $(M_{h})_{\rm
min}$.

\subsection{Virial condition}
\label{sec_vc}

We have seen in Sec. \ref{sec_nd} that the normalized inverse temperature of
an isothermal DM halo is $\eta_v=1.84$.\footnote{Coincidentally,
this value turns out to be very close to the value $\eta_2=1.84$ corresponding
to the
minimum inverse temperature of the classical isothermal spiral (see
Sec. \ref{sec_cigb}).}
Therefore, if we want to make the connection
between the box model and real DM halos, we should consider a value of  $\eta$
equal to $1.84$. It is reassuring to note that $\eta_v=1.84$ is smaller than
$\eta_c\simeq 2.52$, corresponding to the maximum inverse temperature of the
classical isothermal
spiral, implying that there always exists a gaseous (non degenerate) equilibrium state with
$\eta_v=1.84$. Actually, we should not give too much importance on the precise
value of
$\eta_v$. It is sufficient to consider that $\eta_v$
is of the order of unity.  Therefore, we shall take
\begin{equation}
\eta_v\sim 1.
\label{vc1}
\end{equation}
The intersections between the series of equilibria $\eta(\Lambda)$
and the line level $\eta=\eta_v \sim 1$ determine the possible equilibrium
states of our system of self-gravitating fermions. We can generically have three
kinds of solutions: (i) a gaseous solution
(G) which is purely isothermal without quantum core; (ii) a
core-halo solution (CH) with a quantum core (fermion ball) surrounded by a
classical isothermal
atmosphere; (iii) a condensed solution (C) which is an essentially  quantum
object with a tenuous isothermal atmosphere.  The gaseous
solution has been discussed in Sec. \ref{sec_nd}.  The condensed solution is not
physical in the case of large DM halos because it would imply that the halo is
completely condensed, which is not the case. 
This solution only applies to the minimim halo (see Sec. \ref{sec_cdl}). The
core-halo solution is the most important one for our purposes. It is similar to
the gaseous solution at sufficiently large distances but it contains a small
nucleus (fermion ball) at its center. An important question is to determine the
core mass $M_c$ as a function of the halo mass $M_h$.

\subsection{The $M_c-M_h$ relation for the CH solution}
\label{sec_mcmh}

\begin{figure}
\begin{center}
\includegraphics[clip,scale=0.3]{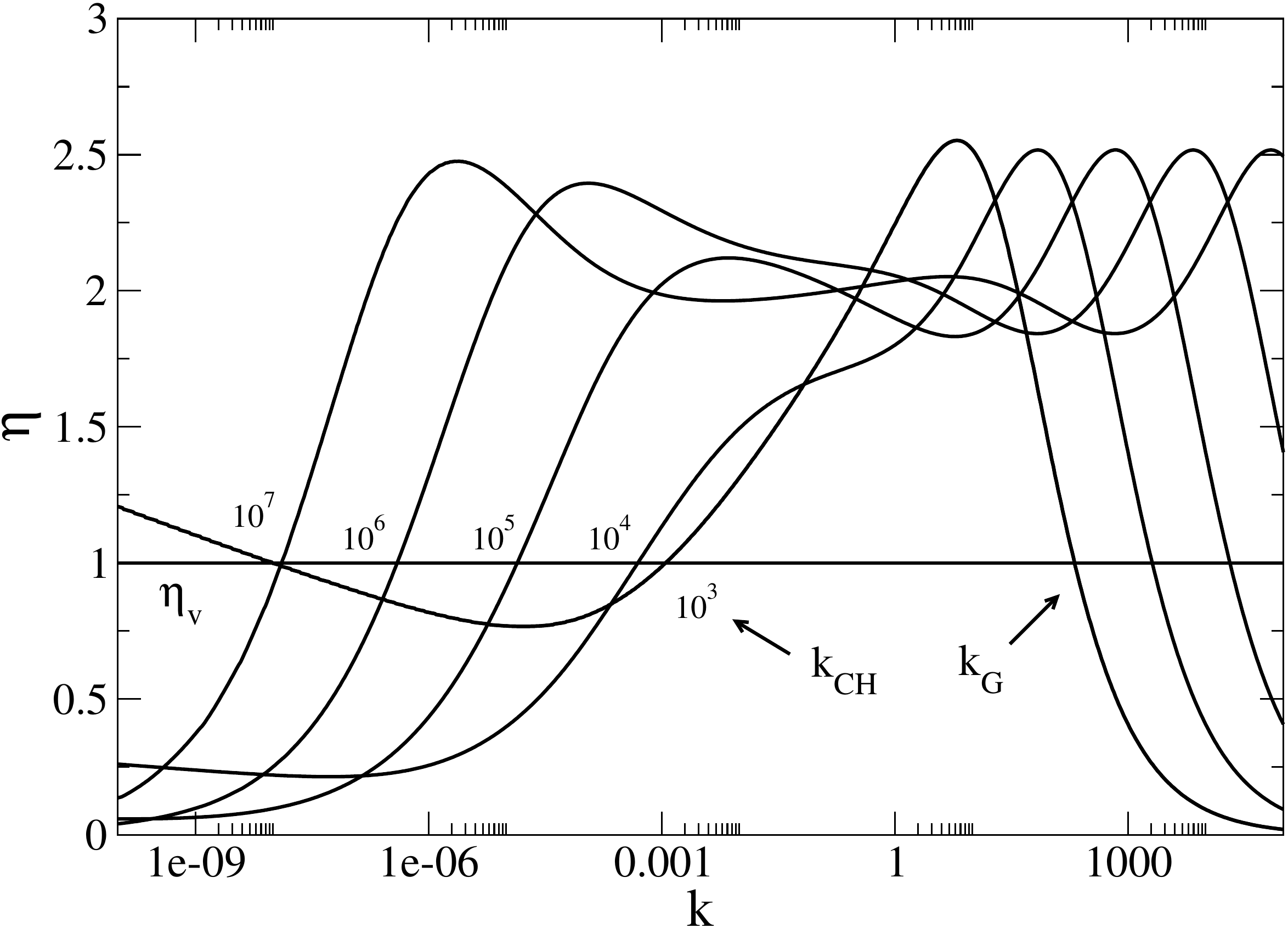}
\caption{Inverse temperature $\eta$ as a
function of $k$ for
different values of $\mu=10^3-10^7$.  The intersection between
the curve $\eta(k)$ and the line $\eta=\eta_v=1$ selects
three solutions (G), (CH) and (C). The core-halo solution (CH) is the one close
to which we have written the value of $\mu$. }
\label{keta}
\end{center}
\end{figure}

The core mass-halo mass relation  $M_c(M_h)$ can be obtained from the box model
as follows. The halo mass $M_h$ determines the value of the degeneracy parameter
$\mu$.
We can then plot the series of equilibria $\eta(\Lambda)$ parametrized
by the concentration parameter $k$ (see Figs. \ref{multimu2PH}, \ref{fel}
and \ref{le5} below). For $\mu>\mu_{\rm CCP}$, the
intersections between the caloric curve $\eta(\Lambda)$ and the virial
condition $\eta=\eta_v\sim 1$ determine three solutions (G), (CH) and (C) (see
Fig. \ref{keta} for an illustration). We
select the core-halo solution (CH) and compute the corresponding concentration
parameter $k=k_{\rm
CH}(\mu,\eta_v)$. The density profile of the core-halo solution is then given by
Eq. (\ref{tf18}). Its central density is
\begin{equation}
\rho_0=\frac{4\pi \sqrt{2}\eta_0 }{(\beta m)^{3/2}}I_{1/2}(k).
\label{mcmh1}
\end{equation}
Introducing the dimensionless variables defined in Sec. \ref{sec_tf}, we get
\begin{equation}
\frac{4\pi \rho_0 R^3}{M}=\frac{\mu}{\eta^{3/2}}I_{1/2}(k).
\label{mcmh2}
\end{equation}
Now, the core-halo configuration can be decomposed into a  fermion ball at
$T=0$ and a classical isothermal atmosphere. For a completely degenerate fermion
ball,
representing
the quantum core of the DM halo, the relation between the core mass
and the central density is given by [see Eqs. (\ref{cdl8})-(\ref{cdl10})]
\begin{equation}
M_c=8.01\, \frac{\hbar^3\rho_0^{1/2}}{G^{3/2}m^4}.
\label{mcmh3}
\end{equation}
Combining Eqs. (\ref{mcmh2}) and (\ref{mcmh3}) we obtain
\begin{equation}
\frac{M_c}{M}=4.07\, \frac{\sqrt{I_{1/2}(k)}}{\mu^{1/2}\eta^{3/4}}.
\label{mcmh4}
\end{equation}
Recalling that
$M=M_h$, $\eta=\eta_v\sim 1$ and $k=k_{\rm
CH}(\mu,\eta_v)$ we get
\begin{equation}
\frac{M_c}{M_h}=4.07\, \frac{\sqrt{I_{1/2}\lbrack k_{\rm
CH}(\mu,\eta_v)\rbrack}}{\mu^{1/2}\eta_v^{3/4}}.
\label{mcmh5}
\end{equation}
Together with the relation (\ref{mu3}) between $\mu$ and $M_h$, Eq.
(\ref{mcmh5}) determines  the
core
mass-halo mass relation $M_c(M_h)$. In principle, the function $I_{1/2}\lbrack
k_{\rm
CH}(\mu,\eta_v)\rbrack$ has to be determined numerically as a function of $\mu$.
However,  it turns out  that, for the
CH solution,   $I_{1/2}\lbrack k_{\rm CH}(\mu,\eta_v)\rbrack$ changes slowly
(logarithmically) with $\mu$. As a result, up to logarithmic corrections (see
Sec. \ref{sec_lc}), it can
be taken as a constant. Therefore, we obtain the scaling
\begin{equation}
\frac{M_c}{M_h}\propto \frac{1}{\mu^{1/2}}.
\label{mcmh5b}
\end{equation}
More precisely,  recalling Eq. (\ref{mu3}), Eq. (\ref{mcmh5}) can be written as
\begin{equation}
\frac{M_c}{(M_{h})_{\rm min}}=1.47\,
\frac{\sqrt{I_{1/2}[k_{\rm CH}(\mu,\eta_v)]}}{\eta_v^{3/4}} \left
\lbrack \frac{M_h}{(M_{h})_{\rm min}}\right
\rbrack^{3/8}.
\label{mcmh6}
\end{equation}
If we take $\eta_v=1$ and $I_{1/2}(k_{\rm CH})=1$ we get 
\begin{equation}
\frac{M_c}{(M_{h})_{\rm min}}=1.47\left \lbrack\frac{M_h}{(M_{h})_{\rm
min}}\right
\rbrack^{3/8}.
\label{mcmh7}
\end{equation}
This relation shows that the core mass $M_c$ scales with the halo mass as
$M_h^{3/8}$. 
The prefactor, obtained from our model, is of order unity implying
that $M_c\sim M_h$ for the minimum halo, as expected. Actually, 
for $M_h=(M_h)_{\rm min}$ we get  $M_c=1.47\, (M_h)_{\rm min}$ in very good
agreement with Eq. (\ref{qwe}). Returning to the original variables, using Eq.
(\ref{cdl15}), we can rewrite Eq. (\ref{mcmh7}) as
\begin{equation}
M_c=3.75\, \frac{\hbar^{3/2}}{m^2} \left (\frac{M_h
\Sigma_0}{G^2}\right )^{3/8}.
\label{mcmh7origin}
\end{equation}
Once we have $M_c$ by Eq. (\ref{mcmh7}) or (\ref{mcmh7origin}) we can get $R_c$
by Eq. (\ref{cdl10}) or (\ref{cdl18}) and $\rho_0$ by Eq.
(\ref{mcmh3}). Explicitly,
\begin{equation}
\label{rc}
R_c=4.51\,\frac{\hbar^2}{Gm^{8/3}M_c^{1/3}}=2.90\,\frac{\hbar^{3/2}}{G^{3/4}
m^2M_h^ { 1/8 } \Sigma_0^ { 1/8 } } ,
\end{equation}
\begin{equation}
\label{r0}
\rho_0=0.0156\, \frac{G^3m^8M_c^2}{\hbar^6}=0.219\,\frac{G^{3/2}m^4M_h^{3/4}
\Sigma_0^{3/4 } } { \hbar^3 } .
\end{equation}

{\it Remark:} In this paper, we 
have defined the halo mass $M_h$ such that
the density at the halo radius $r_h$ is equal to the central density divided by
$4$ (see Appendix \ref{sec_hr}). However, other authors work in terms
of a  halo
mass
$M_v$
defined in another manner as explained in Sec. V. C of \cite{mcmh}. The relation
between $M_h$ and $M_v$ is (see Eq. (146) of Ref. \cite{mcmh})
\begin{eqnarray}
\label{cmhm8}
\frac{M_h}{M_\odot}= 6.01\times 10^{-6}\left (\frac{M_v}{M_{\odot}}\right
)^{4/3}.
\end{eqnarray}
Combining Eqs. (\ref{mcmh7origin}) and (\ref{cmhm8}), we obtain the core mass
--
halo
mass
relation $M_c(M_v)$.
It  exhibits the fundamental scaling $M_c\propto M_v^{1/2}$. This theoretical
scaling, first obtained in the form of Eq. (\ref{mcmh7origin}) in Appendix H of
\cite{clm2},
is
consistent with the scaling found numerically by  Ruffini {\it et al.}
\cite{rar} (they find an exponent equal to $0.52$
instead of $1/2$).

\subsection{Logarithmic corrections}
\label{sec_lc}

\begin{figure}
\begin{center}
\includegraphics[clip,scale=0.3]{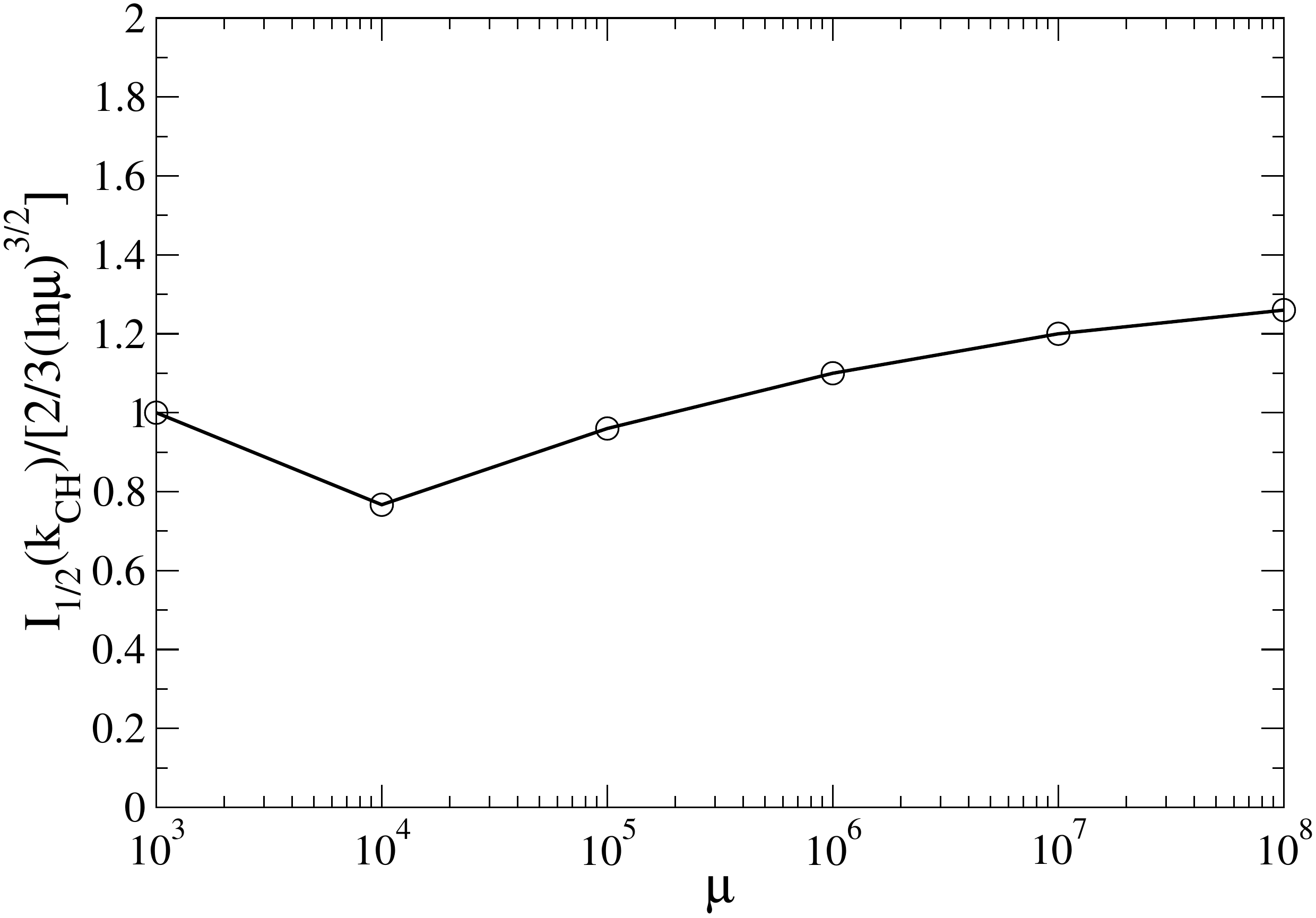}
\caption{This curve confirms that 
$I_{1/2}(k_{\rm CH})\sim \frac{2}{3}(\ln\mu)^{3/2}$ in good approximation.}
\label{muI}
\end{center}
\end{figure}

In the previous section, we have assumed that $I_{1/2}(k_{\rm
CH})$ is approximately constant and we have taken $I_{1/2}(k_{\rm CH})\simeq
1$. A more precise expression of  $I_{1/2}(k_{\rm  CH})$ can be obtained as
follows. We see on Fig. \ref{keta} that $k_{\rm CH}\sim 1/\mu$ for
large values of $\mu$.\footnote{This is not rigorously the case but this
approximation is sufficient for our purposes since $k$ arises in a logarithm.}
Using the
asymptotic expression $I_{1/2}(k)\sim \frac{2}{3}(-\ln k)^{3/2}$ of the Fermi
integral $I_{1/2}(k)$ for $k\rightarrow 0$ [see Eq. (\ref{d9})], we obtain
 $I_{1/2}(k_{\rm CH})\sim \frac{2}{3}(\ln \mu)^{3/2}$. This behavior is
confirmed by the plot of Fig. \ref{muI}. If we take this
logarithmic correction into account we have to multiply $M_c$ [given by Eq.
(\ref{mcmh7}) or (\ref{mcmh7origin})] by the factor
\begin{equation}
{\cal A}=\sqrt{\frac{2}{3}}(\ln\mu)^{3/4}.
\label{logcorr}
\end{equation}
On the other hand, $R_c$ [given by Eq. (\ref{rc})]
has
to be divided by ${\cal A}^{1/3}$ and $\rho_0$ [given
by Eq. (\ref{r0})] has
to be multiplied by ${\cal A}^{2}$.

\subsection{Velocity dispersion tracing relation}

The core mass - halo mass relation can also be obtained from a simple
analytical model of self-gravitating fermions enclosed within a box as detailed
in
Sec. IV of \cite{mcmh}. In that model, the fermion ball is represented by a
polytrope of index $n=3/2$ and the classical isothermal atmosphere is assumed to
be
uniform. Under these approximations, one can compute the energy and the entropy
analytically. The mass of the fermion ball $M_c$ is then
obtained by maximizing the entropy $S(M_c)$ for a given value of $E_h$, $M_h$ and
$r_h$.  This leads to a relation $M_c(M_h)$ similar to that of  Eq.
(\ref{mcmh5b}) with Eq. (\ref{logcorr}) (see Eq.
(123) of Ref. \cite{mcmh}). This relation is obtained from a  thermodynamical
approach (maximum entropy principle) determining the ``most probable'' core
mass $M_c$. It is shown furthermore in
Sec. V of \cite{mcmh} that this relation
is equivalent to the ``velocity dispersion tracing'' relation
\cite{mocz,bbbs,modeldm}
\begin{eqnarray}
\label{ana31b}
v_c^2\sim v_h^2 \quad {\rm or} \quad M_c\sim \frac{R_c}{r_h}M_h
\end{eqnarray}
stating that the
velocity dispersion in the core $v_c^2\sim GM_c/R_c$ is of the same order as the
velocity dispersion in the halo $v_h^2\sim GM_h/r_h$. This is the reason why
Eq. (\ref{mcmh7}) is similar to Eq. (169) of
\cite{mcmh}. It is interesting to note that the prefactors appearing in these
relations  are almost the same although these relations are
obtained from substantially different calculations. Therefore, the present
approach provides
an additional justification of the ``velocity dispersion tracing'' relation from
thermodynamical arguments.

\section{Astrophysical applications}
\label{sec_astapp}

We now consider astrophysical applications 
of our model and discuss several scenarios that are suggested by the
previous results.

\subsection{Minimum halo with $M_h=(M_h)_{\rm
min}$}
\label{sec_min}

The minimum halo has a  mass  $(M_h)_{\rm min}=10^8\,
M_{\odot}$ 
and a radius $(r_{h})_{\rm min}=597\, {\rm pc}$. It corresponds to the
ground state (minimum energy state) of the self-gravitating Fermi gas. A
completely degenerate fermion ball at $T=0$ is equivalent to a polytrope of
index $n=3/2$ (see Figs. \ref{profilepolytrope} and \ref{Vpoly}). This solution
is fully stable
in all statistical ensembles.

\subsection{Ultracompact DM halos with $(M_h)_{\rm
min}<M_h<(M_h)_{\rm CCP}$}
\label{sec_small}

\begin{figure}
\begin{center}
\includegraphics[clip,scale=0.3]{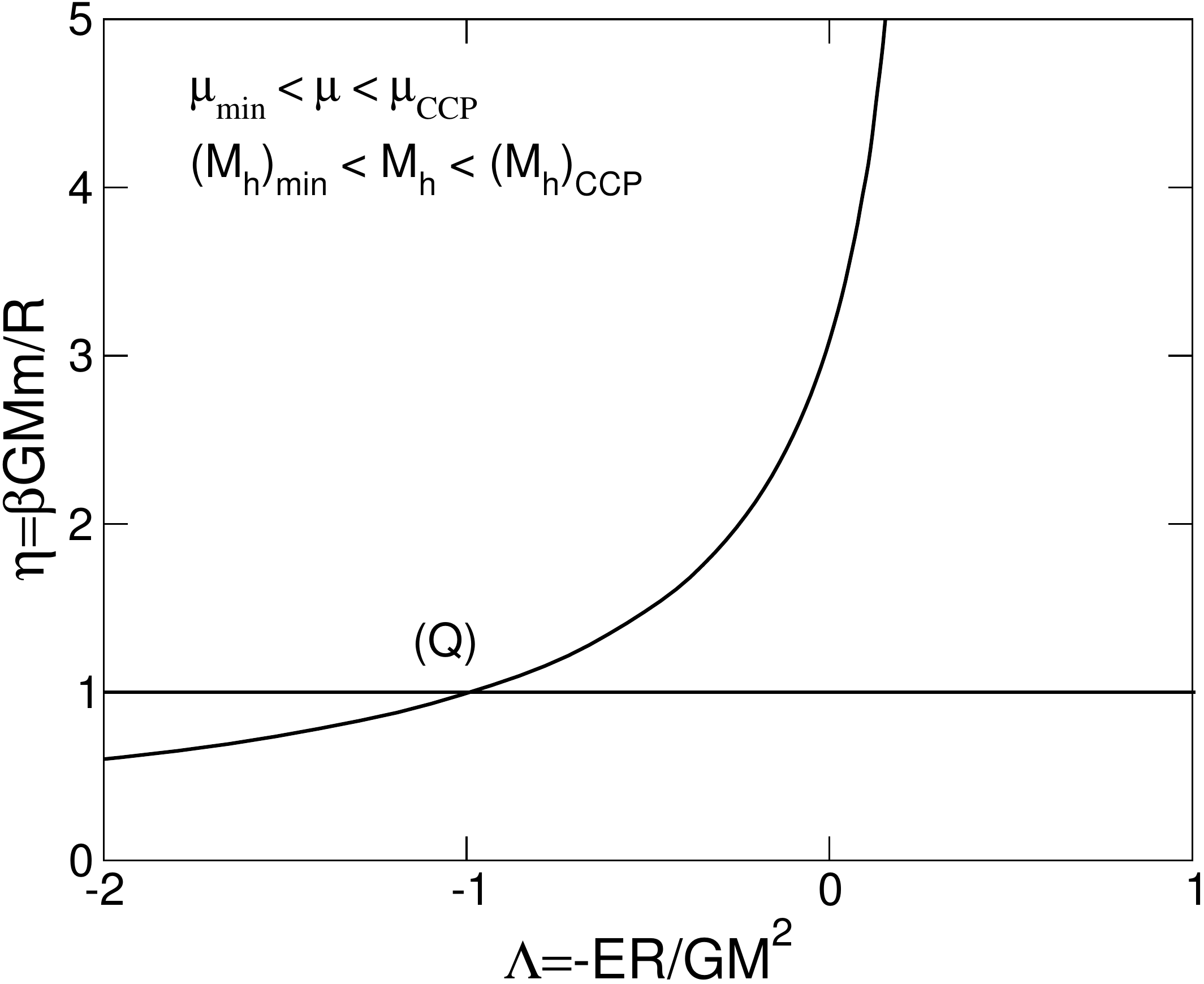}
\caption{Caloric curve of self-gravitating fermions for
$\mu=10$. When $(M_h)_{\rm
min}<M_h<(M_h)_{\rm CCP}$, the
caloric curve is monotonic. The quantum solution (Q) is thermodynamically stable
in all statistical ensembles.}
\label{multimu2PH}
\end{center}
\end{figure}

Ultracompact DM halos have a mass in the range $(M_h)_{\rm
min}=10^8\, M_{\odot}<M_h<(M_h)_{\rm CCP}=6.73\times 10^8\,
M_{\odot}$.  Since $\mu<\mu_{\rm CCP}$ the caloric curve
$\eta(\Lambda)$
is monotonic (see Fig. \ref{multimu2PH}). There is only one equilibrium state with
$\eta_v\sim 1$. It corresponds to a completely degenerate fermion ball
surrounded by a tenuous classical isothermal atmosphere.
This quantum solution (Q) is thermodynamically stable in all
statistical ensembles.

\subsection{Small DM halos with $(M_h)_{\rm
CCP}<M_h<(M_h)_{\rm MCP}$}
\label{sec_large}

\begin{figure}
\begin{center}
\includegraphics[clip,scale=0.3]{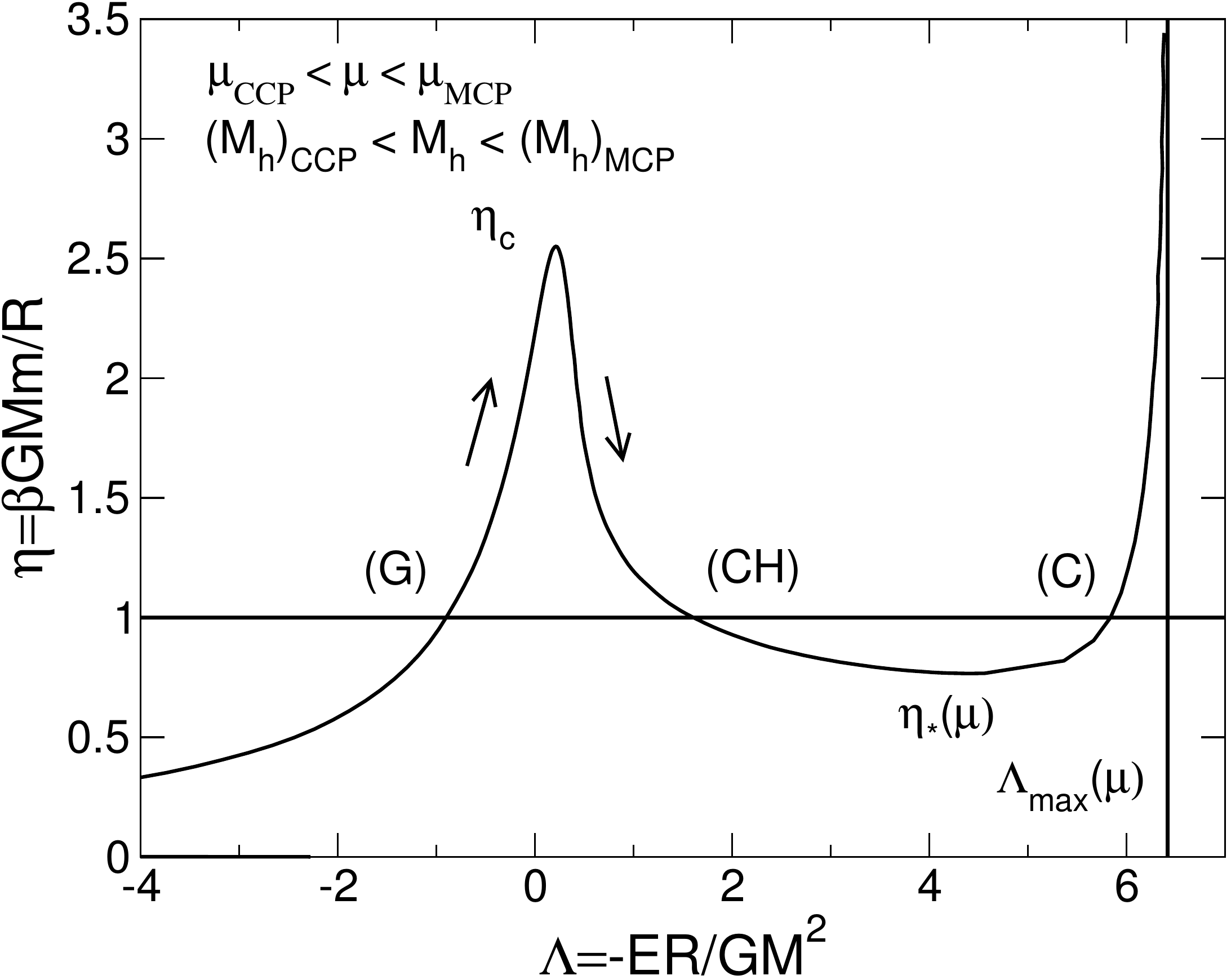}
\caption{Caloric curve of self-gravitating fermions for $\mu=10^3$. When
$(M_h)_{\rm CCP}<M_h<(M_h)_{\rm MCP}$, the caloric curve has an $N$-shape
structure. In the canonical ensemble, the gaseous phase (G) and the
condensed phase (C) are stable while the core-halo phase (CH) is unstable (it
represents a ``critical droplet'' that the system must create to pass from
one phase to the other). In
the microcanonical ensemble, all the equilibrium states are stable. The system
may directly reach the core-halo  phase (CH) through a process of collisionless
violent relaxation. It may also evolve collisionally, following the
arrows, from the gaseous solution (G) to the
core-halo solution (CH). }
\label{fel}
\end{center}
\end{figure}

\begin{figure}
\begin{center}
\includegraphics[clip,scale=0.3]{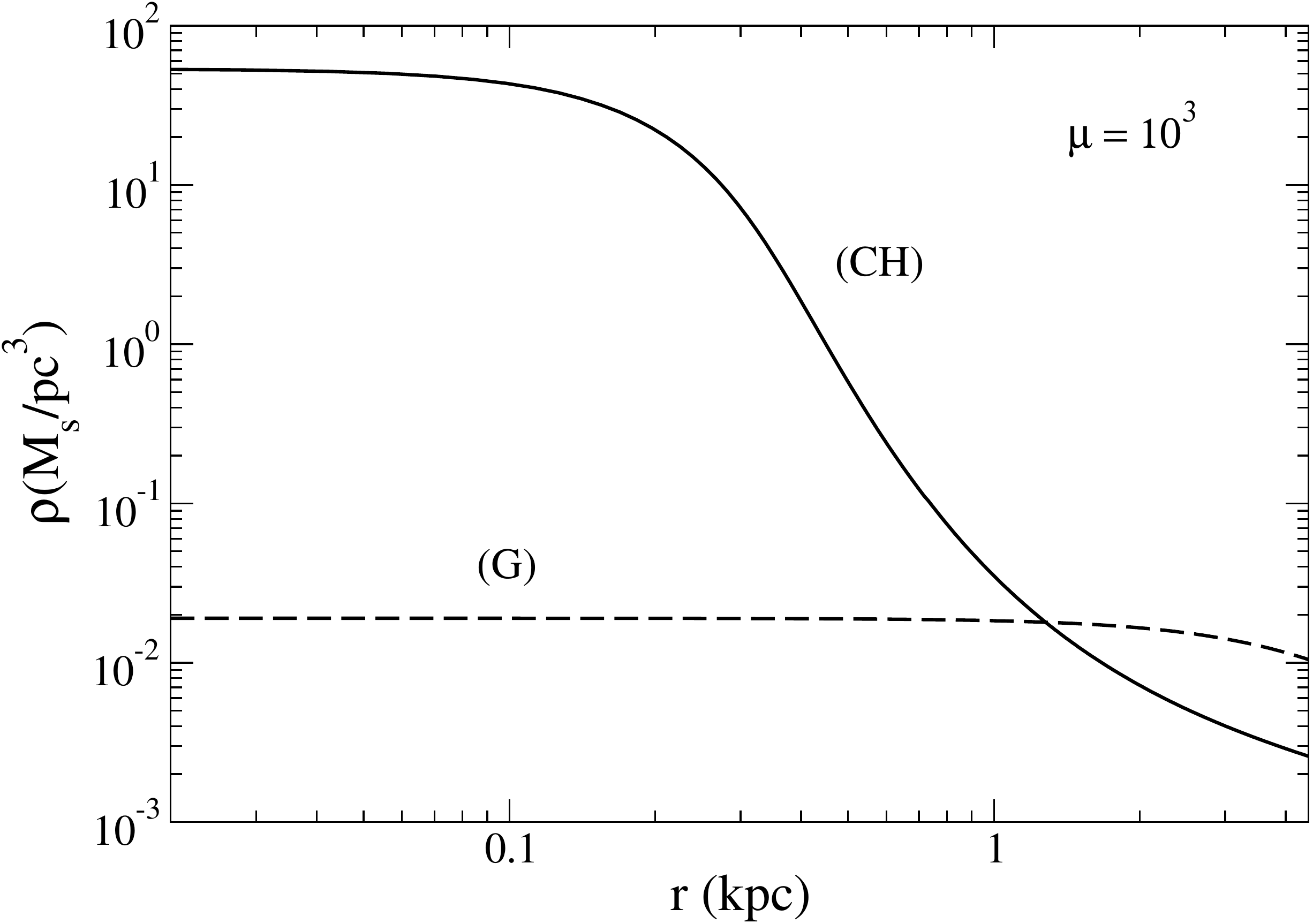}
\caption{Density profile of the core-halo solution (CH) ($k_{\rm CH}=1.12\times
10^{-3}$)
for $\mu=10^3$.
For comparison, we have represented in dashed line the gaseous  solution (G)
which corresponds to a classical 
isothermal halo ($k_G=206$). }
\label{profile1e3}
\end{center}
\end{figure}

Small DM halos have a mass in the range $(M_h)_{\rm CCP}=6.73\times 10^8\,
M_{\odot}<M_h<(M_{h})_{\rm
MCP}=1.08\times 10^{10}\, M_{\odot}$. Specifically, we consider a DM halo
characterized by a degeneracy parameter $\mu=10^3$. It has a mass
$M_h=4.93\times 10^9\, M_{\odot}$ [see Eq. (\ref{mu3})]  and a radius
$r_h=4.46\, {\rm kpc}$ [see Eq. (\ref{mu1})].
The corresponding caloric curve (see Sec. \ref{sec_tf}) is represented in
Fig. \ref{fel}. Since $\mu_{\rm CCP}<\mu<\mu_{\rm MCP}$, the caloric 
curve has an $N$-shape structure (see Sec. \ref{sec_ineq}). The intersection of
this curve with the line $\eta_v\sim 1$ (see Sec. \ref{sec_vc}) determines three
solutions: a gaseous solution (G), a core-halo solution (CH) and a condensed
solution (C) that we do not consider here (see Sec. \ref{sec_vc}). The gaseous
solution (G) has a concentration parameter $k_G=206$ (see Fig. \ref{keta}). The
corresponding density profile is plotted as a dashed line in Fig.
\ref{profile1e3}. It represents a purely
classical isothermal DM halo without quantum core as investigated in Sec.
\ref{sec_nd}.  This solution lies in the region of the caloric curve where the
specific heat is positive ($C=dE/dT>0$).  It is thermodynamically stable in all
statistical ensembles (maximum entropy state at fixed mass and energy and
minimum free energy state at fixed mass). The core-halo
solution (CH) has a concentration parameter $k_{\rm CH}=1.12\times 10^{-3}$ (see
Fig. \ref{keta}). The corresponding density profile (see Sec. \ref{sec_tf}) is
plotted as a solid line in Fig. \ref{profile1e3}. It represents a DM halo with a
 quantum core (fermion ball) of mass $M_c=2.21\times 10^9\, M_{\odot}$, radius
$R_c=389\, {\rm pc}$ and
central density $\rho_0=53.6\, M_{\odot}/{\rm pc}^3$ [we have
used Eq.
(\ref{mcmh7}) 
to obtain $M_c$, Eqs. (\ref{rc}) and (\ref{r0}) to obtain $R_c$ and
$\rho_0$, and we have taken into account the logarithmic correction ${\cal
A}=3.48$ from Eq. (\ref{logcorr})] surrounded by a classical isothermal
atmosphere.
This core-halo solution lies in the region of the caloric curve where the
specific heat is negative ($C=dE/dT<0$). It is thermodynamically unstable in the
canonical ensemble (saddle point of free energy at fixed mass)  but it is
thermodynamically stable  in the microcanonical ensemble (entropy maximum at
fixed mass and energy) which is the relevant ensemble to consider (see Sec.
\ref{sec_kt}).\footnote{The solutions (G), (CH) and (C) that we consider have
different energies but the same temperature. Indeed, the temperature is more
relevant than the energy to characterize a DM halo since,  according to the
virial
theorem, $\eta\sim 1$. However, we stress that we must analyze the
thermodynamical stability of the system in the microcanonical ensemble, not in
the canonical ensemble.}

For small DM halos with $(M_h)_{\rm CCP}<M_h<(M_h)_{\rm MCP}$, the gaseous
solution (G) and the core-halo solution (CH) are both thermodynamically stable
in the microcanonical ensemble (maximum entropy state at fixed mass and
energy). Therefore, they are
both likely to result from a natural evolution in a thermodynamical
sense. Let us consider different scenarios of formation and evolution in
line with the general discussion given in Sec. \ref{sec_kt}:

(i) The core-halo solution (CH) may arise naturally from a process 
of violent collisionless relaxation (following Jeans instability and free fall)
since it is a maximum entropy state in the sense of Lynden-Bell. This is a fast
process taking place on a dynamical timescale. If the evolution is
collisionless, the system remains in that state.

(ii) The gaseous  solution (G) may also arise naturally from a 
process of violent collisionless relaxation since it is a maximum entropy state
in the sense of Lynden-Bell. Then, there are two possibilities:

(ii-a) If the evolution is collisionless, the system remains in that state.

(ii-b) If the evolution is collisional, the system may evolve along the series
of equilibria (see Fig. \ref{fel}). Indeed, because of collisions\footnote{These
collisions between DM particles are not two-body gravitational encounters
because the relaxation time would be too long \cite{clm1,clm2}, but they can have another
origin such as short-range interactions (SIDM model) like in,
e.g.,
Ref. \cite{balberg}.} 
and evaporation the central density increases and the
energy decreases. The temperature first decreases in the region of positive
specific heat ($C=dE/dT>0$) then increases in the region of
negative specific heat ($C=dE/dT<0$). The whole series of equilibria represented
in Fig. \ref{fel} is stable in the microcanonical
ensemble. Therefore, if the DM halo evolves
adiabatically
under the effect of collisions, it  can progressively pass from the gaseous
solution (G) to the core-halo solution (CH). This is a slow relaxation taking place on a secular timescale.
This may be a mechanism  -- alternative to violent relaxation --
which explains how the system reaches the core-halo solution (CH).

In conclusion, small DM halos with $(M_h)_{\rm CCP}<M_h<(M_h)_{\rm MCP}$ can be in two types of configuration:

(I) The gaseous solution (G) coinciding with the classical isothermal sphere.
This simple solution is consistent with the observations because we have shown
in Sec. III.C of \cite{modeldm} that an isothermal DM
halo is almost indistinguishable from the observational Burkert profile.

(II) The core-halo solution (CH) made of a quantum core (fermion ball) 
of mass   $M_c=2.21\times 10^9\, M_{\odot}$ and radius $R_c=389\, {\rm pc}$
surrounded by a classical isothermal atmosphere. The quantum  core cannot mimic
a SMBH because it is too big (its radius $R_c=389\, {\rm pc}$ is
much larger than its Schwarzschild
radius $R_S=2GM/c^2=2.11\times 10^{-4}\, {\rm pc}$). However, it
can represent a large quantum bulge made of DM. This quantum bulge may possibly
exist at present at the center of certain galaxies or may have
existed in the past as a temporary state, and has disappeared since then.
Indeed, a large bulge may provide a favorable environment for triggering the
formation of a SMBH that can then grow by accretion. The final outcome of this
scenario would then be a classical isothermal halo containing either a quantum
bulge
(large fermion ball) or a SMBH that would be the remnant of the original bulge.

\subsection{Large DM halos with $M_h>(M_h)_{\rm
MCP}$}
\label{sec_vlh}

\begin{figure}
\begin{center}
\includegraphics[clip,scale=0.3]{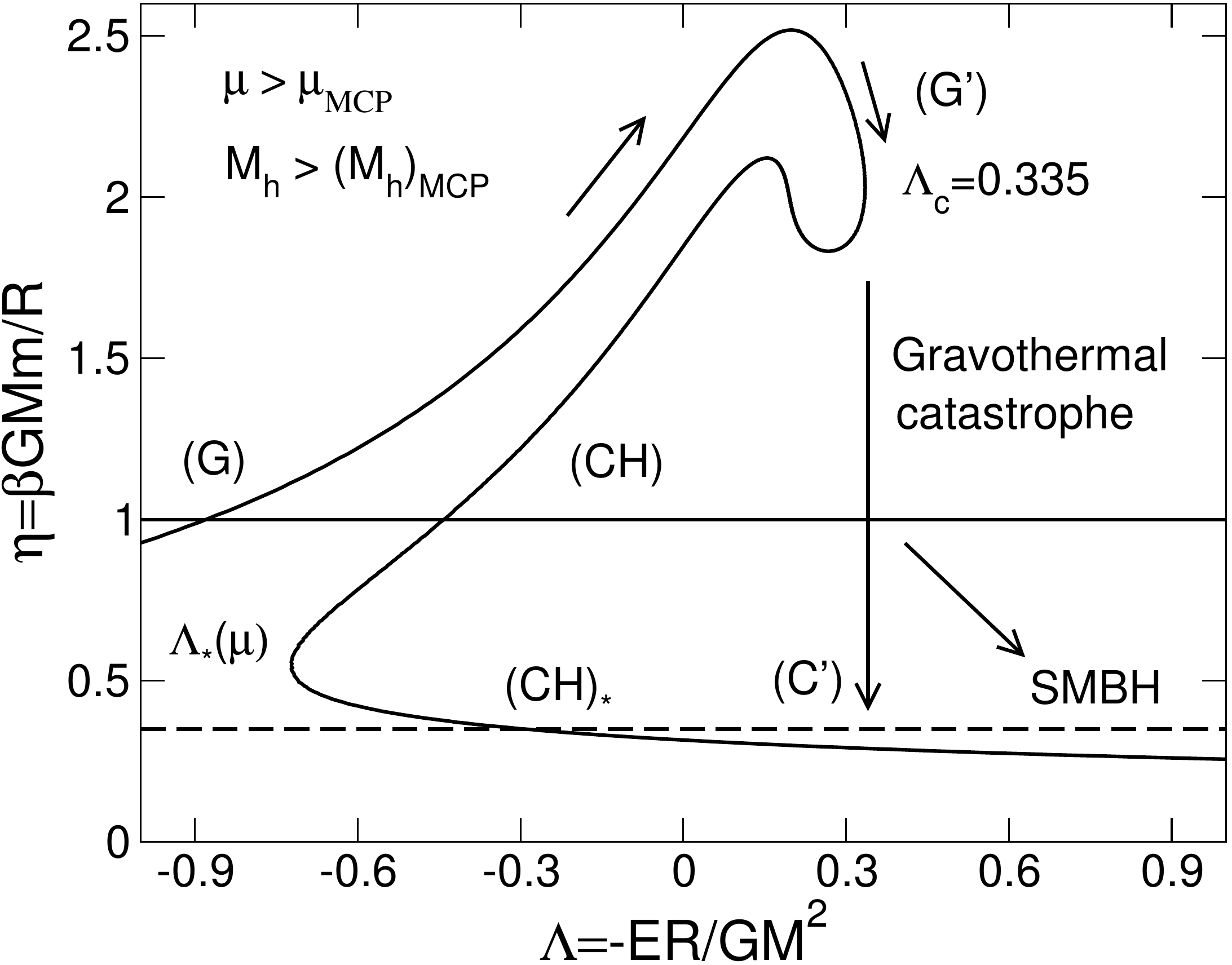}
\caption{Caloric curve of self-gravitating fermions for $\mu=10^5$. For
$M_h>(M_h)_{\rm MCP}$, the caloric curve has a
$Z$-shape structure (dinosaur's neck). In the microcanonical ensemble, the
gaseous phase (G) and (G') before the first turning point of energy  and
the condensed phase (C') and (C) after the last turning point of energy are
stable. By contrast, the core-halo phase (CH) in the intermediate branch between
the first
and the last turning points of energy is unstable. The
system can evolve collisionally in the gaseous phase (G) and (G') up
to
the turning point of energy $E_c$ and collapse in
the condensed phase
(C') (see arrows). This corresponds to the gravothermal
catastrophe  \cite{lbw} arrested by quantum effects.
Another
possibility is that the gravothermal catastophe triggers a dynamical
instability of general relativistic origin leading to the formation of a
SMBH \cite{balberg} (see Fig.
\ref{fbbh} below).}
\label{le5}
\end{center}
\end{figure}

\begin{figure}
\begin{center}
\includegraphics[clip,scale=0.3]{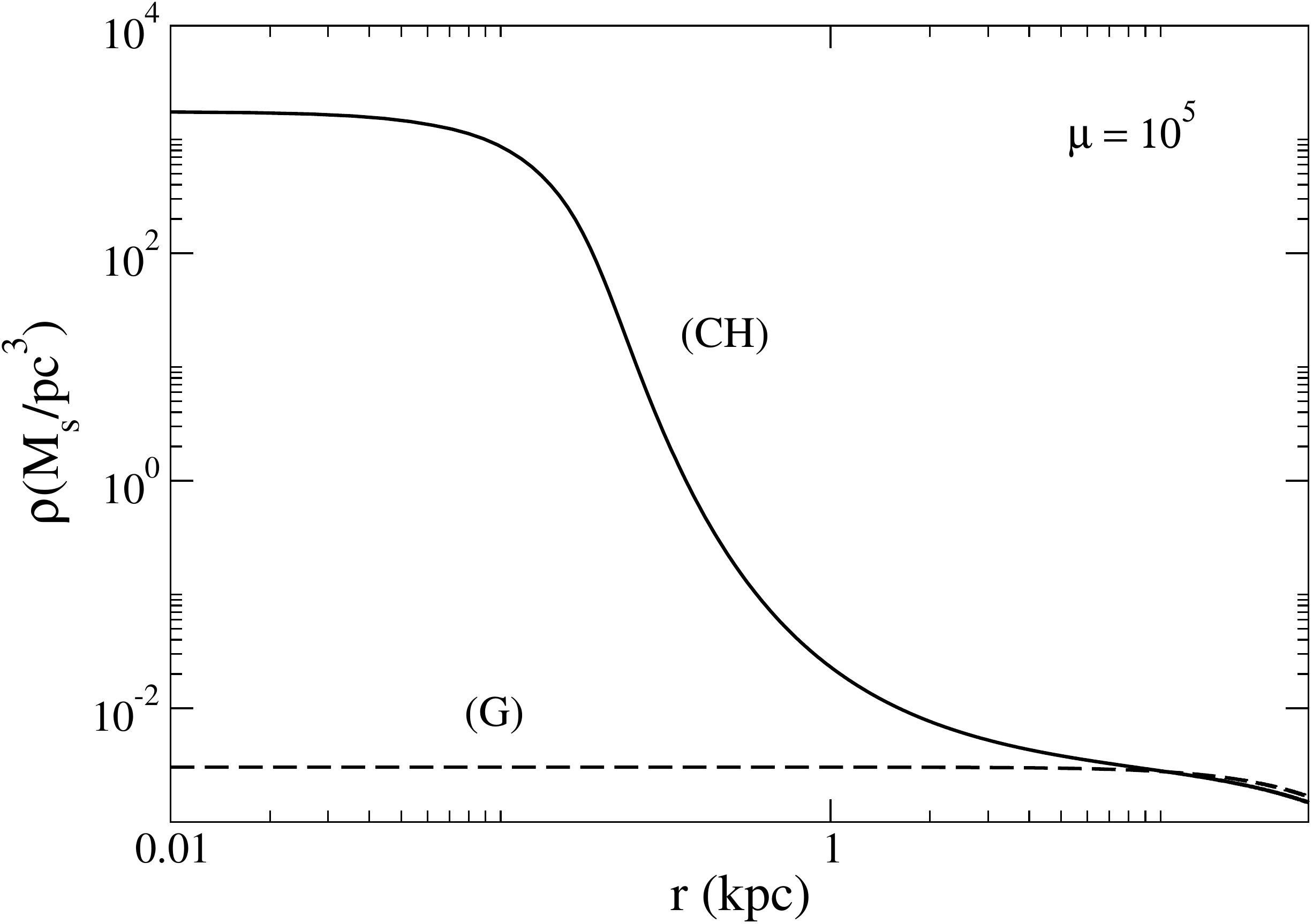}
\caption{Density profile of the core-halo (CH) solution ($k_{\rm CH}=1.38\times
10^{-5}$)
for $\mu=10^5$.
For comparison, we have represented in dashed line the gaseous (G) solution
which corresponds to a classical 
isothermal halo ($k_G=2.05\times 10^4$).}
\label{profile1e5}
\end{center}
\end{figure}

Large DM halos have a mass $M_h>(M_{h})_{\rm
MCP}=1.08\times 10^{10}\, M_{\odot}$. Specifically, we consider a DM halo
characterized by a degeneracy parameter $\mu=10^5$. It has a mass
$M_h=1.96\times 10^{11}\, M_{\odot}$ [see Eq. (\ref{mu3})] and a radius
$r_h=28.1\, {\rm kpc}$
[see Eq. (\ref{mu1})]. The
corresponding caloric curve (see Sec. \ref{sec_tf})  is represented in
Fig. \ref{le5}. Since $\mu>\mu_{\rm MCP}$,
the caloric curve has a $Z$-shape structure similar to a dinosaur's neck (see
Sec. \ref{sec_ineq}). As before, the intersection of this curve with the  line
$\eta_v\sim 1$ (see Sec. \ref{sec_vc}) determines two physical solutions: a
gaseous solution (G) and a core-halo solution (CH).  The gaseous solution (G)
has a concentration parameter $k_G=2.05\times 10^4$ (see Fig. \ref{keta}). The
corresponding density profile is plotted as a 
dashed line in Fig. \ref{profile1e5}. It
represents a  purely classical isothermal DM halo of mass $M_h$ and radius $r_h$
without quantum core as investigated in Sec. \ref{sec_nd}. It lies in a region
of positive specific heat.  It is thermodynamically stable in all
statistical ensembles
(maximum entropy state at fixed mass and energy and minimum free energy state
at fixed mass). The core-halo solution
(CH) has a concentration parameter $k_{\rm CH}=1.38\times 10^{-5}$ (see Fig.
\ref{keta}). The corresponding density profile (see Sec. \ref{sec_tf}) is
plotted as a solid line in Fig. \ref{profile1e5}. It represents a DM halo with a
quantum core (fermion ball) of mass $M_c=1.29\times 10^{10}\, M_{\odot}$, radius
$R_c=216\, {\rm pc}$ and central density $\rho_0=1820\, M_{\odot}/{\rm pc}^3$
[we have used Eq. (\ref{mcmh7}) to obtain $M_c$, Eqs.
(\ref{rc}) and
(\ref{r0}) to obtain $R_c$ and $\rho_0$, and we have taken into account the
logarithmic correction ${\cal A}=5.10$ from Eq. (\ref{logcorr})] surrounded
by
a classical isothermal atmosphere of mass $\sim M_h$ and radius $r_h$. Since
this
solution is located between the first and the last turning points of energy, it
is thermodynamically unstable in all statistical ensembles (saddle point of
entropy at fixed
mass and energy and saddle point of free energy at fixed mass). Note that it
lies in a region of the caloric curve with a positive specific heat, showing
that a positive specific heat does not  necessarily imply that the system is
stable.

For large DM halos with $M_h>(M_h)_{\rm MCP}=1.08\times 10^{10}\,
M_{\odot}$ the gaseous solution (G) is thermodynamically stable but the
core-halo solution (CH) is thermodynamically unstable. Therefore, the gaseous
solution is likely to result from a natural evolution in a thermodynamical sense
while the core-halo solution (CH) should not be
observed.\footnote{This statement could be alleviated by the following
considerations. On the one hand, the core-halo solution would be stable if
$\eta_v$
is smaller than the value $\eta_v=1$ that we have somehow arbitrarily chosen
(see the dashed
line in Fig. \ref{le5}). For example,
the core-halo solution (CH)$_*$ located
just after the last turning point of
energy $\Lambda_*$ is stable.  On the other hand, it is always possible that
the process of
{\it incomplete relaxation} \cite{lb,incomplete} leads to a Vlasov stable
core-halo profile
that is
not of the Lynden-Bell (or Fermi-Dirac) type. Indeed, we have seen that all DFs
$f=f(\epsilon)$ with $f'(\epsilon)<0$ are dynamically stable in Newtonian
gravity even those that are
thermodynamically unstable.} 
Let us consider different scenarios of formation and evolution in line
 with the general discussion given in Sec. \ref{sec_kt}.

The gaseous solution (G) may arise naturally from a process of violent
collisionless relaxation (following Jeans instability and free fall) since it is
a maximum entropy state in 
the sense of Lynden-Bell. This is a fast
process taking place on a few dynamical times. Then, there are two
possibilities:

(a) If the evolution is collisionless, the system remains in that state.

(b) If the evolution is collisional, the system may slowly evolve along the
series of equilibria (see Fig. \ref{le5}). The beginning of the collisional
evolution is similar to that described previously. The temperature first
decreases in the region of positive specific heat ($C=dE/dT>0$) then increases
in the region of negative specific heat ($C=dE/dT<0$). However, when the system
reaches the turning point of energy (corresponding to the minimum energy $E_c$)
it becomes thermodynamically unstable 
and undergoes the gravothermal catastrophe \cite{lbw}. At that point, there are
several possibilities:

(b.1) We first assume that the gravothermal catastrophe is eventually halted by
quantum mechanics (Pauli's
exclusion principle) and that the system reaches an equilibrium state. This
takes the system from the gaseous phase (G') to the
condensed phase (C') in which only a fraction (typically $\sim 1/4$) of the mass
of the DM halo forms a compact fermion ball while the rest of the mass constitutes a
hot halo. The hot halo has a uniform density so
that it is strongly held by the box (see Fig. 16 in \cite{ijmpb}).  As discussed
in
\cite{ijmpb,acf}, if we remove the box, the halo should be expelled
at large distances in a process reminiscent  of a supernova explosion
\cite{pomeau1,pomeau2,pomeaulettre}. This is because the
collapse of the core heats the
halo which thus extends at large distances. Although
this mechanism could be at work
for fermion stars such as white dwarfs and neutron stars, it may
not be relevant for DM halos. Therefore, we shall prefer the following
scenarios.

(b.2) We assume that the gravothermal catastrophe is eventually halted by 
quantum mechanics as before, but the system does not reach the equilibrium
solution (C'). It may reach an out-of-equilibrium core-halo structure (CH)$_{\rm
out}$ that is not
described by the Fermi-Dirac distribution. 
This out-of-equilibrium state (CH)$_{\rm
out}$  may be made of a slowly evolving quantum core
surrounded by a classical atmosphere that is not as much extended as the
classical atmosphere of the equilibrium solution (C'). Actually, in this
scenario, the
initial isothermal halo (at criticality) is essentially left undisturbed. Since
the solution (CH)$_{\rm
out}$ is an
out-of-equilibrium structure, we
expect that the core-halo mass relation $M_c(M_h)$ is different from the one
predicted by Eq. (\ref{mcmh7}). In particular, the quantum core resulting from
the gravothermal catastrophe should be more compact and more massive than the
quantum core composing the (CH) solution.  The occurrence of
this out-of-equilibrium state (CH)$_{\rm
out}$ is due to the fact that
the exchange of energy between the core and the halo, and the process of
thermalization, may take a very long time. Therefore, the equilibrium state
(C') of scenario (b.1) may not be reached on relevant timescales.

(b.3) Finally, we assume that the halo undergoes a
gravothermal catastrophe at $E_c$ but we consider another evolution in which
quantum mechanics cannot prevent gravitational collapse (the
validity of
this hypothesis is
considered in Sec. \ref{sec_critbh}). This scenario, already advocated in
Refs. \cite{clm1,clm2}, is based on
the SIDM model of Balberg {\it et al.}
\cite{balberg} who
developed
the idea of an ``avalanche-type contraction'' towards a SMBH initially suggested by
Zeldovich and
Podurets \cite{zp}, improved by Fackerell {\it et al.} \cite{fit}, and
confirmed numerically  by Shapiro and Teukolsky
\cite{st2,st3,st4}. The
initial stage of the gravothermal catastrophe is well-known. The core collapses
and reaches high densities and high
temperatures while the halo is not sensibly affected by the collapse of the
core and maintains its initial structure. Now,
Balberg {\it et al.} \cite{balberg} argue that during the gravothermal
catastrophe, when the central density and the temperature increase above a
critical value, the system
undergoes a dynamical instability of general relativistic
origin leading to the
formation of a SMBH
on a
dynamical time scale. Only the central region of the DM halo (not its outer
part) is affected by this collapse so the final outcome of this scenario is
a classical isothermal halo  at criticality containing a central SMBH.

In conclusion, large DM halos with $M_h>(M_h)_{\rm MCP}$ can be in three types of configuration:

(I) A purely classical isothermal halo (G) without quantum core.

(II) An out-of-equilibrium core-halo solution (CH)$_{\rm out}$, resulting from
the gravothermal catastrophe, which is different
from the solution (CH) that is unstable or from the solution (C') that is
unphysical. It is made of a compact (small and
massive) quantum core surrounded by a classical isothermal atmosphere at
criticality. 

(III) A classical isothermal halo at
criticality containing a SMBH resulting from the
gravothermal catastrophe followed by a dynamical instability of general
relativity origin.

It
is also possible that, following the gravothermal
catastrophe, the system first forms a fermion ball then a SMBH (see
Secs. \ref{sec_critbh} and \ref{sec_gr}).

{\it Remark:} The scenarios (II) and (III) may be particularly interesting
especially if we account for tidal effects. Indeed, it has been shown in
\cite{clm1,clm2} that the King profile at criticality (i.e. at the verge of the
gravothermal
catastrophe) is very close to the observational Burkert profile (see, e.g.,
Figs.
18 and 27 of \cite{clm1} and Fig. \ref{densityLOG}).
Therefore, the
structure of large DM halos could
consist of a fermion ball or a SMBH surrounded by an envelope with a
marginal (critical)
King profile
unaffected by the collapse of the core \cite{clm1,clm2}. The conditions for
forming a SMBH at the center of a DM halo are discussed in Sec. \ref{sec_critbh}
based on the results of Alberti and Chavanis \cite{caf,acf}.

\subsection{Criterion for the existence of a SMBH  at the center of a galaxy}
\label{sec_critbh}

According to the above scenario, the formation of a SMBH at the center of a galaxy is possible only if the system can experience the gravothermal catastrophe and if, during core collapse, the core can reach sufficiently high densities and high temperatures to trigger a general relativistic dynamical instability leading to the formation of a SMBH. This may happen in sufficiently large systems. By contrast, in small systems, quantum mechanics (Pauli's exclusion principle for fermions) prevents the gravothermal catastrophe and leads to a large fermion ball (bulge) instead of a SMBH. In conclusion, a SMBH can form only if the degeneracy parameter $\mu$ is sufficiently large so that the gravothermal catastrophe is efficient. Therefore, we expect that DM halos harbor a  SMBH if $\mu\gg \mu_{\rm MCP}=2670$  i.e.
\begin{equation}
\label{crit1}
M_h\gg (M_h)_{\rm MCP}=1.08\times 10^{10}\, M_{\odot}
\end{equation}
and we expect that DM halos harbor a large quantum bulge (fermion ball) in the
opposite 
case.\footnote{Large DM halos may contain a SMBH but they should not contain a
fermion ball because the core-halo solution (CH) is thermodynamically unstable.
By contrast, small DM halos may contain a large quantum bulge (fermion ball) but
they should not contain a SMBH because the gravothermal catastrophe is inhibited
by quantum mechanics.}

This result is qualitatively consistent with the conclusion 
reached by Ferrarese \cite{ferrarese} on the basis of observations. She found
that black holes can form only in sufficiently large galaxies, above a typical
mass $\sim 5\times 10^{11}\, M_{\odot}$. This limit may correspond to the
microcanonical critical point $(M_h)_{\rm MCP}$ of our model. To facilitate
further comparisons, using Eq. (\ref{mu2}), we rewrite this criterion
as\footnote{Equation (\ref{crit1lit}) is in good agreement with
the criterion $H>8.24$ obtained in Appendix H of \cite{clm2} where $H$ is
defined by Eq. (E4) of that paper.}
\begin{equation}
\label{crit1lit}
M_h\gg (M_h)_{\rm MCP}=483\left (\frac{\hbar^3\Sigma_0^{3/4}}{G^{3/2}m^4}\right
)^{4/5}.
\end{equation}

\begin{figure}
\begin{center}
\includegraphics[clip,scale=0.3]{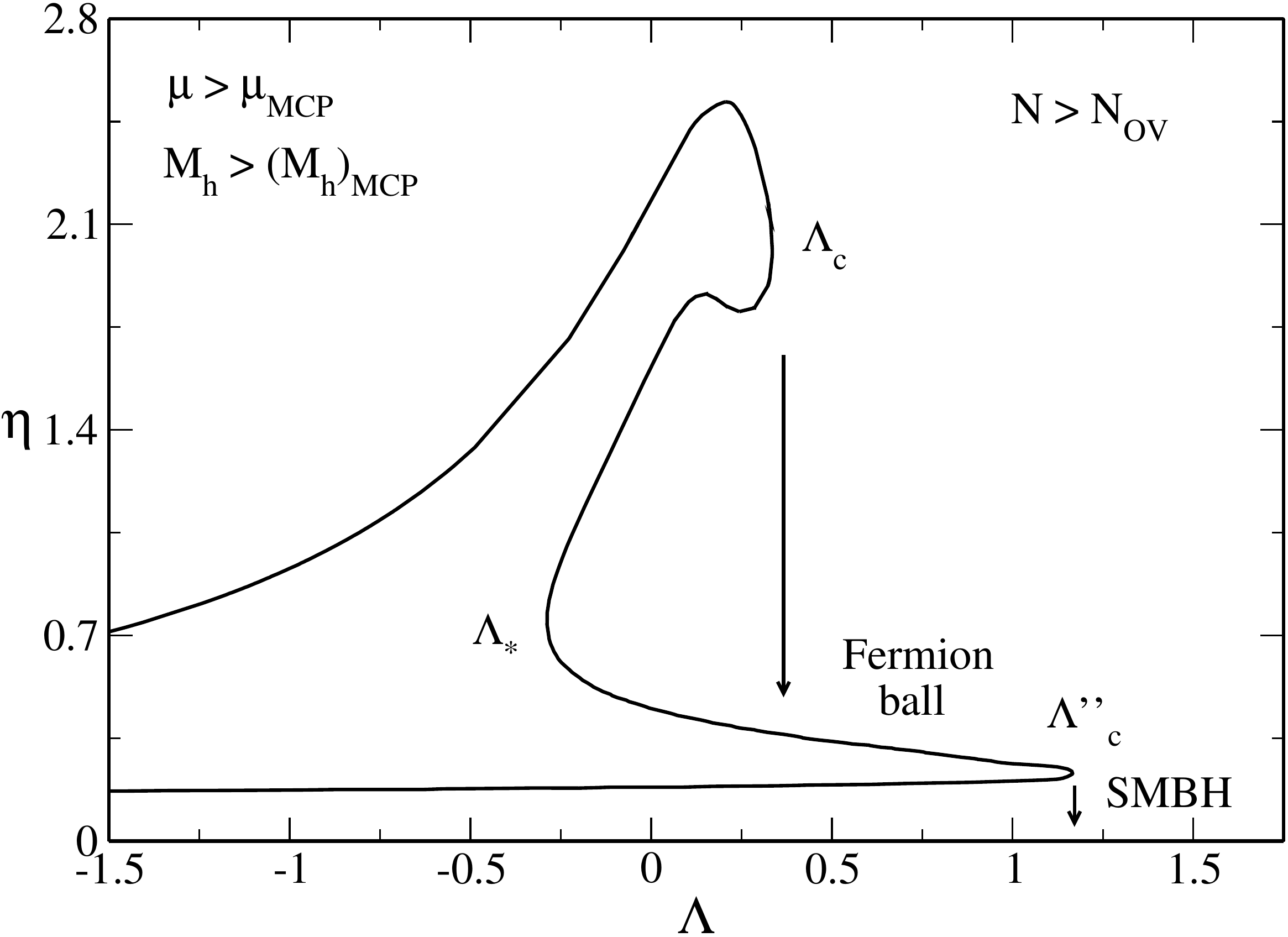}
\caption{Caloric curve for $M_h>(M_h)_{\rm MCP}$ and
$N>N_{\rm OV}$ (from \cite{caf,acf}). As energy decreases, the system first
experiences a gravothermal catastrophe at $\Lambda_c$ leading to a fermion ball
stabilized by quantum degeneracy, then a gravitational collapse at
$\Lambda_c''$ leading to a SMBH. They are both surrounded by
a classical isothermal envelope.}
\label{fbbh}
\end{center}
\end{figure}

Actually, things are more complicated than the scenario just exposed. 
Indeed, as shown by Alberti and Chavanis \cite{acf,caf}, when general relativity
is taken into account, the caloric curves of the self-gravitating Fermi gas
depend not only on $\mu$, but also on the value of the particle number $N$ with
respect to $N_{\rm OV}$. When $N<N_{\rm OV}$, the caloric curve is
similar to the one reported in Fig. \ref{le5}. In particular, there is an
equilibrium state for any value of the energy since quantum mechanics (Pauli's
exclusion principle) can prevent gravitational collapse even at $T=0$. By
contrast, when
$N>N_{\rm OV}$, a new turning point of energy appears \cite{caf,acf}
as shown in Fig.
\ref{fbbh}. In that case,
there is no equilibrium state below a minimum energy $E_c''$ and the system
collapses
towards a black hole. These results suggest that the correct criterion for the
existence of a 
SMBH at the center of a galaxy is that $\mu\gg \mu_{\rm MCP}$ (in order to
trigger the gravothermal catastrophe) and $N>N_{\rm OV}$ (in order to have a
gravitational collapse towards a SMBH). The first condition yields Eq.
(\ref{crit1lit}). It signals the instability of the core-halo solution (CH)
with respect to the gravothermal catastrophe. If we approximate the second
condition by $M_h>M_{\rm OV}$, where $M_{\rm OV}$ is given by Eq.
(\ref{cdl19}),
we obtain the condition
\begin{equation}
\label{crit1b}
M_h>M_{\rm OV}=0.384\, \left (\frac{\hbar
c}{G}\right )^{3/2}\frac{1}{m^2}.
\end{equation}
If $(M_h)_{\rm MCP}<M_h<M_{\rm OV}$ we expect that the halo 
experiences the gravothermal catastrophe but does not form a SMBH. It may rather
form an out-of-equilibrium fermion ball (CH)$_{\rm out}$ (scenario b.2). By
contrast, if $M_h>M_{\rm OV}$,  the halo may form either  an out-of-equilibrium
fermion ball (CH)$_{\rm out}$ (scenario b.2) or a SMBH (scenario
b.3). A necessary condition to form a SMBH is that $M_{\rm OV}>(M_h)_{\rm MCP}$.
This yields 
\begin{equation}
\label{crit1bw}
m>383\, \frac{\hbar^{3/4}\Sigma_0^{1/2}G^{1/4}}{c^{5/4}}=0.278\, {\rm eV/c^2}.
\end{equation}
This condition is always fulfilled in practice.

For $m=165\, {\rm eV}/c^2$ we find that $(M_h)_{\rm MCP}=1.08\times
10^{10}\, M_{\odot}$ and $M_{\rm OV}=2.30\times
10^{13}\,
M_{\odot}$. In that case, the OV mass is very
large, much larger than the mass $M_h= 10^{11}\, M_{\odot}$ of a DM halo
comparable to the Milky Way. As a result, the gravothermal catastrophe should be
stopped
by quantum mechanics and a SMBH cannot be formed. This suggests that the Milky
Way
contains an out-of-equilibrium fermion ball rather
than a SMBH.
However, if we consider a larger fermion mass $m\sim 1\, {\rm keV}/c^2$ we find 
$(M_h)_{\rm MCP}=3.38\times 10^{7}\, M_{\odot}$ and $M_{\rm OV}=6.26\times
10^{11}\, M_{\odot}$, which are closer to the conditions required to form a
SMBH (see, however, the Remark below).

{\it Remark:} It is natural to expect that the gravitational
collapse at $E''_c$ leads to a SMBH of mass $M_{\rm OV}$ because the instability
of the DM halo occurs precisely at the moment where the core mass becomes
critical ($M_c=M_{\rm
OV}$) \cite{acf,rarnew}. In that case, we find for $m=165\, {\rm eV}/c^2$ and
$m\sim
1\, {\rm keV}/c^2$ that the SMBH mass would be $M_{\rm OV}=2.30\times 10^{13}\,
M_{\odot}$ and $M_{\rm OV}=6.26\times 10^{11}\, M_{\odot}$ respectively. These
very large masses are not consistent with the observations of SMBHs. The OV
mass is more relevant if the fermion has a larger mass $m$ as considered in
Sec. \ref{sec_smbh}. For
example, for $m=54.6\, {\rm keV/c^2}$, we get $M_{\rm OV}=2.10\times 10^8\,
M_{\odot}$ which is of the order of the mass of SMBHs observed in AGNs. On the
other hand, for $m=386\, {\rm keV/c^2}$
we get $M_{\rm OV}=4.2\times 10^6\, M_{\odot}$ which is of the order of the 
mass of Sagittarius
A$^*$. In that case, according to the scenario (III) discussed above, the
Milky Way
could
consist of a SMBH of mass $M_{\rm
OV}=4.2\times 10^6\, M_{\odot}$ (Sagittarius
A$^*$) resulting from the gravothermal catastrophe surrounded by an envelope
with
a marginal King profile similar to the Burkert profile  (see Sec. \ref{sec_gr}).

\subsection{Application to the Milky Way}
\label{sec_bulge}

\begin{figure}
\begin{center}
\includegraphics[clip,scale=0.3]{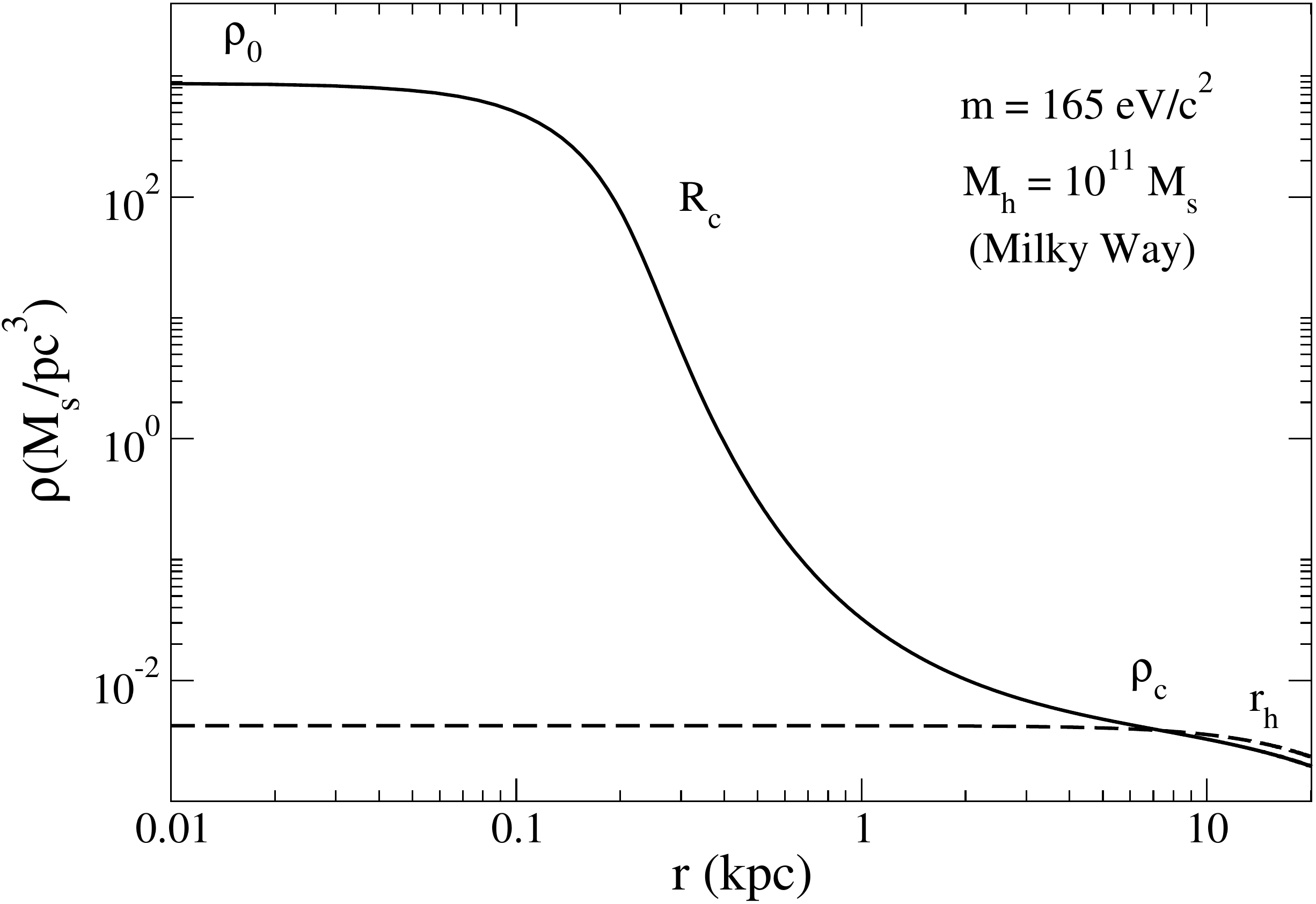}
\caption{Density profile of the core-halo (CH) solution ($k_{\rm CH}=5.21\times
10^{-5}$) for a DM halo of mass $M_h=10^{11}\,
M_{\odot}$ (Milky Way) up to the halo radius $r_h=20.1\, {\rm kpc}$.
For comparison, we have represented in dashed line the gaseous solution (G) 
which corresponds to a classical 
isothermal halo ($k_G=8.85\times 10^3$). }
\label{profileMW}
\end{center}
\end{figure}

\begin{figure}
\begin{center}
\includegraphics[clip,scale=0.3]{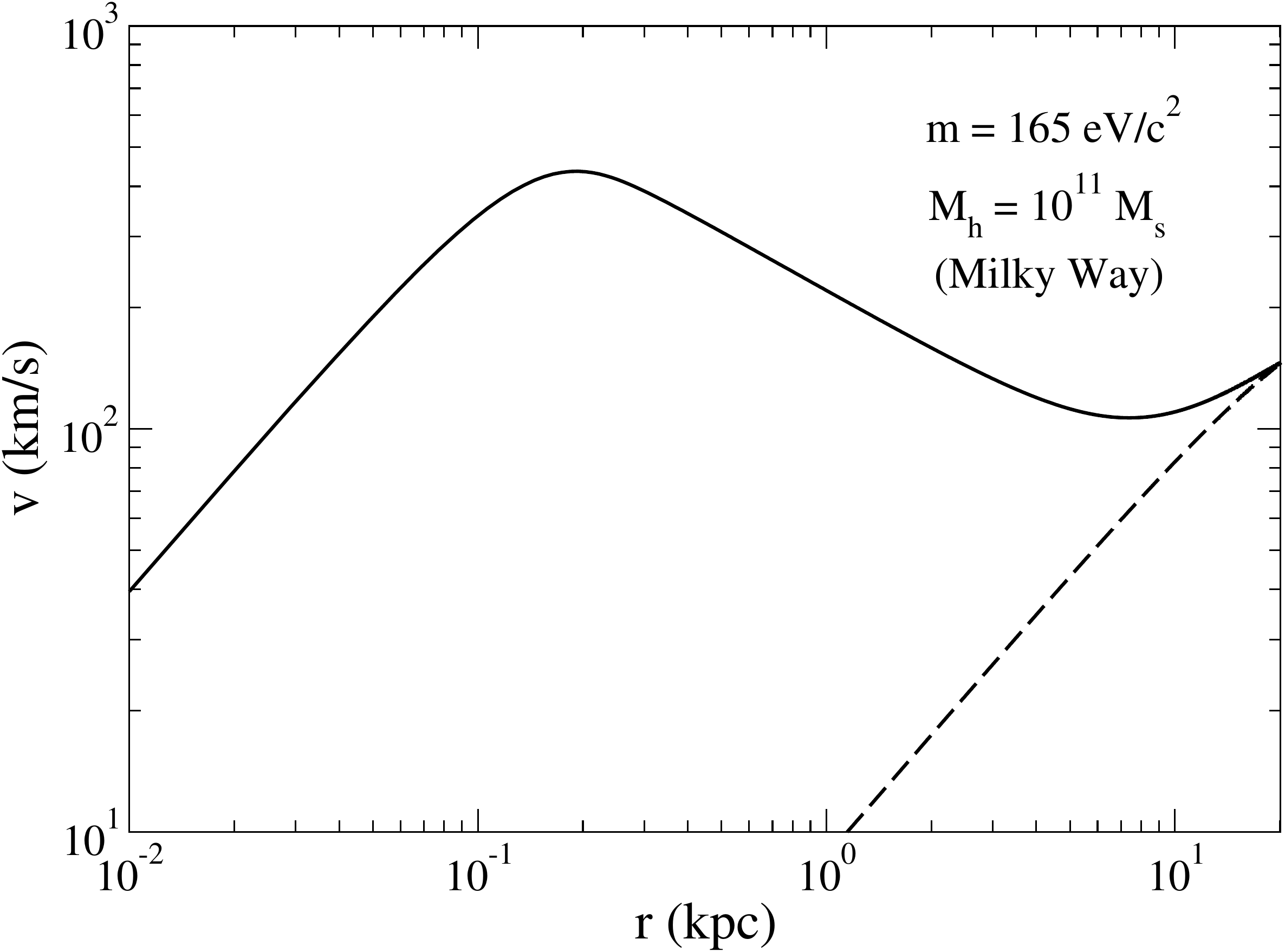}
\caption{Velocity profile of the core-halo (CH) solution ($k_{\rm CH}=5.21\times
10^{-5}$) for a DM halo of mass $M_h=10^{11}\,
M_{\odot}$ (Milky Way) up to the halo radius $r_h=20.1\, {\rm kpc}$.
For comparison, we have represented in dashed line the gaseous (G) solution
which corresponds to a classical 
isothermal halo ($k_G=8.85\times 10^3$).}
\label{profileMWvitesse}
\end{center}
\end{figure}

We now specifically apply our fermionic model to the Milky Way. We consider a DM
particle mass $m=165\, {\rm eV/c^2}$ so that the minimum halo has a mass 
$(M_{h})_{\rm min}=10^8 \, M_{\odot}$ 
and a radius $(r_h)_{\rm min}=597\, {\rm pc}$ (see Sec. \ref{sec_cdl}). To be
specific, we consider a DM halo of mass $M_h=10^{11}\, M_{\odot}$ (corresponding
to $M_v\sim 10^{12}\, M_{\odot}$) and radius $r_h=20.1\, {\rm kpc}$ similar to
the one that surrounds our Galaxy (see Sec. \ref{sec_nd}). Using
Eq. (\ref{mu3}) we find that the corresponding degeneracy parameter is
$\mu=4.31\times 10^4$. The corresponding caloric curve has a
$Z$-shape structure like in Fig. \ref{le5}. The gaseous solution (G)
corresponding
to a purely classical
isothermal halo is plotted as a dashed line in Figs. \ref{profileMW} and
\ref{profileMWvitesse}. Then,
considering the
core-halo solution (CH) and using Eqs. 
(\ref{cdl18}), (\ref{mcmh3}), (\ref{mcmh7}) and (\ref{logcorr}), we find that
the DM
halo should contain a quantum core of mass $M_c=9.45\times 10^9\, M_{\odot}$,
radius $R_c=240\, {\rm pc}$ and central density $\rho_0=983\, M_{\odot}/{\rm
pc}^3$.\footnote{Comparatively, for $m=1\, {\rm keV}/c^2$ we get
$\mu=4.42\times 10^7$, $M_c=4.30\times 10^8\, M_{\odot}$, $R_c=5.84\, {\rm pc}$
and $\rho_0=3.71\times 10^6\,
M_{\odot}/{\rm
pc}^3$. }  The density and velocity profiles
given by Eqs. (\ref{tf35b}) and (\ref{tf35c}) are
represented as a solid lines  in
Figs. \ref{profileMW} and \ref{profileMWvitesse}. Clearly, the fermion ball is
too extended to mimic a black hole. It is more likely to represent a large
quantum bulge as discussed in Sec. \ref{sec_large}.

The halo mass  $M_h=10^{11}\, M_{\odot}$ is above the microcanonical 
critical point  $(M_h)_{\rm MCP}=1.08\times 10^{10}\, M_{\odot}$. The gaseous
solution (G) is thermodynamically stable  and could
result from a process of violent relaxation. The core-halo solution (CH) is
thermodynamically unstable  and should not be
observed. It should be replaced by  an
out-of-equilibrium core-halo structure (CH)$_{\rm out}$ with a compact quantum
core as discussed in Sec. \ref{sec_vlh}. However, since we are relatively close
to the microcanonical
critical point, the core-halo solution (CH) may be marginally relevant,
especially if $\eta_v$ is smaller than expected, e.g., if we select the
solution
(CH)$_*$ of Fig. \ref{le5} (see footnote 25).

{\it Remark:}  We may wonder if there is evidence of a
large quantum bulge at the center of the Milky Way.  If we make the analogy with
bosonic models of DM halos, the fermion ball is the equivalent of the soliton
\cite{ch2,ch3}.
The mass and size of the fermion ball that we find correspond to the typical
mass and size of the solitons that have been predicted theoretically or observed
in numerical simulations of BECDM \cite{ch2,ch3}. In addition, De Martino {\it
et al.}
\cite{martino} have suggested that a large
soliton, forming a quantum bulge of mass $\simeq 10^{9}\,
M_{\odot}$ and radius $\simeq 100 \, {\rm pc}$, may be present at the center
of the Milky Way and  may
account for the observed dispersion velocity peak (see the
discussion in Sec. VII.E. of
\cite{modeldm}).  If this result is confirmed, we could argue that
this quantum bulge may correspond to a fermion ball rather than a bosonic
soliton since our model yields the same characteristic mass and
radius.\footnote{In that case, we must also add a primordial black hole in the
model in order to account
for the observation of a large central mass at the center of the
Milky Way corresponding to Sgr
A$^*$.}

\subsection{Problems with a fermionic model involving a mass $m=165\, {\rm
eV/c^2}$ or $m\sim
1\, {\rm
keV/c^2}$}
\label{sec_pro}

In this section, we mention some problems with a fermionic DM halo model
involving a ``small'' fermion mass $m=165\, {\rm eV/c^2}$  or $m\sim 1\, {\rm
keV/c^2}$:

(i) Arg\"uelles {\it et al.} \cite{krut} show in their Figs. 3
and 4 that a fermion
mass $m\sim 0.6 \, {\rm keV/c^2}$ is not consistent with the structure of the
Milky Way. A much larger mass $m\sim 48 \, {\rm keV/c^2}$ is necessary to
reproduce the rotation curve of the Milky Way.  Therefore, the study of 
Arg\"uelles {\it et al.}
\cite{krut} rules out the possibility to have a large DM bulge of mass $\simeq
10^{9}\, M_{\odot}$ and radius $\simeq 100 \, {\rm pc}$ at the center
of the Milky Way. This is in contradiction with our claim and with the
claim of De Martino {\it et al.} \cite{martino} that a large DM bulge may
account for the dispersion velocity peak observed in the Milky
Way.\footnote{As noted by C. Arg\"uelles (private
communication), De Martino {\it et al.} \cite{martino} have to add ``by hand'' a
Plummer component of bulge stars to reduce the central dispersion because
otherwise the BECDM model overestimates the data. On the other
hand, Bar {\it et al.} \cite{bbes} argue in their Sec. III that the central mass
component could well be due to ordinary baryonic matter rather than a DM
soliton.} It would be
extremely important to
clarify this issue. 

(ii) A warm dark matter fermionic particle with a mass $m<3 \, {\rm keV/c^2}$ is
ruled out by cosmological observations \cite{chatterjee,carena}. Likewise, a
boson of
mass $m=1.44\times 10^{-22}{\rm eV/c^2}$ which produces results similar to a
fermion
of mass  $m=165\, {\rm eV/c^2}$ (the soliton in the BEC model being
the
counterpart of the
fermion ball) is in tension with cosmological observations such as the Lyman
$\alpha$ forest (it is one or two orders of magnitude smaller than the required
value) \cite{hui}. This suggests that a minimum
halo of mass $(M_h)_{\rm min}=0.39\times 10^6\, M_{\odot}$ (Willman I) leading
to a fermion mass
$m\sim 1 \, {\rm keV/c^2}$ and a boson mass $m=9.22\times
10^{-21}{\rm eV/c^2}$ \cite{abrilph}  may be more
relevant
than a  minimum
halo of mass $(M_h)_{\rm min}=10^8\, M_{\odot}$  (Fornax) leading
to a fermion mass $m=165\, {\rm eV/c^2}$  and a boson mass $m=1.44\times
10^{-22}{\rm eV/c^2}$ (see \cite{mcmh} for the determination of the DM
particle mass).

(iii) The OV mass $M_{\rm OV}=2.30\times 10^{13}\,
M_{\odot}$ or $M_{\rm OV}=6.26\times 10^{11}\, M_{\odot}$ associated with a
fermion of mass $m=165\, {\rm eV/c^2}$ or $m\sim 1\, {\rm keV/c^2}$ seems to be
too large to be of much astrophysical interest (see the discussion in
Sec. \ref{sec_critbh}).

\section{Can a fermion ball mimic a SMBH?}
\label{sec_smbh}

\subsection{Sagittarius A*}
\label{sec_sagittarius}

The detailed study of the motion of S-stars near the Galactic center has
revealed the presence of a very massive central object, Sagittarius A* (Sgr A*).
This central object is usually associated with a SMBH
of mass $M=4.2\times 10^6\, M_{\odot}$ and Schwarzschild radius $R_S=4.02\times
10^{-7}\, {\rm pc}$. Whatever the object may be, its radius must be smaller than
$R_{\rm P}=6\times 10^{-4} \, {\rm pc}$ ($R_{\rm P}=1492\, R_S$), the S2
star pericenter \cite{gillessen}. Similar objects are expected to reside at the
centers of most spiral and elliptical galaxies, in active galactic nuclei (AGN).
Although it is commonly believed that these objects are SMBHs
\cite{gillessen,nature,reid,genzel}, this is not yet established on a firm
observational basis in all cases. Some authors have proposed that such objects
could be fermion balls \cite{bvn,bvr,bmtv,btv,rar,krut} or
boson stars \cite{torres2000,guzmanbh} that could mimic a SMBH.

Let us consider
this possibility in the framework of the fermionic model. More precisely, let us
investigate if a fermion ball can mimic a SMBH at the
center of the Galaxy.

\subsection{Standard Fermi-Dirac distribution}

Bilic   {\it et al.} \cite{btv} developed a general relativistic model of
fermionic DM halos at finite temperature with a fermion mass $m=15\, {\rm
keV/c^2}$ that describes both the center and the halo of the Galaxy in a
unified manner. The density profile has a core-halo structure with a quantum
core (fermion ball) and a classical isothermal atmosphere. By using the usual
Fermi-Dirac distribution and choosing parameters so as to fit observational data
at large distances, they found
a fermion ball of mass $M_c=2.27\times 10^6\, M_{\odot}$ and radius
$R_c=18\, {\rm mpc}$.\footnote{The fermion ball is weakly general relativistic
because $M_c=2.27\times 10^6\,
M_{\odot}\ll M_{\rm OV}=2.78\times 10^9\, M_{\odot}$ [see Eq. (\ref{cdl19})].}
Unfortunately, its radius is larger by a
factor $100$ than the bound
$R_P=6\times
10^{-4}\, {\rm pc}$ set by observations \cite{gillessen}. This is why
Bilic and coworkers abandoned this fermion ball scenario (R. Viollier, private
communication). The same
problem was encountered later by Ruffini  {\it et al.} \cite{rar} who developed a similar
 model with a fermion mass $m\sim 10\, {\rm
keV/c^2}$.

\begin{figure}
\begin{center}
\includegraphics[clip,scale=0.3]{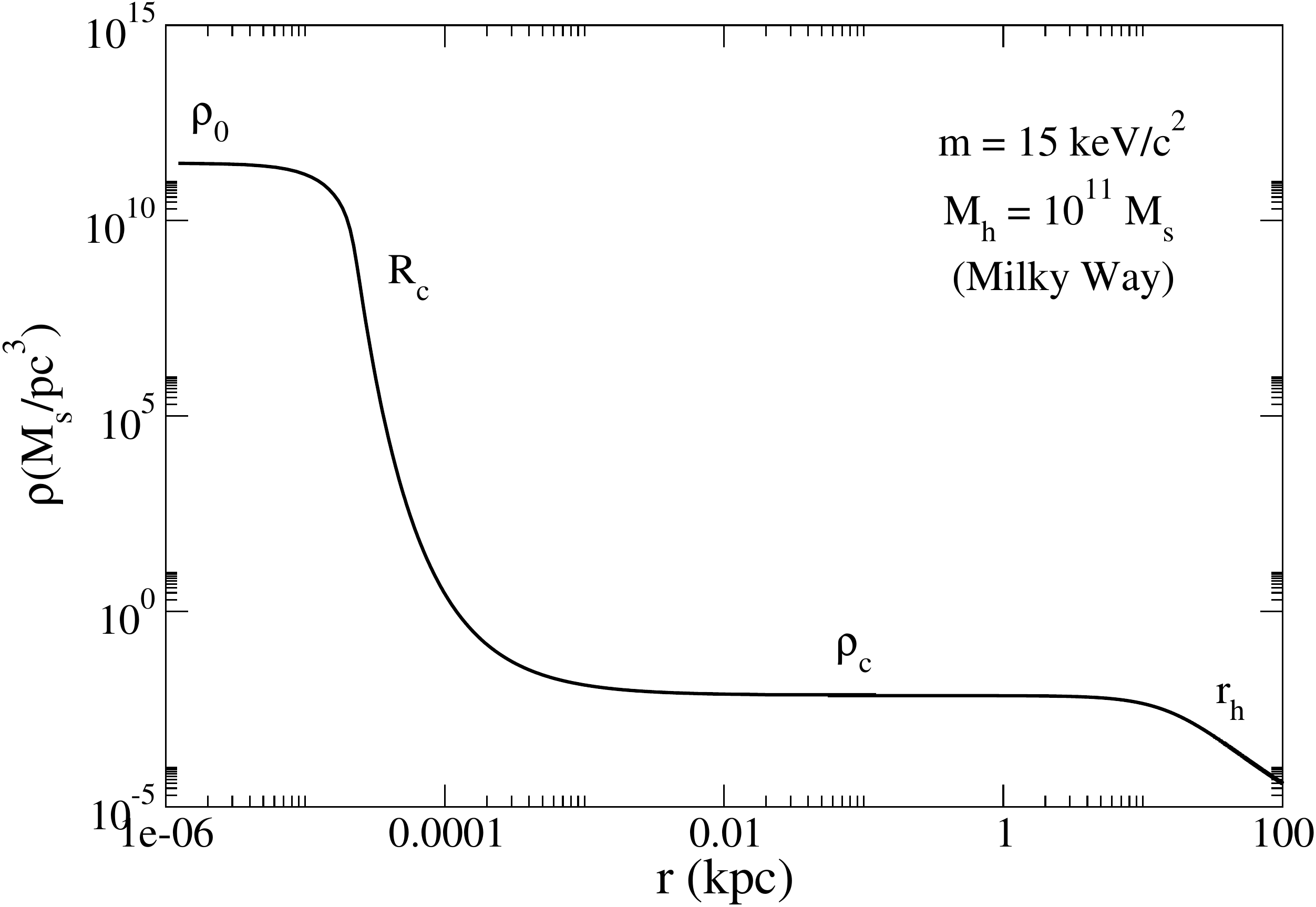}
\caption{Density profile of a DM halo
of mass
$M_h=10^{11}\, M_{\odot}$ (Milky Way) assuming a fermion mass $m=15\,
{\rm keV/c^2}$. }
\label{profileBH}
\end{center}
\end{figure}

\begin{figure}
\begin{center}
\includegraphics[clip,scale=0.3]{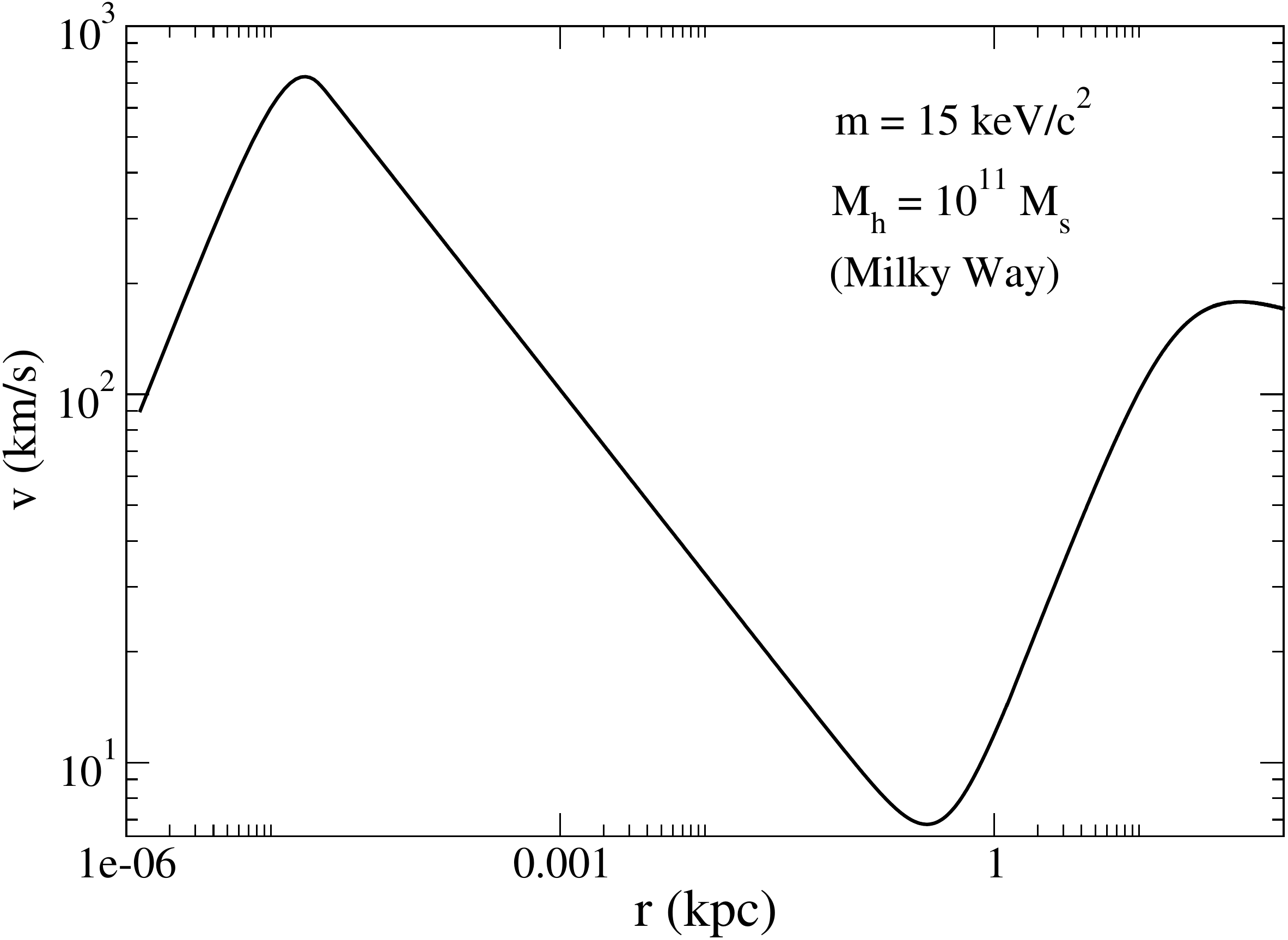}
\caption{Velocity profile of a DM
halo of mass
$M_h=10^{11}\, M_{\odot}$ (Milky Way) assuming a fermion mass $m=15\,
{\rm keV/c^2}$.}
\label{profileBHvitesse}
\end{center}
\end{figure}

Let us check that their results are consistent with our
analytical box model. Following Bilic   {\it et al.} \cite{btv}, we take a DM particle mass
$m=15\, {\rm keV/c^2}$. The corresponding minimum halo (see Sec. \ref{sec_cdl}) has a mass $(M_{h})_{\rm
min}=54.0 \, M_{\odot}$ and a radius $(r_h)_{\rm min}=0.439\, {\rm pc}$. If we
consider a DM halo of
mass $M_h=10^{11}\, M_{\odot}$  and radius $r_h=20.1\, {\rm kpc}$ similar to the
one that
surrounds our Galaxy (see Sec. \ref{sec_nd}) we find that the 
corresponding degeneracy parameter is $\mu=2.94\times 10^{12}$ [see Eq.
(\ref{mu3})].
Considering the core-halo solution (CH) and using Eqs. (\ref{cdl18}),
(\ref{mcmh3}), (\ref{mcmh7}) and (\ref{logcorr}), we find that this DM halo
should
contain a quantum
core of mass $M_c=2.39\times 10^6\,
M_{\odot}$, radius $R_c=22.7\, {\rm mpc}$ and central density $\rho_0=2.95\times 10^{11}\,
M_{\odot}/{\rm
pc}^3$  in
good agreement with the numerical
results of  Bilic   {\it et al.} \cite{btv} and Ruffini  {\it et al.}
\cite{rar}.  
The corresponding density and velocity profiles given by Eqs.
(\ref{tf35b}) and (\ref{tf35c}) are
represented in Figs.
\ref{profileBH} and \ref{profileBHvitesse}. They are in good
agreement  with Fig. 3 of Bilic   {\it et al.} \cite{btv}  and Figs. 1 and 3 of
Ruffini  {\it et al.} \cite{rar}. Therefore, our semi-analytical model [see
in particular Eq. (\ref{mcmh7})] can
reproduce their numerical results.

\begin{figure}
\begin{center}
\includegraphics[clip,scale=0.3]{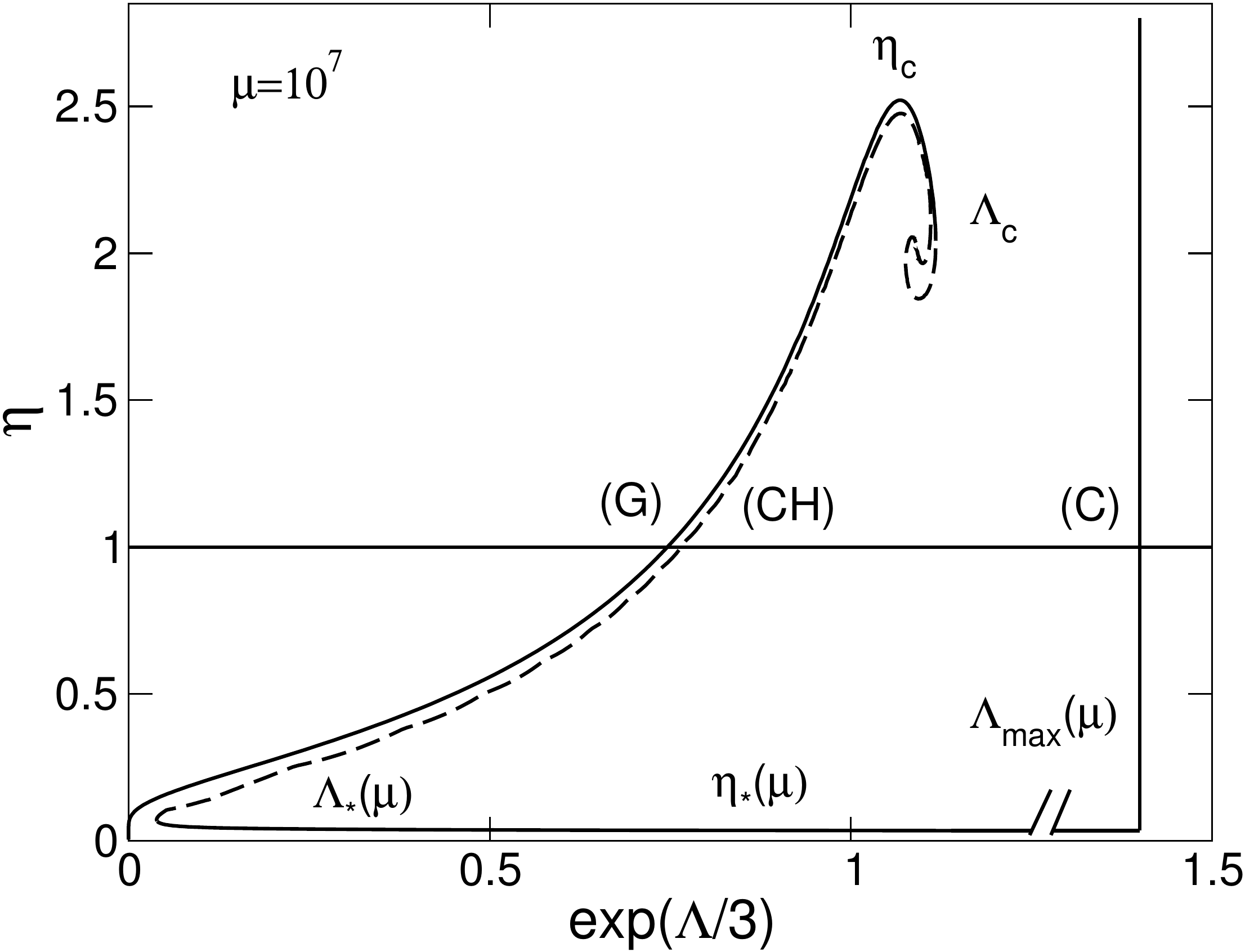}
\caption{Caloric curve for $M_h\gg (M_h)_{\rm MCP}$.}
\label{limitefermi}
\end{center}
\end{figure}

Let us now discuss the thermodynamical stability of the core-halo solution 
considered by Bilic   {\it et al.} \cite{btv} and Ruffini  {\it et al.}
\cite{rar}. The caloric curve (see Sec. \ref{sec_tf})
corresponding to $\mu=2.94\times 10^{12}$ is similar to the one represented in
Fig.
\ref{limitefermi}. For large values of $\mu$, a spiral appears in the caloric
curve at the location of the ``head'' of the dinosaur. As $\mu$ increases, the
spiral winds
more and more before unwinding. For $\mu\gg 1$ the direct and reversed
spirals are very close to each other. The intersections of the 
caloric curve with the  line $\eta_v\sim 1$ (see Sec. \ref{sec_vc}) determines
two physical solutions a before: a gaseous solution (G) and a core-halo solution
(CH).  The gaseous solution represents a  purely
classical isothermal DM halo of mass $M_h$ and radius $r_h$ without quantum core
as investigated in Sec. \ref{sec_nd} (see Figs. \ref{profileisotherme}
and \ref{Viso}). This solution lies on the caloric curve in
the region of positive specific heat. This solution (G) is thermodynamically
stable in all statistical ensembles (maximum entropy state at fixed mass and
energy and minimum free energy state at fixed mass).
The core-halo solution (CH) (see Sec. \ref{sec_tf})
represents a DM halo with a quantum core (fermion ball) of mass $M_c=2.39\times
10^{6}\, M_{\odot}$, radius $R_c=22.7\, {\rm mpc}$ and central
density $\rho_0=2.95\times 10^{11}\,
M_{\odot}/{\rm pc}^3$ surrounded by a classical isothermal
atmosphere of mass $\sim M_h=10^{11}\, M_{\odot}$ and radius $r_h=20.1\, {\rm
kpc}$. The corresponding density and velocity profiles are plotted in
Figs. \ref{profileBH} and \ref{profileBHvitesse}. Since this solution lies
between the
first and the last turning points of
energy, it is thermodynamically unstable in all statistical ensembles (saddle
point of
entropy at fixed mass and energy and saddle point of free energy at fixed mass).
This solution lies in a region of the caloric curve with a positive specific
heat.

The discussion about the thermodynamical stability of the 
core-halo solution (CH) is essentially the same as in Sec. \ref{sec_vlh}. The
main differences are the followings:

(i) According to the Poincar\'e-Katz \cite{poincare,katzpoincare1} criterion, 
the system loses more and more modes of stability, one at each turning point of
energy, as we progress clockwise into the spiral. However, when the spiral
unwinds the modes of stability are progressively regained. Indeed, one mode of
stability is regained at each turning point of energy as we follow the spiral
anticlockwise.\footnote{See \cite{ijmpb} and Appendix C of \cite{acf} for
a detailed discussion of the Poincar\'e-Katz criterion.} As a
consequence, the core-halo solutions that lie on the spiral are very unstable
since they have several modes of instability. We note, however, that the
core-halo solution (CH) has only one mode of instability as before.

(ii) The energy of the core-halo solution (CH) almost coincides with the energy 
of the gaseous solution (G).This is because their external structure is exactly
the same. The core-halo solution (CH) only differs from the gaseous solution (G)
by the presence of a small core with a small mass, a small radius and a very
high density (see Fig. \ref{profileBH}). The core and the halo are separated by
a large plateau
where the density is approximately constant.\footnote{See Sec. V of
\cite{modeldm} for a detailed discussion of the structure of quantum DM
halos involving a quantum core, a plateau, and a classical isothermal
atmosphere. The
core-halo profiles of DH halos with a ``small'' $\mu$ do not show a plateau
while an extended plateau is present in DM halos with $\mu\gg\mu_{\rm MCP}$.} 
Therefore, when $\mu\gg \mu_{\rm MCP}$, the core-halo
solution (CH) almost coincides with the gaseous solution (G) except that it
contains a small nucleus (fermion ball). For smaller values of $\mu$, the
plateau is reduces and finally disappears. For example, in Fig.
\ref{profile1e5}, the core-halo solution (CH) does not show a very pronounced
separation between the quantum core and the halo. Furthermore, 
the halo is perturbed by the presence of the core (unlike in Fig.
\ref{profileBH}). As a result,  the energy of the core-halo and gaseous
solutions on the caloric curve of Fig. \ref{le5} are relatively different
(unlike in Fig.
\ref{limitefermi}).

In conclusion, the models of 
 Bilic   {\it et al.} \cite{btv} and Ruffini  {\it et al.} \cite{rar} that are
based on the standard Fermi-Dirac DF lead to DM halos with a core-halo structure
made of a small quantum core (fermion ball) of mass $M_c=2.39\times 10^6\,
M_{\odot}$ and radius
$R_c=22.7\, {\rm mpc}$ surrounded
by a classical isothermal atmosphere. The core and the halo are separated by an
extended plateau. The quantum core describes a very compact central object not
very different from Sagittarius A*. However, the quantum core is not small
enough to account for the observational constraints. Furthermore, this core-halo
configuration is thermodynamically unstable so it is not expected to result from
a natural evolution (in the sense of Lynden-Bell). Therefore, the original
models of  Bilic   {\it et al.} \cite{btv} and Ruffini  {\it et al.} \cite{rar} 
have to be rejected.\footnote{The claim that the core-halo
solution of Refs. \cite{btv,rar} is thermodynamically unstable was first made in
\cite{clm2,modeldm,acf}.}

\subsection{Fermionic King model}

More recently, Arg\"uelles {\it et al.} \cite{krut} considered the general
relativistic fermionic King model accounting for a tidal
confinement.\footnote{The fermionic King 
model was heuristically introduced by Ruffini and Stella \cite{stella} as a
generalization of the classical King model \cite{king}. It was also introduced 
independently by Chavanis \cite{mnras} who derived it from a kinetic theory
based on the fermionic Landau equation. The nonrelativistic fermionic King model
was studied by Chavanis {\it et al.}
\cite{clm2} who showed that the density profiles typically have a core-halo
structure with a quantum core (fermion ball) and a tidally truncated isothermal
halo leading to flat rotation curves. They also studied the  caloric curves and
the thermodynamical stability of the equilibrium states. The name ``fermionic
King model'' was
introduced in \cite{clm2,clmh}.} 
They applied this model to the Milky Way and determined the parameters by
fitting the core-halo profile to the observations. For  a fermion mass $m=48\,
{\rm keV/c^2}$ they obtained a fermion ball of mass $M_c=4.2\times 10^6\,
M_{\odot}$ and radius
$R_c=R_P=6\times 10^{-4}\, {\rm pc}$ which, this time, is consistent with the
observational constraints.\footnote{The fermion ball is weakly general
relativistic because $M_c=4.2\times 10^6\,
M_{\odot}\ll M_{\rm OV}=2.71\times 10^8\, M_{\odot}$ [see Eq. (\ref{cdl19})].}

Let us see if their results are consistent with our analytical box
model. Following Arg\"uelles {\it et al.} \cite{krut}, we take a DM particle
mass $m=48\, {\rm keV/c^2}$. The corresponding minimum halo
(see Sec. \ref{sec_cdl}) has a mass  $(M_{h})_{\rm min}=1.30 \, M_{\odot}$ and a
radius $(r_h)_{\rm min}=0.0683\, {\rm pc}$. If we consider a DM halo of mass
$M_h=10^{11}\, M_{\odot}$
and radius
$r_h=20.1\, {\rm kpc}$ similar to the one that surrounds our Galaxy (see Sec.
\ref{sec_nd}) we find that the corresponding degeneracy parameter is
$\mu=3.09\times 10^{14}$ [see Eq. (\ref{mu3})]. Considering the core-halo
solution (CH) and using 
Eqs. (\ref{cdl18}), (\ref{mcmh3}), (\ref{mcmh7}) and (\ref{logcorr}), we find
that this halo
should contain a
quantum core of mass $M_c=2.61\times 10^5\, M_{\odot}$, radius $R_c=2.13\,
{\rm mpc}$ and central density $\rho_0=3.87\times 10^{13}\, M_{\odot}/{\rm
pc}^3$. Our analytical
results are {\it not} consistent with
the results of Arg\"uelles {\it et al.} \cite{krut} because 
we find that the mass $M_c$ of the fermion ball is about $10$ times smaller than
their
value. Since our analytical model is consistent with the results of Bilic {\it
et al.} \cite{btv} and  Ruffini {\it et al.} \cite{rar} that are based on the
usual Fermi-Dirac DF but not with the results of Arg\"uelles {\it et al.}
\cite{krut} that are based on the fermionic King model we deduce that the
difference comes from the fact that tidal effects -- not taken into account
in our analytical model -- are important  (a priori,
the
difference does not come from general relativity effects which are small as we
have indicated in footnote 36).

Therefore, in order to obtain accurate results, it is important to use the
fermionic King model \cite{clm2,krut} instead of the usual fermionic model
\cite{ijmpb,btv,rar}. Arg\"uelles {\it et al.} \cite{krut} managed to
fit
the density profile and the rotation curve of the Milky Way with the 
fermionic King distribution and argued that a fermion ball can mimic the effect
of a SMBH. This
scenario is very attractive because it can explain the whole structure of the
galaxy, the supermassive central object and the isothermal halo, by a single
DF (the fermionic King model \cite{stella,mnras}).

\begin{figure}
\begin{center}
\includegraphics[clip,scale=0.3]{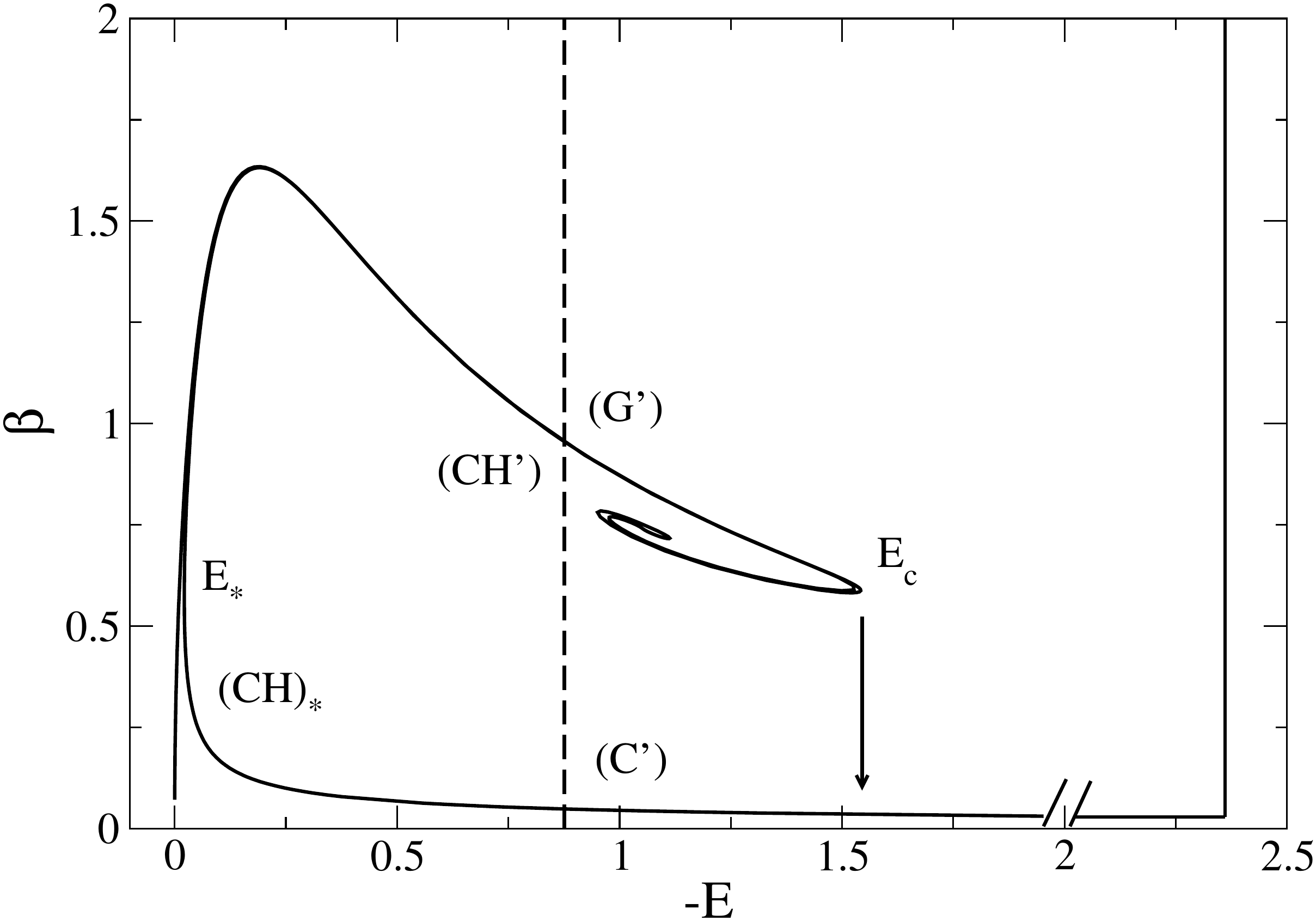}
\caption{Caloric curve of the fermionic King model for large
DM halos (from  \cite{clm2}). The core-halo solution (CH') considered in
\cite{clm2} is unstable. The  core-halo solution (CH)$_*$ considered in
\cite{rarnew} has a similar structure and is stable (being located after
the last turning point of energy $E_*$). It may account for the structure of the
Milky Way in which the fermion ball mimics a SMBH.}
\label{mutresgrandprolongehenon}
\end{center}
\end{figure}

\begin{figure}
\begin{center}
\includegraphics[clip,scale=0.3]{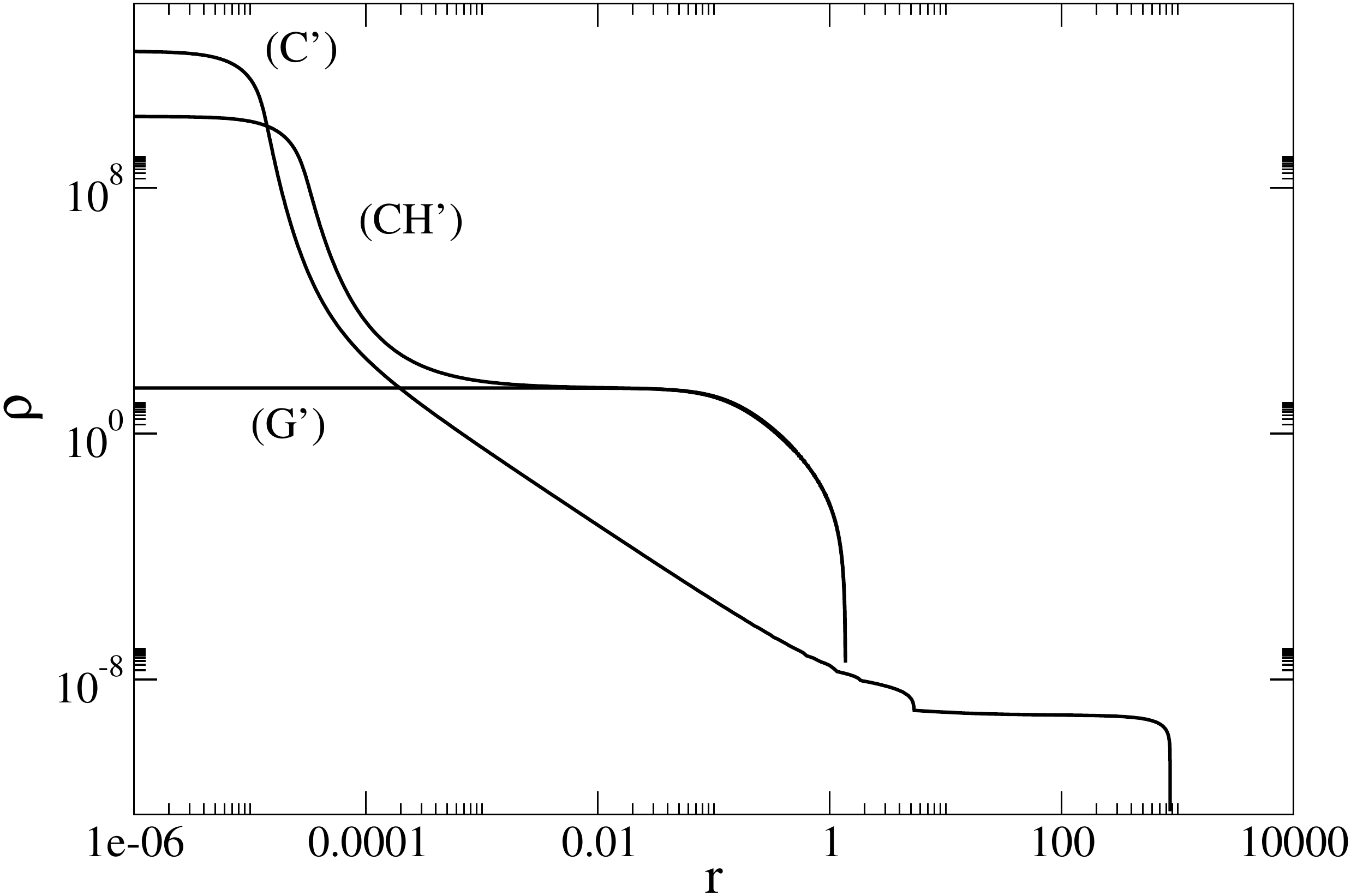}
\caption{Density profiles of the gaseous (G'), core-halo (CH')
and condensed (C') solutions identified in Fig. \ref{mutresgrandprolongehenon}
(from \cite{clm2}).}
\label{rhoABCmugrand}
\end{center}
\end{figure}

Let us now discuss the thermodynamical stability 
of the core-halo solution considered by Arg\"uelles {\it et al.} \cite{krut}. 
The
caloric curves of the fermionic King model in Newtonian gravity for arbitrary
values of $\mu$ were first studied by Chavanis {\it et al.} \cite{clm2}.  The
caloric curve corresponding to a large value of $\mu$ (i.e. having the
characteristics of the Milky Way) is plotted in Fig.
30 of
\cite{clm2}. In that paper, we have focused on the density profiles of the
solutions located in the region of the spiral (see Fig. 44 of \cite{clm2}). For
a given energy in that region, we
found  a gaseous solution (G'), a core-halo solution (CH') and a condensed
solution (C').  These results are reproduced in Figs.
\ref{mutresgrandprolongehenon} and \ref{rhoABCmugrand} for convenience. The
gaseous solution (G') corresponds to
the classical isothermal sphere. Since it lies before the first turning point of
energy, it is thermodynamically stable in the microcanonical ensemble (maximum
entropy state at fixed mass and energy).  The condensed solution (C') is also
stable in the microcanonical
ensemble because it lies after the last turning point of energy. However, this
solution is not astrophysically relevant because it has a too extended halo that
is not consistent with the structure of DM halos (see Fig. \ref{rhoABCmugrand}).
The core-halo solution (CH') is
similar to the solution found by  Ruffini {\it et al.} \cite{rar} which  was
claimed to reproduce the structure of the Milky Way. It consists in a large
nondegenerate isothermal atmosphere harboring a small ``fermion ball''  with a
high density, a large mass and a small radius that could mimic a SMBH. Since
this solution lies between the first and the last turning points of energy, it
is thermodynamically unstable in the microcanonical ensemble (saddle point of
entropy at fixed mass and energy). Therefore, we concluded in \cite{clm2} that
this type of solution is not likely to result from a natural evolution and,
consequently, we questioned the possibility  that a fermion ball could  mimic a
central SMBH.

However, in our analysis, we did not consider the stable solution (CH)$_*$
located just after the turning point of energy $E_*$, believing that
this solution would be unreachable by a natural
evolution or that it would
look like the solution (C') which has a too extended halo.  Recently, 
Arg\"uelles {\it et al.} \cite{rarnew} computed the caloric curves of
the
fermionic King model in general relativity. For not too negative
energies\footnote{For smaller energies, it becomes crucial to
take general relativity into account (see Sec. \ref{sec_gr}).
In that case, a new
turning point of energy appears which was first evidenced by 
Alberti and Chavanis \cite{caf,acf} in the framework of the box model (see
Fig. \ref{fbbh}). Below this critical energy the system collapses
towards a SMBH as discussed in Sec. \ref{sec_gr}.} they obtained a caloric
curve
similar to the one represented in Fig.
\ref{mutresgrandprolongehenon}.
They confirmed the instability of the (CH')  solutions in the region of the
spiral previously considered by Chavanis {\it et al.} \cite{clm2} but they also
investigated the solution (CH)$_*$ close to $E_*$ and showed that this solution
actually
corresponds to the density profile obtained in their previous work
\cite{krut} which provides a good agreement with the structure of the
Milky Way. Since this solution is located after the last
turning point of energy it is thermodynamically stable in the microcanonical
ensemble. This is a very interesting result because it shows that the core-halo
structure found by Arg\"uelles {\it et al.} \cite{krut}   is
thermodynamically
stable and can, therefore, arise from a natural evolution. 

In conclusion, when we use the ordinary Fermi-Dirac DF, the core-halo 
solution purported to reproduce the structure of the Milky Way is
thermodynamically
unstable  but when we
use
the fermionic King model, this
core-halo solution is thermodynamically stable.
Therefore, this core-halo configuration may result from a natural evolution in
the sense of Lynden-Bell. This gives further support to the scenario according
to which    a fermion ball could mimic a SMBH
at the centers of the
galaxies.

{\it Remark:} The discovery \cite{rarnew} that 
the core-halo solution (CH)$_*$ with a compact fermion ball mimicking a SMBH is
thermodynamically stable is a very important result. However,  it does not prove
that this structure will effectively arise from a natural evolution. The reason
is
that violent relaxation is in general incomplete \cite{lb,incomplete}. In
particular, the fluctuations of the gravitational potential that are the engine
of the collisionless relaxation can die out before the system has reached
statistical equilibrium. Therefore, it is not
clear if violent relaxation can produce
this type of structures with a very high central density.\footnote{It may be
easier to form core-halo configurations with a very high central density if the
fermions are self-interacting and if the Fermi-Dirac equilibrium state results
from a collisional evolution of nongravitational origin as
discussed in Sec. \ref{sec_kt}.} In order to
vindicate
this scenario, the next step would be to perform direct numerical simulations
of collisionless fermionic matter to see if they spontaneously generate fermion
balls with the characteristics of SMBHs. Indeed, it is not
clear why the system should spontaneously reach an equilibrium state that is
just in the bend
after the turning point of energy $\Lambda_*$. The purely gaseous solution
(G')
without a quantum core, which is also a
maximum entropy state, may be easier to reach through a violent relaxation
process and is consistent with the observations. However, it does not account
for a massive central object at the centers of the galaxies. In that case,
we either have to introduce a primordial SMBH ``by hand'' or advocate a
scenario of gravitational collapse such as the one discussed in the following
section. 

\subsection{General relativistic collapse towards a SMBH}
\label{sec_gr}

\begin{figure}
\begin{center}
\includegraphics[clip,scale=0.3]{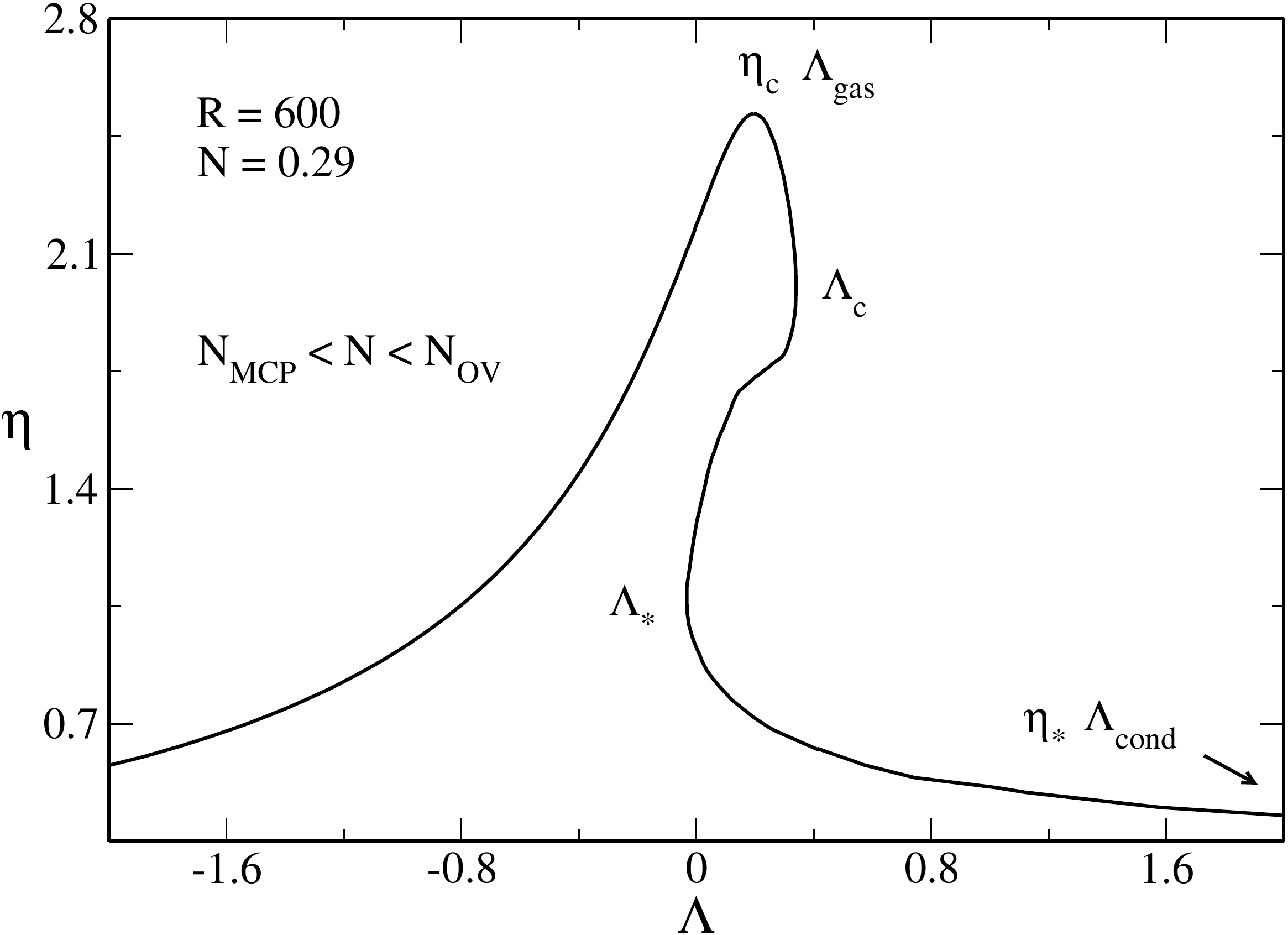}
\includegraphics[clip,scale=0.3]{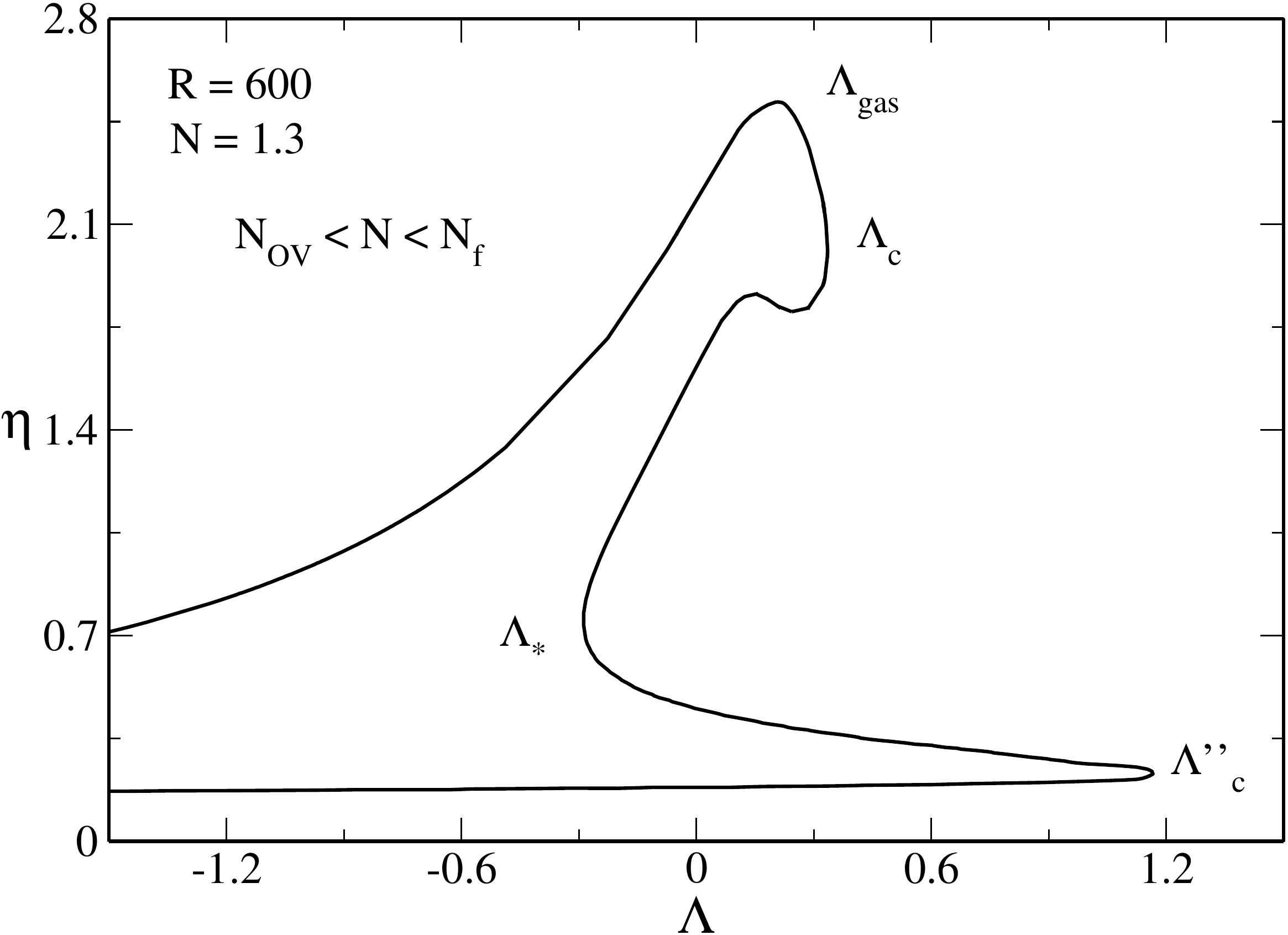}
\includegraphics[clip,scale=0.3]{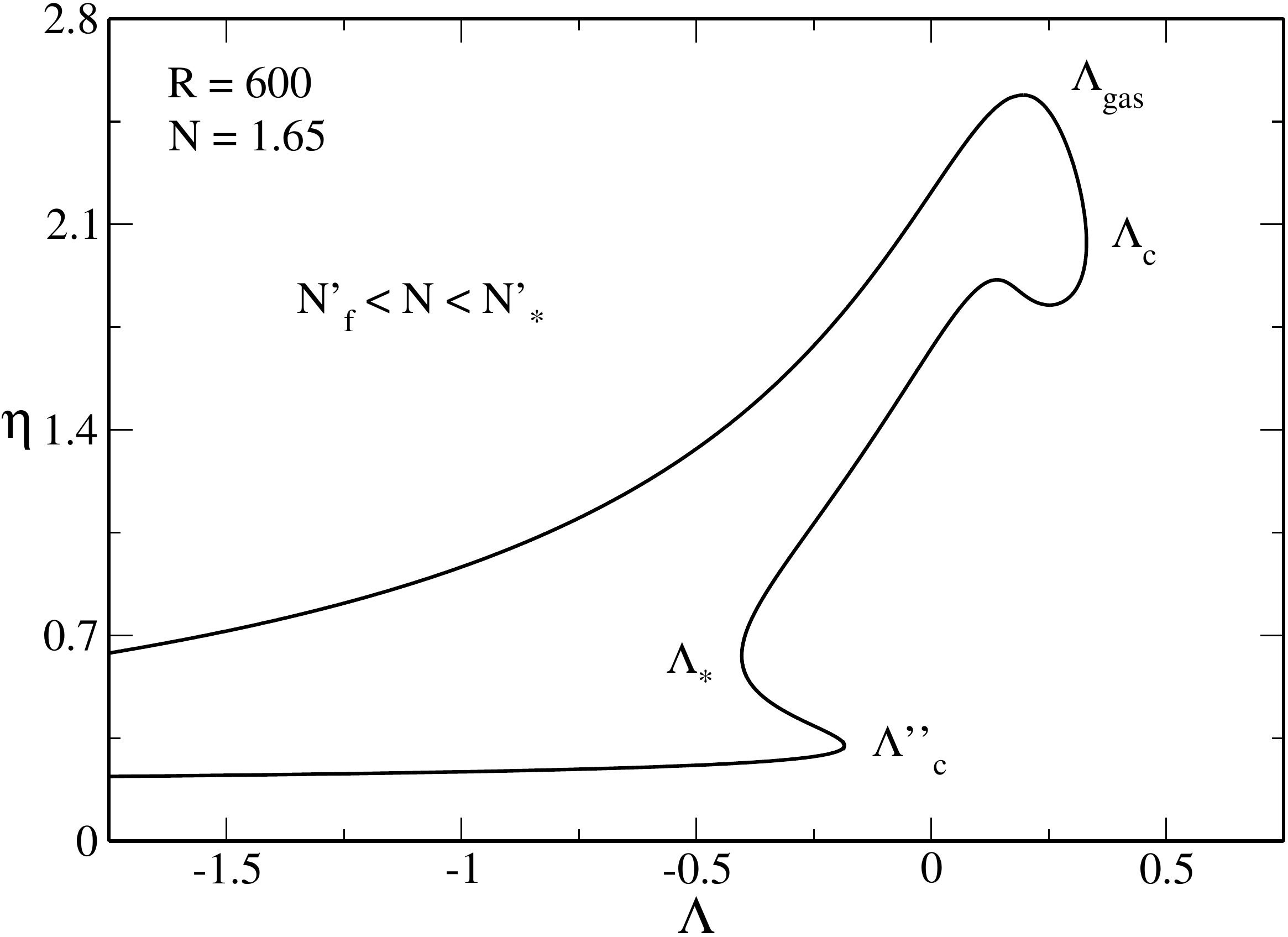}
\includegraphics[clip,scale=0.3]{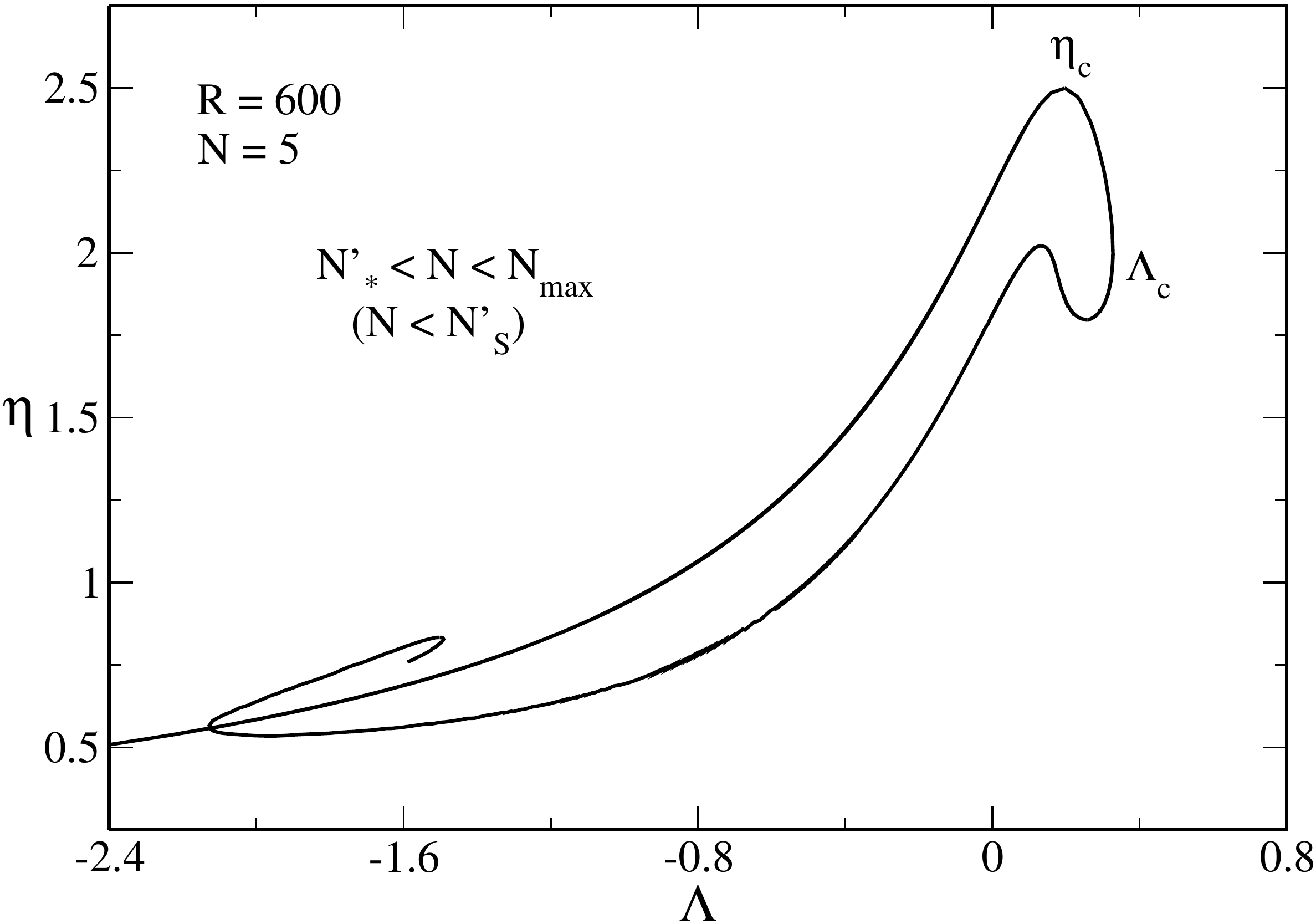}
\caption{Caloric curve of the general relativistic Fermi gas in a box as a
function of the particle number $N$ (adapted from \cite{acf}). For $N<N_{\rm
OV}$, the gravothermal catastrophe at $E_c$ leads to a fermion ball surrounded
by a hot halo. For $N_{\rm OV}<N<N'_*$ the system first takes a quantum
core-halo structure resulting from the  gravothermal catastrophe at $E_c$ (as
before) then collapses towards a SMBH at
$E''_c$. For $N>N'_*$ the condensed branch disappears so that only the collapse
at $E''_c$ towards a SMBH is possible. [These caloric curves
are valid for relatively small DM halos.
For larger halos a spiral develops in the head of the dinosaur but the
phenomenology remains the same].}
\label{total}
\end{center}
\end{figure}

For a fermion mass $m=48\, {\rm keV/c^2}$, the mass $M_h=10^{11}\, M_{\odot}$ of
the Milky Way is larger than the OV mass $M_{\rm OV}=2.71\times 10^8\,
M_{\odot}$, so
we have to take into account general relativity effects in the caloric curve. As
first shown by  Alberti and Chavanis \cite{caf,acf} for box-confined systems,
and
recovered by Arg\"uelles {\it et al.} \cite{rarnew} for tidally
truncated models,
relativistic effects create a new turning point of energy in
the caloric
curve at which the
condensed branch terminates (see Fig. \ref{fbbh}). Below
$E''_c$ the system collapses towards
a black hole. 
As we have seen previously, two stable equilibrium states are relevant 
in the structure of DM halos: the gaseous solution (G') equivalent to the
classical isothermal sphere and the core-halo solution (CH)$_*$ which
contains a fermion ball mimicking a
SMBH. Only direct numerical simulations can tell us which metaequilibrium state
will be reached in practice from a violent collisionless relaxation.
Since these
numerical results are not available yet, we shall consider the two
possibilities.  If the system were truly collisionless, the DM halo would remain
in the metaequilibrium state (G') or (CH)$_*$ for ever. In order to be more
general, we consider
below the possibility that the system slowly evolves dynamically due to
collisions and evaporation. There are then two situations to consider:

(A) Suppose that violent relaxation selects the gaseous solution (G'). 
On a secular timescale, the system follows the upper series of equilibria from
point (G') to the point of minimum energy $E_c$. At that point, it becomes
thermodynamically unstable and undergoes a gravothermal catastrophe up to point
(C') where the collapse is stopped by quantum mechanics, leading to
the formation of a fermion ball.
Then, if the energy keeps decreasing, the system  follows the lower series of
equilibria up to the point of minimum energy $E''_c$ where it becomes
thermodynamically and dynamically unstable (in a general relativistic sense) and
collapses towards a SMBH.\footnote{This requires that the core
mass increases until it reaches the critical OV value. The increase of the core
mass may take place through an accretion process.} As discussed in \cite{acf},
there
are two possible evolutions: (i) If the particle number $N$ is below a critical
value
$N'_*$, then $\Lambda_c<\Lambda_c''$ and the system is first arrested by quantum
mechanics before collapsing
towards a BH.  (ii) If the particle number $N$ is above a critical value $N'_*$,
then $\Lambda_c>\Lambda_c''$ and the system directly collapses towards a
BH. These two possibilities are illustrated in Fig. \ref{total}.

(B) Suppose that violent relaxation selects the core-halo 
solution (CH)$_*$ where the fermion ball mimics a SMBH. On a secular timescale,
the system follows the series of equilibria from point (CH)$_*$ to the point
of minimum energy $E''_c$. At that point, it becomes thermodynamically and
dynamically unstable (in a general relativistic sense) and collapses towards a
SMBH.

In the two cases, the ultimate fate of the system 
is to form a SMBH surrounded by an envelope. This
picture may be just qualitative
because it is not clear if the lower branch of equilibrium states is
astrophysically relevant. Indeed, we have
indicated that the envelope of the solutions (C') is too much extended to match
the
characteristics of DM halos. Therefore, the collisional evolution of the system
from point (G') or from point (CH)$_*$ up to the formation of a SMBH at
$E''_c$ may
involve out-of-equilibrium states (CH)$_{\rm out}$ instead of following the
series of equilibrium
states (C').

For a fermion mass $m=48\, {\rm keV/c^2}$, the OV mass $M_{\rm OV}=2.71\times
10^8\, M_{\odot}$ is too large to account for the mass of a SMBH like Sgr A$^*$
at the center of the Milky Way. Either the mass of the SMBH resulting from
gravitational collapse is smaller than $M_{\rm OV}$ or there is no
gravitational collapse and Sgr A* is a fermion ball (CH)$_*$ as suggested
by Arg\"uelles {\it et al.} \cite{rarnew}. Therefore, a
fermion ball is favored in medium size galaxies like the Milky Way. However, for
very large halos it is
shown by Alberti and Chavanis \cite{acf} that the condensed branch disappears
(see the last panel of Fig. \ref{total}).\footnote{It would be
interesting to determine precisely the
condition of disappearance of the condensed branch in the framework of the
relativistic fermionic King model, i.e., the value of $N'_*$.} In that case,
there is no solution with a fermion ball such as (CH)$_*$ and the system 
necessarily collapses towards a SMBH. Therefore, medium size galaxies
($N<N'_*$) like the Milky
Way may harbor a fermion ball of mass $M=4.2\times 10^6\, M_{\odot}$ while
very large galaxies ($N>N'_*$) like
ellipticals may harbor a SMBH of mass $M_{\rm OV}=2.71\times
10^8\, M_{\odot}$ that could even grow by accretion. This
could account for the mass of SMBHs in AGNs like the one
recently photographed in M87 ($M_h\sim 10^{13}\, M_{\odot}$ and $M_{\rm BH}\sim
10^{10}\, M_{\odot}$).

For a fermion mass $m=386\, {\rm keV/c^2}$, the OV mass $M_{\rm
OV}=4.2\times 10^6\, M_{\odot}$ is comparable to the mass of Sgr A$^*$.
Furthermore,  the caloric
curve is similar
to the one reported in the last panel of Fig. \ref{total} and there is no
possibility to have a solution  (CH)$_*$ involving a fermion ball. In that
case, the Milky Way could have undergone a gravitational collapse leading to
a SMBH of mass $M_{\rm OV}=4.2\times 10^6\, M_{\odot}$. The
halo surrounding the SMBH is left undisturbed and could correspond to a marginal
classical King profile which gives a good
agreement with the Burkert profile (see Ref. \cite{clm1} and Fig.
\ref{densityLOG}).

\subsection{Potential problems with a DM model involving a fermion mass $m=48\,
{\rm keV/c^2}$ or $m=386\, {\rm keV/c^2}$}
\label{sec_pot}

In Sec. \ref{sec_mh} we have determined the mass $m$ of the DM particle by
arguing that the smallest halo observed in the universe (``minimum halo'') with
a typical mass $M\sim 10^8\, M_{\odot}$ and a typical radius $R\sim 1\, {\rm
kpc}$
(Fornax) represents the ground state of the self-gravitating Fermi gas at $T=0$.
This yields $m=165\, {\rm eV/c^2}$. This value (previously given in Appendix D
of \cite{abrilphas}) is of the order
of magnitude of the fermion mass obtained by other authors \cite{du,rsu,bbbk}
using
more detailed comparisons with observations.\footnote{Domcke and
Urbano \cite{du} model dSphs as a completely degenerate fermionic system and
find that $m=200\, {\rm eV/c^2}$ provides the best fit to observations of
velocity dispersion. Randall {\it et al.} \cite{rsu} show that self-gravitating
fermions under full-degeneracy do not fit well the velocity dispersion data of
some local dwarfs and introduce by hand a Boltzmannian tail (i.e. finite
temperature effects) in order to better reproduce the data.  They find good
agreement for $70\, {\rm eV/c^2} <m<400\, {\rm eV/c^2}$. Bar {\it et al.}
\cite{bbbk} study the globular cluster timing problem in Fornax assuming that
the core is a completely degenerate fermion ball. They find $m=135\, {\rm
eV/c^2}$ but point out that this mass violates the Lyman-alpha limit.}
Alternatively, Arg\"uelles {\it et
al.} \cite{krut,rarnew}
determined the mass of the fermionic DM particle in such a way that the
fermion ball that composes the core-halo structure of a large DM halo like
the Milky Way, obtained
in the framework of the fermionic King model, mimics the effect of a SMBH at the
center of the Galaxy. This leads to a much larger mass
$m=48\, {\rm keV/c^2}$.\footnote{If we use the nonrelativistic mass-radius
relation (\ref{cdl10}) of a fermion ball at $T=0$ and take $M_c=4.2\times
10^{6}\, M_{\odot}$ and $R_c=6\times 10^{-4}\,
{\rm pc}$, corresponding to the characteristics of the massive object at the
center of our Galaxy (see Sec. \ref{sec_sagittarius}), we get $m=54.6\, {\rm
keV/c^2}$.} In very recent works, Becerra-Vergara {\it
et al.} \cite{bvetal1,bvetal2} showed that the gravitational potential of a
fermion ball (with a particle mass $m=56\, {\rm keV/c^2}$) leads to a better fit
of the orbits of all the $17$ best resolved S-stars orbiting Sgr A$^*$
(including the S2 and G3 objects) than the one obtained by the
central SMBH model.

A possible problem with this model is the following. If the DM particle had a
mass $m=48\,
{\rm keV/c^2}$, the minimum halo (ground state) would be too small: it would
have a mass  $(M_{h})_{\rm min}=1.30 \, M_{\odot}$
and a
radius $(r_h)_{\rm min}=0.0683\, {\rm pc}$. This would imply the formation of
structures at very small scales, up to  $\sim 1 \, M_{\odot}$. Therefore,
DM halos should exist up to very small scales, like in the CDM model.
Indeed, (bosonic or fermionic) quantum  models with a large
particle mass $m$ behave essentially  as CDM. This is not what we observe.
There is
apparently no DM halos with a mass below $\sim 10^8\, M_{\odot}$, leading to the
missing
satellite problem \cite{satellites1,satellites2,satellites3}. This is why
quantum models of DM with a {\it small} particle mass have been introduced.
Namely, they have been introduced
precisely in order to have a ground state (minimum halo) with a typical
mass $M\sim 10^8\, M_{\odot}$ and a typical radius $R\sim 1\, {\rm kpc}$,
corresponding to dSphs like Fornax, not smaller.
Accordingly, a fermionic model with $m=48\,
{\rm keV/c^2}$ may not be able to solve the missing
satellite problem. 

\begin{figure}
\begin{center}
\includegraphics[clip,scale=0.3]{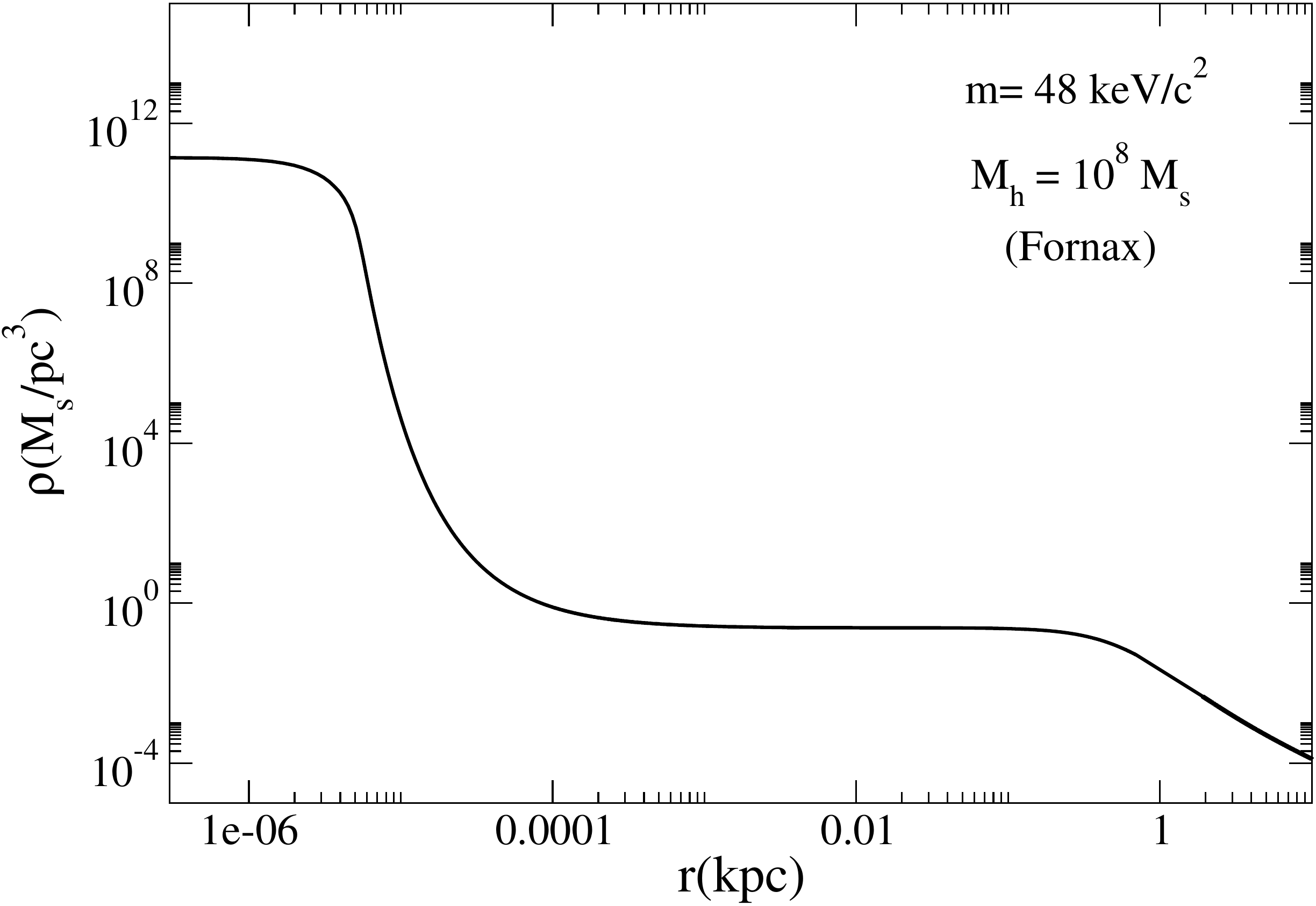}
\caption{Density profile
of a DM halo of mass $M_h=10^{8}\,
M_{\odot}$ (Fornax) assuming that the fermion mass is $m=48\, {\rm
keV/c^2}$. It presents a core-halo structure.}
\label{fornaxBH}
\end{center}
\end{figure}

\begin{figure}
\begin{center}
\includegraphics[clip,scale=0.3]{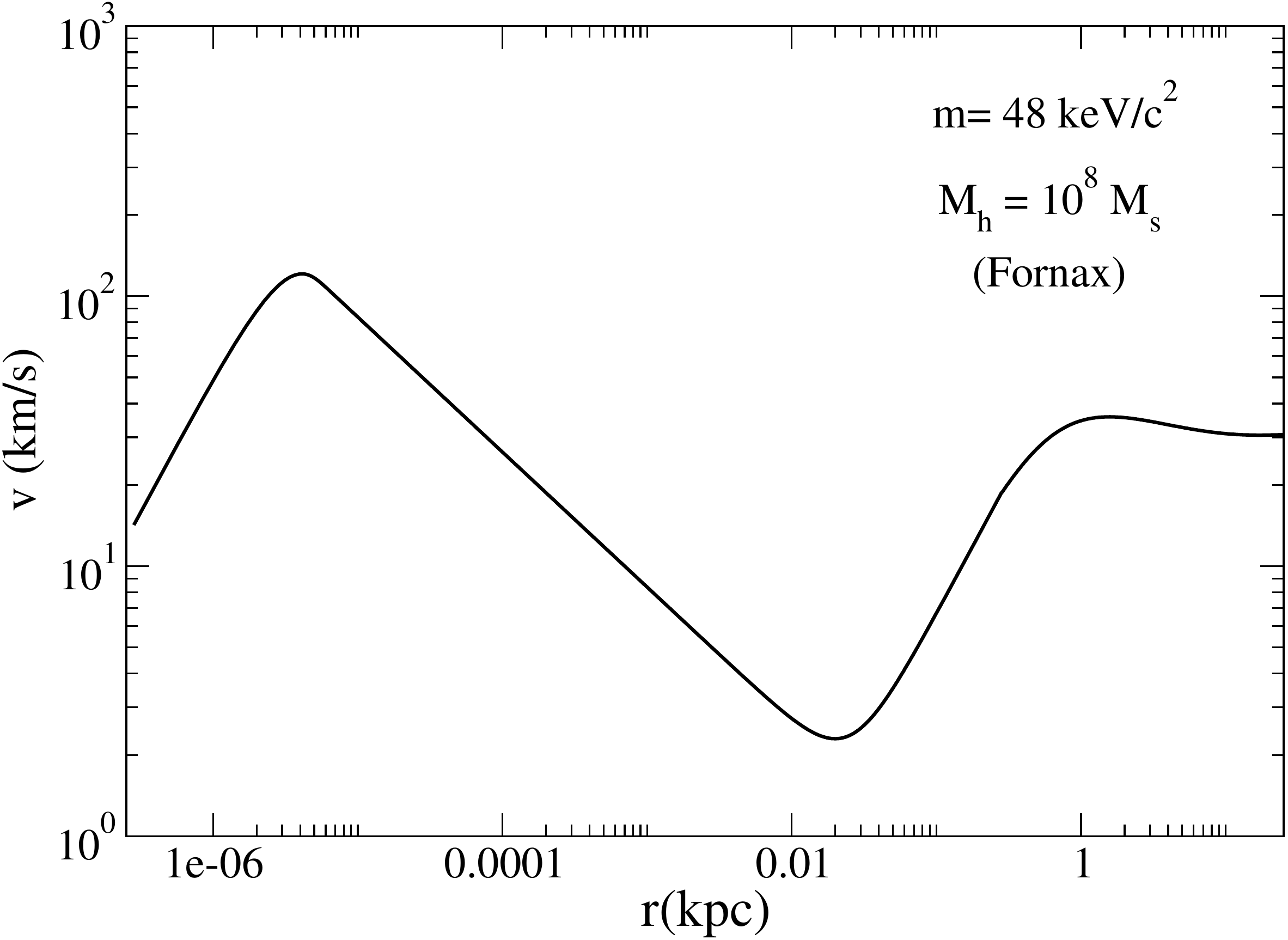}
\caption{Velocity profile
of a DM halo of mass $M_h=10^{8}\,
M_{\odot}$ (Fornax) assuming that the fermion mass is $m=48\, {\rm
keV/c^2}$.}
\label{fornaxBHvitesse}
\end{center}
\end{figure}

If we disregard this difficulty, another consequence 
of the model of Arg\"uelles {\it et al.} \cite{krut,rarnew} is that dSphs should
have a very pronounced core-halo structure (since they do not correspond to the
ground state
of the self-gravitating Fermi gas). For
example, a compact DM halo of mass $M_h=10^{8}\, M_{\odot}$ (Fornax) should have
a core-halo structure with a small central fermion
ball (possibly mimicking an intermediate mass BH) and an atmosphere. Using 
Eqs. (\ref{cdl18}), (\ref{mcmh3}), (\ref{mcmh7}) and (\ref{logcorr}), we find
that this DM halo
should contain a
quantum core of mass $M_c=1.57\times 10^4\, M_{\odot}$,
radius $R_c=5.42\,
{\rm mpc}$ and central density $\rho_0=1.40\times 10^{11}\, M_{\odot}/{\rm
pc}^3$.\footnote{Interestingly, these
values obtained from our
semi-analytical model  [see
in particular Eq. (\ref{mcmh7})] are comparable to the values  obtained
numerically in
\cite{krut2}.} The corresponding
density and velocity profiles are plotted in Figs. \ref{fornaxBH} and
\ref{fornaxBHvitesse}. To our
knowledge, this
core-halo structure has not been observed in ultracompact DM halos. dSphs are
rather expected to correspond to pure fermion balls at $T=0$ (possibly
surrounded by a tenuous atmosphere). Therefore, they are expected to have a
profile similar to Figs. \ref{profilepolytrope}
and \ref{Vpoly} instead of Figs. \ref{fornaxBH} and
\ref{fornaxBHvitesse}. It would be
extremely important to clarify this issue by
applying the model of Arg\"uelles {\it et al.} \cite{krut} to ultracompact halos
in order to determine which of the two scenarios (the scenario of Arg\"uelles
{\it et al.} \cite{krut,rarnew} with $m=48\,
{\rm keV/c^2}$ or the one developed in the present paper with $m=165\, {\rm
eV/c^2}$ or $m\sim 1\, {\rm keV}/c^2$) is the most relevant.

There is also a problem related to the validity of the
Fermi-Dirac (or Lynden-Bell) DF as discussed further in Sec. \ref{sec_tqc}.
Indeed, for
a large fermion mass $m\gg 1\, {\rm
keV}/c^2$, the DM halo is essentially classical except in a
very small quantum core (fermion ball). Away from the core, we
should recover the NFW profile
leading to cusps. It is precisely in order to avoid these cusps that quantum
models of DM with a {\it small} particle mass $m\lesssim 1\, {\rm
keV}/c^2$ have been introduced.

\section{Possible solutions to an apparent paradox related to the universal surface density of DM halos}
\label{sec_paradox}

\subsection{The apparent paradox}
\label{sec_ap}

The mass-radius relation of a 
completely degenerate  fermion ball (ground state of the self-gravitating Fermi
gas at $T=0$) is given by [see Eq. (\ref{cdl10})]
\begin{equation}
\label{cdl10bis}
R=0.114\, \frac{h^2}{G m^{8/3} M^{1/3}}.
\end{equation}
The radius decreases like $M^{-1/3}$ as the mass increases. Therefore, if we
identify $M$ with the halo mass $M_h$ and $R$ with the halo radius $r_h$, this
result is in contradiction with the universality of the surface density of DM
halos [see Eq (\ref{nd13})]  implying that the radius increases with the
mass as $M^{1/2}$ [see Eq. (\ref{nd14})]. A similar problem arises in the BECDM
model.

This apparent paradox was pointed out by the author at several occasions in the
case of fermions and bosons (see, e.g., Appendix F of Ref. \cite{clm2}, the
Introduction of Ref. \cite{chavtotal} and Appendix L of \cite{modeldm}). It has
also been recently emphasized by  Deng {\it et al.} \cite{deng} and 
Burkert \cite{burkertfdm} in the case of bosons. A possible implication of
this paradox is that the quantum (fermionic and bosonic) models of DM are ruled
out because they are not consistent with the constraint from Eq. (\ref{nd13}).
This is essentially the conclusion reached by Deng {\it et al.} \cite{deng} and 
Burkert \cite{burkertfdm} for the BECDM model. Below, we
discuss several possible solutions to this apparent paradox that were suggested
in \cite{modeldm} for bosonic DM and that can be straightforwardly adapted to
fermionic DM.

{\it Remark:} The constant surface density of DM halos
$\Sigma_0=\rho_0r_h=141_{-52}^{+83}\, M_{\odot}/{\rm
pc}^2$ may be explained by the logotropic model developed
in \cite{epjp,lettre,jcap,pdu,graal} which involves a logotropic envelope
instead of an isothermal one. This model not only explains why the
surface density of DM halos is constant but it also determines its universal
value
(in agreement with the observations) in terms of fundamental constants
$\Sigma_0^{\rm th}=0.01955\, c\sqrt{\Lambda}/G=133\, M_{\odot}/{\rm
pc}^2$ without adjustable parameter.
At the same
time, in a cosmological context, it correctly  accounts for the accelerating
expansion of the Universe with
a single dark fluid.

\subsection{Model I: purely gaseous solution}
\label{sec_model1}

A first possible solution to this 
problem is that DM halos do not have a quantum core such as a fermion ball or
such as a soliton (in the BECDM model). Indeed, the DM halos could be in the
purely gaseous phase (G) corresponding to the classical isothermal sphere (see
Sec. \ref{sec_nd}). This solution is always thermodynamically stable (maximum
entropy state) so it represents the most probable state of the system.
Furthermore, it is always possible to satisfy the constraint from Eq.
(\ref{nd13}) by adapting the temperature [see Eq. (\ref{nd14})]. This leads to
the mass-radius relation from Eq. (\ref{nd14}). As shown in \cite{modeldm},
the classical isothermal distribution (without quantum core) is fully consistent
with the observational Burkert profile and can therefore represent a satisfying
description of DM halos. It is nevertheless crucial to take quantum mechanics
into account in the case of ultracompact DM halos with a small mass,
corresponding to dSphs like Fornax. This leads to the Model I  of Ref.
\cite{modeldm} in which the DM halos are purely isothermal (without quantum
core) except near the ground state. More precisely:

(i) At $(M_h)_{\rm min}$ the DM halo is completely degenerate
 (see Sec. \ref{sec_cdl}). The values of $M_h$ and $r_h$ for this minimum halo
are consistent with the constraint from Eq. (\ref{nd13}).

(ii) Ultracompact DM halos with a mass $(M_h)_{\rm min}\le M_h\le (M_h)_{\rm
CCP}$ have 
a quantum core surrounded by a tenuous isothermal atmosphere. The presence of a
small isothermal halo allows us to satisfy the constraint from Eq. (\ref{nd13})
as discussed in Sec. VI of \cite{modeldm} for BECDM halos. All the profiles
constructed in Sec. VI of \cite{modeldm} satisfy the constraint from
Eq. (\ref{nd13}). The same results apply to fermionic DM halos.

(iii) DM halos with a mass $M_h\ge (M_h)_{\rm CCP}$ are purely 
isothermal without quantum core. Indeed, as shown in \cite{modeldm} for BECDM
halos, if we enforce the constraint from Eq. (\ref{nd13}) in Model I we find
that the core mass {\it decreases} as the halo mass increases so that large DM
halos are essentially classical without quantum core.

\begin{figure}
\begin{center}
\includegraphics[clip,scale=0.3]{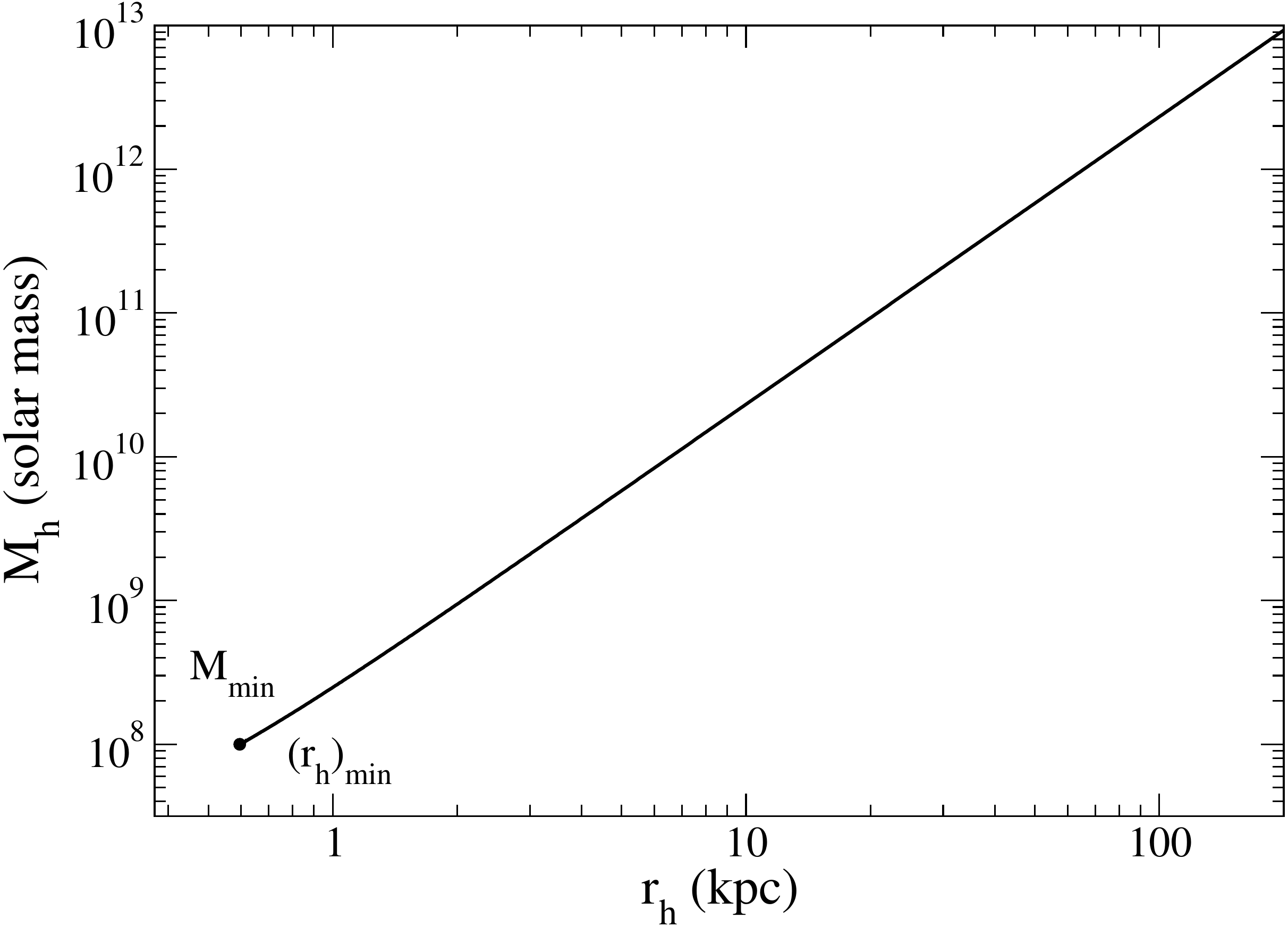}
\caption{Mass-radius relation of
fermionic DM halos. In
Model I, quantum mechanics (Pauli's exclusion principle) is important
only close to the ground state (bullet) where the halos have the form of a
fermion ball surrounded by a tenuous isothermal atmosphere. Larger DM halos
are purely isothermal without a quantum core. Note that the halo mass-radius
relation from Eq. (\ref{nd14}) remains valid for large halos in Models II and
III
since the quantum core mass $M_c$ is always much smaller than the halo mass
$M_h$ (see Sec. \ref{sec_nd}).}
\label{massradius}
\end{center}
\end{figure}

This model leads to the mass-radius relation reported in Fig. 16 of
\cite{modeldm} and reproduced in Fig. \ref{massradius} (adapted to fermions).
It coincides
with the classical law from Eq. (\ref{nd14}) except at small halo masses.
Quantum mechanics just determines the ground state of DM halos at $(M_h)_{\rm
min}=10^8\, M_{\odot}$ and $(r_h)_{\rm min}=597\, {\rm pc}$.

This scenario does not account for the presence of a compact object, such as a
SMBH, at the centers of the galaxies. Of course, we can always add ``by hands''
a primordial SMBH at the center of a classical isothermal halo but this is
almost assuming the result. In order  to explain self-consistently the presence
of a SMBH at the center of the galaxies, we can consider the following
scenarios.

\subsection{Model II: core-halo solution with a large quantum bulge}
\label{sec_model2}

Another possibility to solve the paradox of Sec. \ref{sec_ap} and ``save'' the
quantum core-halo solution (CH) from Sec. \ref{sec_app}  is to assume that the
constraint from Eq. (\ref{nd13}) should be replaced by
\begin{equation}
\Sigma_0=\rho_c r_h=141_{-52}^{+83}\, M_{\odot}/{\rm
pc}^2,
\label{nd13mod}
\end{equation}
where $\rho_c$ is not the true central density $\rho_0$ but rather an 
``apparent'' central density. It corresponds to the density at the separation
between the quantum core and the classical halo in configurations such as those 
from Fig. \ref{profileMW}. Similarly, $r_h$ is the radius at which $\rho_c$
(instead of $\rho_0$)
is divided by $4$. The idea underlying this replacement is that observations may
not be able to resolve the presence of a quantum bulge at the centers of
galaxies. Therefore, we have to distinguish between the true central density
$\rho_0$ (which is the central density of the quantum core) and the apparent
central density $\rho_c$ (which is the ``central'' density of the classical
isothermal halo surrounding the quantum core). Similarly, we have to distinguish
the
halo radius $r_h$ which typically corresponds to the distance where the apparent
central density $\rho_c$ is divided by $4$ from the core radius $R_c$ which is
of the order of the distance where the core central density $\rho_0$ is divided
by $4$. The distinction between these quantities is explicitly shown in
Fig. \ref{profileMW}. It is
clear that the radius of large DM halos is given by
$r_h$ not by the quantum core radius $R_c$. This leads to Model II of 
\cite{modeldm} in which the DM halos of large mass have a core-halo structure.
More precisely:

(i) At $(M_h)_{\rm min}$ the DM halo is completely degenerate (see
Sec. \ref{sec_cdl}). The values of $M_h$ and $r_h$ for this minimum halo are
consistent with the constraint from Eq. (\ref{nd13}).

(ii) Ultracompact DM halos with a mass $(M_h)_{\rm min}\le M_h\le (M_h)_{\rm
CCP}$ have a quantum 
core surrounded by a tenuous isothermal atmosphere. The presence of a small
isothermal halo allows us to satisfy the constraint from Eq. (\ref{nd13}) as
discussed in Sec. VI of \cite{modeldm} for BECDM halos. All the profiles
constructed in Sec. VI of \cite{modeldm} satisfy the constraint from Eq.
(\ref{nd13}). The same results apply to fermionic DM halos.

(iii) Small DM halos with a mass $(M_h)_{\rm CCP}\le M_h\le (M_h)_{\rm MCP}$
have a
core-halo 
structure made of a large quantum bulge surrounded by a classical isothermal
halo. The core mass $M_c$  increases with the halo mass $M_h$ according to Eq.
(\ref{mcmh7}). The constraint from Eq. (\ref{nd13})  is satisfied provided
that we replace the central density $\rho_0$ by the apparent central density
$\rho_c$, i.e., provided that we use Eq. (\ref{nd13mod}). From the outside (i.e.
considering the external structure of the DM halo and ignoring the quantum
bulge), the system looks like a classical isothermal sphere. This leads to the
mass-radius relation from Eq. (\ref{nd14}) which is
consistent with the observations.

(iv) For large DM halos with a mass  $M_h\ge (M_h)_{\rm MCP}$, the quantum
bulge
is replaced either by a small out-of-equilibrium quantum core or by a SMBH.
In
that case, the
replacement of Eq. (\ref{nd13}) by  Eq. (\ref{nd13mod}) is even more justified.
What is relevant is not the central density of the compact object but rather the
density of the classical halo at the contact with this object.\footnote{The
same comment holds if the quantum bulge in scenario (iii) has led to the
formation of a SMBH by accretion of the gas (see Sec. \ref{sec_large}).
}

This model leads to the mass-radius relation from Fig. \ref{massradius}.
It coincides with the classical law from Eq. (\ref{nd14})
except at small halo masses. Quantum mechanics determines the ground state of DM
halos  at $(M_h)_{\rm
min}=10^8\, M_{\odot}$ and $(r_h)_{\rm min}=597\, {\rm pc}$.  It also implies
the existence of a large quantum bulge of mass $M_c$ that increases with the
halo mass $M_h$ according to Eq. (\ref{mcmh7}). For very large halos, the
fermionic bulge is replaced by  an out-of-equilibrium compact quantum
core or by a SMBH.

These arguments may solve, or alleviate, the problem reported in Sec.
\ref{sec_ap}. The crucial point is to know if observations are able to
detect a large quantum bulge of typical mass $M_c=9.45\times 10^9\,
M_{\odot}$ and size $R_c=240\, {\rm pc}$ at the center of the Milky Way that is
predicted by our model. This
possibility is discussed in Sec. \ref{sec_bulge}.

\subsection{Model III: core-halo solution mimicking a SMBH}

Finally, we note that the apparent paradox reported in Sec. \ref{sec_ap} does
not arise
in the model of Arg\"uelles {\it et
al.} \cite{krut,rarnew} where the fermion ball mimics a SMBH of negligible
extent. Indeed, in that case, there is a clear separation between the quantum
core and the classical isothermal halo as depicted in Fig. \ref{profileBH}.
It is
clear that the central density to consider in Eq. (\ref{nd13}) is not the
central density $\rho_0$ of the fermion ball but rather the density $\rho_c$ of
the plateau that connects the fermion ball to the classical isothermal halo.
Similarly, the halo radius $r_h$ is not the radius where the central density
is divided by $4$ (which would coincide with the radius of the quantum
core $R_c$) but the radius where the density of the plateau is divided by
$4$.

In model III, the DM halo behaves from the outside as a 
classical isothermal halo but harbors a tiny massive fermion ball mimicking a
SMBH. The halo mass-radius relation is similar to that reported in Fig.
\ref{massradius} except that it starts at a much smaller minimum halo mass
$(M_h)_{\rm min}=1.30\, M_{\odot}$ corresponding to $(r_h)_{\rm
min}=0.0683\, {\rm pc}$.

\section{Thermal or quantum core?}
\label{sec_tqc}

In this paper, we have assumed that a fermionic DM halo reaches a statistical
equilibrium state described by the Fermi-Dirac
DF (see Sec. \ref{sec_kt}). The same assumption was made by
other authors \cite{dvs1,rarnew,btv,du,rsu}. However, this is a strong
assumption and the establishment of a statistical equilibrium state for
self-gravitating systems is far from trivial. 

Let us first consider a collisionless system of classical 
self-gravitating particles. According to the statistical theory of Lynden-Bell
\cite{lb}, it
should reach a Fermi-Dirac-like DF, reducing to the classical isothermal DF in
the dilute limit. The corresponding density profile has a core due to effective
thermal effects (in the sense of Lynden-Bell). However, this prediction is
{\it not}
consistent with numerical simulations of 
classical collisionless self-gravitating systems. Indeed, such simulations lead
to NFW profiles \cite{nfw} presenting a $r^{-1}$ central cusp, not a core. This
demonstrate that, for classical collisionless self-gravitating systems, the 
Lynden-Bell
prediction does not work in the central part of the system.\footnote{It does not
work well neither in the outer part of the system since it predicts a density
profile
decaying as $r^{-2}$ instead of $r^{-3}$. However, we have previously argued
that the difference is not very strong and that it can even be reduced if we
take
into account tidal effects \cite{clm1,clm2}.} Therefore, classical collisionless
self-gravitating systems are not in a maximum entropy state.
 Since
observations show that DM halos
possess a core rather than a cusp, we conclude that DM halos are either quantum
or collisional, two features that are not accounted for in NFW numerical
simulations \cite{nfw} .

Let us now consider a collisionless system of quantum self-gravitating 
particles (fermions). If the fermion mass is small ($m\lesssim 1\, {\rm
keV}/c^2$), as assumed in Sec. \ref{sec_astapp}, DM halos should harbor a large
quantum bulge of radius $R_c\gtrsim 240\,
{\rm pc}$ (Model II) according to  the Lynden-Bell
prediction. In that case,
the $r^{-1}$ cusps are prevented
by the Pauli exclusion principle which forbids high densities. As a result, the
classical cusp is replaced by a large quantum bulge. It is possible
that, for quantum systems with
a small fermion mass $m\lesssim 1\, {\rm keV}/c^2$, the Lynden-Bell prediction
works well in the central part of the system.
This is not in contradiction with NFW numerical  simulations \cite{nfw} since
they
do
not take into account quantum mechanics. Quantum  effects may
facilitate the collisionless relaxation of the system towards a maximum entropy
state. Therefore, when  $m\lesssim 1\, {\rm
keV}/c^2$, quantum effects can solve the
core-cusp problem.

By contrast, if the fermion
mass is large ($m\gg 1\, {\rm
keV}/c^2$), as assumed in Sec. \ref{sec_smbh}, the quantum core predicted by
the Lynden-Bell theory is very small
($R_c\le 6\times 10^{-4}\, {\rm pc}$). 
In the main part of the DM halo excluding the tiny
fermion ball at the very center (i.e. for $r>R_c$) the system is essentially in
the classical
regime. In that case,
we should recover the NFW profile which displays a $r^{-1}$ cusp while
Arg\"uelles {\it et al.} \cite{krut} find a classical
isothermal
profile with a thermal core. This is because their model assumes that the 
Lynden-Bell DF is valid everywhere (even in the classical region) while we have
just seen that the Lynden-Bell DF is not valid in a classical system.
Therefore, when $m\gg 1\, {\rm
keV}/c^2$ (e.g. $m\sim 50\, {\rm
keV}/c^2$), quantum effects cannot solve the
core-cusp problem if the system is collisionless.

One possibility to solve this problem and ``save'' the scenario of
Arg\"uelles
{\it et al.} \cite{krut} 
is to assume that
the fermions are self-interacting and that the evolution of DM halos is
collisional.\footnote{In this respect, Yunis
{\it et al.} \cite{yunis} have taken into account self-interaction in fermionic
DM halos with $m=48\, {\rm keV/c^2}$ and shown that  the extended hydrostatic
equilibrium equations for tidally truncated systems (which account for such
interactions) do not spoil the rotation curve fittings for typical
cross-sections.} In that case, the Fermi-Dirac DF is
established through a collisional relaxation of nongravitational
origin, not through
a collisionless relaxation  (see Sec. \ref{sec_kt}). In the classical
(nonquantum) regime, collisions lead to an isothermal core
of size $r_h$ instead of a $r^{-1}$ cusp. The classical core is due to thermal
effects like in the SIDM model. This is not in contradiction
with NFW numerical  simulations \cite{nfw} since they do not take into account
self-interaction and collisions among the particles. In
the quantum $+$ collisional regime, we should both have a quantum core of size
$R_c$ and an
isothermal core of size $r_h$. Therefore, quantum
and/or thermal (collisional) effects can solve the core-cusp problem. In
particular,
collisions can establish a Maxwell-Boltzmann DF for classical particles and
a 
Fermi-Dirac DF for fermions.

\section{Summary and conclusions}
\label{sec_conclusion}

In this paper, we have developed a predictive model 
of fermionic DM halos. We have considered different scenarios depending on the
fermion mass $m$ and on the DM halo mass $M_h$. We have discussed the case of a
collisionless evolution and the case of a collisional  evolution
of nongravitational origin possibly due
to SIDM. Below, we recall our basic
assumptions, summarize our main results, present synthetic
phase diagrams, and
conclude.

\subsection{Assumptions}

We have used the following observational results:

(i) The surface density of DM halos has 
a universal value $\Sigma_0=141\, M_{\odot}/{\rm
pc}^2$ \cite{kormendy,spano,donato}.\footnote{If this observational result were
not (exactly) valid, our model could be generalized but it would depend on more
parameters.}

(ii) There exists a minimum halo of mass $(M_h)_{\rm min}\sim 10^8\, M_{\odot}$
corresponding to dSphs like Fornax. Observations reveal that there is no
DM halo below this typical mass.\footnote{We have taken this value in order to
be consistent with our previous papers. It is possible that this mass is
overestimated. Some authors \cite{dvs1,dvs2,vss,vsedd,vs2,vsbh} (see also Refs.
\cite{clm2,abrilph}) argue that the minimum halo mass is $(M_h)_{\rm
min}=0.39\times 10^6\, M_{\odot}$, corresponding to Willman I. Numerical
applications  have also be given in that case for a
comparison.}

We have made the following assumptions:

(i) We assumed that DM is made of fermions and that 
DM halos are in a statistical equilibrium state of the self-gravitating
Fermi gas (see Sec. \ref{sec_sgfgb}). This statistical equilibrium state
may result
from a process of collisionless violent relaxation in the sense of Lynden-Bell 
or, possibly, from a collisional relaxation of nongravitational
origin if the fermions are self-interacting
(see Sec. \ref{sec_kt}). Note that this is a strong assumption. It is possible
that DM halos are in an out-of-equilibrium state in which case predictions
become more complicated (or even impossible).

(ii) We assumed that the minimum halo of mass $(M_h)_{\rm min}=10^8\,
M_{\odot}$ (Fornax)
is completely degenerate, i.e., it corresponds to the ground state of the
self-gravitating Fermi gas (see Sec. \ref{sec_cdl}). This automatically
determines the fermion mass $m=165\, {\rm
eV/c^2}$.\footnote{This prediction could be refined by taking a
possibly more relevant value of the minimum halo mass but the order of magnitude
of $m$ should be correct up to a factor $10$. For example, if we take 
$(M_h)_{\rm min}=0.39\times 10^6\,
M_{\odot}$ (Willman I) we get $m\sim 1\, {\rm keV/c^2}$ and $(r_h)_{\rm
min}=33\, {\rm
pc}$ (see Sec. \ref{sec_mh}). A fermion mass in the keV scale (or slightly
larger) is more consistent with the constraints coming from cosmological
observations such as the Lyman
$\alpha$ forest (see Sec. \ref{sec_pro}).} We then
find that the radius of the minimum halo is
determined by the constraint from Eq. (\ref{nd13}) giving $(r_h)_{\rm
min}=597\, {\rm pc}$ [see Eq. (\ref{cdl17})]. With this assumption, there is
no free (undetermined) parameter in our model. In this sense, it is completely
predictive.

(iii) We assumed that the external structure of large DM halos ($M_h\gg
(M_h)_{\rm min}$) 
is described by the classical isothermal distribution (see Sec. \ref{sec_nd}).
This corresponds to the nondegenerate Fermi-Dirac DF, or to the approximate form
of the Fermi-Dirac DF at sufficiently large distances where the density is low.
This classical isothermal distribution is in agreement with the Burkert profile
(see Sec. III.C of \cite{modeldm}). Combining the properties of classical
isothermal spheres with the constraint from Eq. (\ref{nd13}) we obtained the
mass-radius relation $M_h=1.76\, \Sigma_0 r_h^2$ [see Eq. (\ref{nd14})]. In
principle, this relation
is valid for $M\gg (M_h)_{\rm min}$. In practice, it is fulfilled as soon as
$M_h$ is slightly larger than $(M_h)_{\rm min}$ (see Fig. \ref{massradius}).

(iv) We considered the possibility that large DM halos, in addition 
to their classical isothermal envelope of mass $M_h$ and radius $r_h$, contain a
quantum core (fermion ball) of mass $M_c$ and radius $R_c$. In other words, we
assumed that large DM halos may be partially degenerate (see Sec.
\ref{sec_app}).

\subsection{Methodology}

The structure of fermionic DM halos, and the mass $M_c$ of the quantum
core, are obtained by maximizing the Fermi-Dirac entropy at fixed mass and
energy (microcanonical ensemble). To solve this problem, we proceeded as
follows. We considered the thermodynamics of a gas of self-gravitating fermions
in a box of radius $R=r_h$ containing a mass $M=M_h$ (see Sec.
\ref{sec_sgfgb}). The caloric curve depends on a unique parameter $\mu$ which is
a measure of the mass $M_h$ of the DM halo [see Eq. (\ref{mu3})]. We then used
the fact
that the DM halos are virialized so that the dimensionless inverse temperature $\eta=\beta GMm/R$ is
of order unity. The intersection between the line $\eta_v=1$ and the caloric curve
$\eta(\Lambda)$ determines the possible equilibrium states of the system 
consistent with the virial condition. For  $M_h<(M_h)_{\rm CCP}$ there is only
one solution. However, for $M_h>(M_h)_{\rm CCP}$ we found
two
relevant solutions:

(i) A purely gaseous solution (G) without quantum core corresponding 
to the classical isothermal sphere \cite{chandrabook}. Its density profile is 
consistent with the observational Burkert profile (see Sec.
III.C of \cite{modeldm}). This corresponds to Model I of Sec. \ref{sec_model1}.

(ii) A core-halo solution (CH) 
with a quantum core surrounded by a classical isothermal halo. The core is
relatively large ($M_c=9.45\times 10^9\, M_{\odot}$ and $R_c=240\, {\rm pc}$ for
the Milky Way) so it can represent a quantum bulge. The core mass increases with
the halo mass as $M_h^{3/8}$ [see Eq. (\ref{mcmh7})]. This corresponds to Model
II of Sec. \ref{sec_model2}.\footnote{de Vega and coworkers
\cite{dvs1,dvs2,vss,vsedd,vs2,vsbh}  only considered the gaseous
(nondegenerate) solution (G). A merit of our study is to have evidenced a
bifurcation
above the canonical critical point $\mu_{\rm CCP}$  yielding a new branch of
solutions (CH) possessing a quantum core.}

The gaseous solution (G) is always 
thermodynamically stable. The core-halo solution
(CH) is thermodynamically
stable  for
$M_h<(M_h)_{\rm MCP}$ and unstable for $M_h>(M_h)_{\rm MCP}$ (as explained
before, we consider the thermodynamical stability in the microcanoncal
ensemble).

The formation of DM halos arises 
from a process of collisionless violent relaxation. The metaequilibrium state
resulting from violent relaxation must be a maximum entropy state in the sense
of Lynden-Bell. It turns out that the Lynden-Bell DF coincides with the
Fermi-Dirac DF. For $M_h<(M_h)_{\rm MCP}$ the gaseous solution (G) and the
core-halo solution (CH) are both entropy maxima in the sense of Lynden-Bell.
They could naturally arise from a process of violent relaxation. For
$M_h>(M_h)_{\rm MCP}$ only the gaseous solution (G) is a maximum entropy state
in the sense of Lynden-Bell. Therefore, violent relaxation should lead to the
gaseous solution (G), not to the core-halo solution (CH). Actually, there exist
stable core-halo solutions (CH)$_*$ close to the last turning point of energy
that may be physically relevant. Violent relaxation may
also lead to a stable quasistationary state, resulting from incomplete
relaxation, that
is not a maximum entropy state.

If the DM halos are collisionless, 
they remain in the state resulting from violent relaxation. If the DM halos
are collisional, they follow the series of equilibria determined by the caloric
curve towards states of higher and higher density. When $M_h<(M_h)_{\rm
MCP}$  the system can evolve from the gaseous solution (G) to the core-halo
solution (CH). For $M_h>(M_h)_{\rm MCP}$  the system can follow the
series of equilibria from the gaseous solution (G) up to the point of minimum
energy $E_c$. At that point, it becomes thermodynamically unstable and undergoes
a gravothermal catastrophe. If the DM halo is small enough ($M_h<M_{\rm OV}$),
the gravothermal
catastrophe stops when the core becomes degenerate. In that case, gravitational
collapse is prevented by quantum mechanics (Pauli's exclusion principle). The
system
achieves a core-halo state with a small quantum core. This state does not
correspond to a state of statistical equilibrium such as solution (C') which
would have a too extended halo [scenario (b.1)] but it could
be an
out-of-equilibrium structure (CH)$_{\rm out}$ [scenario
(b.2)]. Alternatively,  if the DM halo is large enough ($M_h>M_{\rm OV}$), the
gravothermal
catastrophe can lead to the formation of a SMBH by the
mechanism discussed in Secs. \ref{sec_vlh}, \ref{sec_critbh} and \ref{sec_gr}
[scenario (b.3)].

{\it Remark:} For small halos $M_h<(M_h)_{\rm MCP}$, the
core-halo solution (CH) is stable (see Fig. \ref{fel}) and the
scaling
$M_c\propto M_h^{3/8}$ from Eq. (\ref{mcmh7}) is reliable. However,
for large DM halos $M_h>(M_h)_{\rm MCP}$, this is no more the case because the
core-halo solution (CH) is unstable and is replaced by an out-of-equilibrium
core-halo solution (CH)$_{\rm out}$ (see Fig. \ref{le5}). It is possible, in
that case, that DM halos of the same mass
$M_h$ may contain cores of different masses $M_c$ depending on their evolution.
The core mass $M_c$ may evolve from a small value $M_c^*$ corresponding to
the begining of the condensed branch at $\Lambda_*$ up to the
value $M_{\rm OV}$ corresponding to the end of the condensed branch at
$\Lambda''_c$ at which it becomes unstable and collapses
towards a SMBH (see Fig. \ref{fbbh}).

\subsection{Results for $m=165 \, {\rm eV}/c^2$}

For a fermion mass $m=165 \, {\rm eV}/c^2$, we obtained the following
results:

(i) There exists a minimum halo of mass  $(M_h)_{\rm min}=10^8\, M_{\odot}$ 
and radius $(r_h)_{\rm min}=597\, {\rm pc}$ corresponding to the ground
state ($T=0$) of the self-gravitating Fermi gas. This DM halo is a purely
quantum object (fermion ball) without atmosphere. It is completely degenerate.
It is equivalent to a polytrope of index $n=3/2$. It is fully stable.
Quantum mechanics (Pauli's
exclusion principle) determines the minimum mass and the minimum radius of fermionic
DM halos.\footnote{Actually, the mass and the size of the DM
halos should be
determined by a theory  of structure formation. The first stage of this theory
is the Jeans instability, leading to the formation of clumps in the linear
regime. When the density of the clumps has grown significantly, we enter in the
nonlinear regime of structure formation where the overdensity regions experience
free fall, violent relaxation, nonlinear Landau damping, merging and accretion,
leading to the DM halos
that we observe today.} This minimum halo can be assimilated with
ultracompact dSphs like Fornax.

(ii) For $(M_h)_{\rm min}=10^8\,
M_{\odot}<M_h<(M_h)_{\rm CCP}=6.73\times 10^8\, M_{\odot}$,
the caloric curve is monotonic ($\mu<\mu_{\rm
CCP}$; see Fig. \ref{multimu2}). There is only one solution with $\eta\sim 1$: A
quantum object (Q) corresponding to a fermionic ball surrounded
by a tenuous isothermal atmosphere.
This equilibrium state is fully stable. This
situation may describe dSphs.
Even if collisions allow the system to evolve along the series of equilibria,
no instability occurs.

(iii) For $(M_h)_{\rm
CCP}=6.73\times 10^8\, M_{\odot}<M_h<(M_h)_{\rm
MCP}=1.08\times 10^{10}\, M_{\odot}$, the caloric curve has
an $N$-shape structure ($\mu_{\rm CCP}<\mu<\mu_{\rm MCP}$; see Fig. \ref{fel}).
There are two physical solutions with $\eta\sim 1$: A gaseous
solution (G) corresponding to a purely classical isothermal halo without quantum core
and a core-halo (CH) solution with a quantum core surrounded by a massive
atmosphere.  The fermion ball may mimic a large bulge but not a SMBH because it is too much extended (see
Sec. \ref{sec_bulge}). The gaseous
solution and the core-halo solution are both stable.
If the system evolves adiabatically along the
series of equilibria under the effect of collisions, it can pass from the gaseous
solution to the core-halo solution without collapsing. The gravothermal catastrophe
is prevented by quantum mechanics. This situation may describe small and medium
spiral galaxies. They may have a
core-halo
structure made of a quantum core (representing a bulge) and an isothermal atmosphere.
The bulge may provide a favorable environment to induce the formation
of a SMBH on a long timescale by an accretion process (see Sec.
\ref{sec_large}).

(iv) For $M_h>(M_h)_{\rm MCP}=
1.08\times 10^{10}\, M_{\odot}$, the caloric
curve has a $Z$-shape structure ($\mu>\mu_{\rm MCP}$; see Fig. \ref{le5}). There
are
two physical solutions for a given value of $\eta\sim 1$ as before. The gaseous
solution (G) is stable  while the core-halo
solution (CH) is unstable [we must also keep in mind the potentially relevant
solution (CH)$_*$]. In principle, only the
gaseous solution may result from a process of violent collisionless relaxation
because the core-halo solution is not a maximum entropy state in the sense of
Lynden-Bell. If, starting from the gaseous phase, the
system evolves adiabatically along the series of equilibria  under the effect of collisions,  it
can undergo a gravothermal catastrophe at the point of minimum energy $E_c$.
Then,
there
are two possibilities:

(iv-a) For $(M_h)_{\rm MCP}<M_h<M_{\rm OV}=2.30\times
10^{13}\, M_{\odot}$ the gravothermal catastrophe is stopped by quantum
degeneracy. This leads to a possibly out-of-equilibrium small quantum core
(CH)$_{\rm out}$ (different from a large quantum bulge) surrounded by an
envelope. In that case, the core mass -- halo mass relation
from Eq. (\ref{mcmh7}) may not be valid anymore.

(iv-b) For $M_h>M_{\rm OV}=2.30\times
10^{13}\, M_{\odot}$ a new turning point of
energy occurs in the caloric curve \cite{caf,acf} below which the
condensed branch disappears and the core of the DM halo
collapses towards a SMBH of mass $M_{\rm OV}$ (presumably). If
$M_{\rm OV}<M_h<M'_*$ the DM halo may either harbor a fermion ball or
a SMBH. If $M_h>M'_*$ there is no condensed branch at all and the DM halo
cannot harbor a fermion ball. It can just harbor a SMBH of mass $M_{\rm OV}$.

This
situation may apply to large spiral and elliptical galaxies. Therefore,
large spiral and elliptical galaxies are expected to contain a small
quantum core or a SMBH resulting from the gravothermal catastrophe instead of a
large quantum bulge.\footnote{We have argued
that the Milky Way may contain a large quantum bulge in agreement with certain
observations despite the fact that $M_h>(M_h)_{\rm MCP}$. This is because its
mass $M_h=10^{11}\, M_{\odot}$ is close to the
microcanoncal critical point $(M_h)_{\rm MCP}= 1.08\times
10^{10}\, M_{\odot}$, especially if the fermion mass is larger
(recall that $(M_{h})_{\rm MCP}=4.21\times 10^{7}\, M_{\odot}$     for
$m\sim
1\, {\rm keV}/c^2$). Therefore, the large quantum bulge may be (marginally)
stable.} During the gravothermal catastrophe, their envelope is left undisturbed
and should correspond to a marginal King profile (if we take into account tidal
effects) which is in good agreement with the Burkert profile (see
Refs. \cite{clm1,clm2} and Fig. \ref{densityLOG}).

 \begin{figure}
\begin{center}
\includegraphics[clip,scale=0.3]{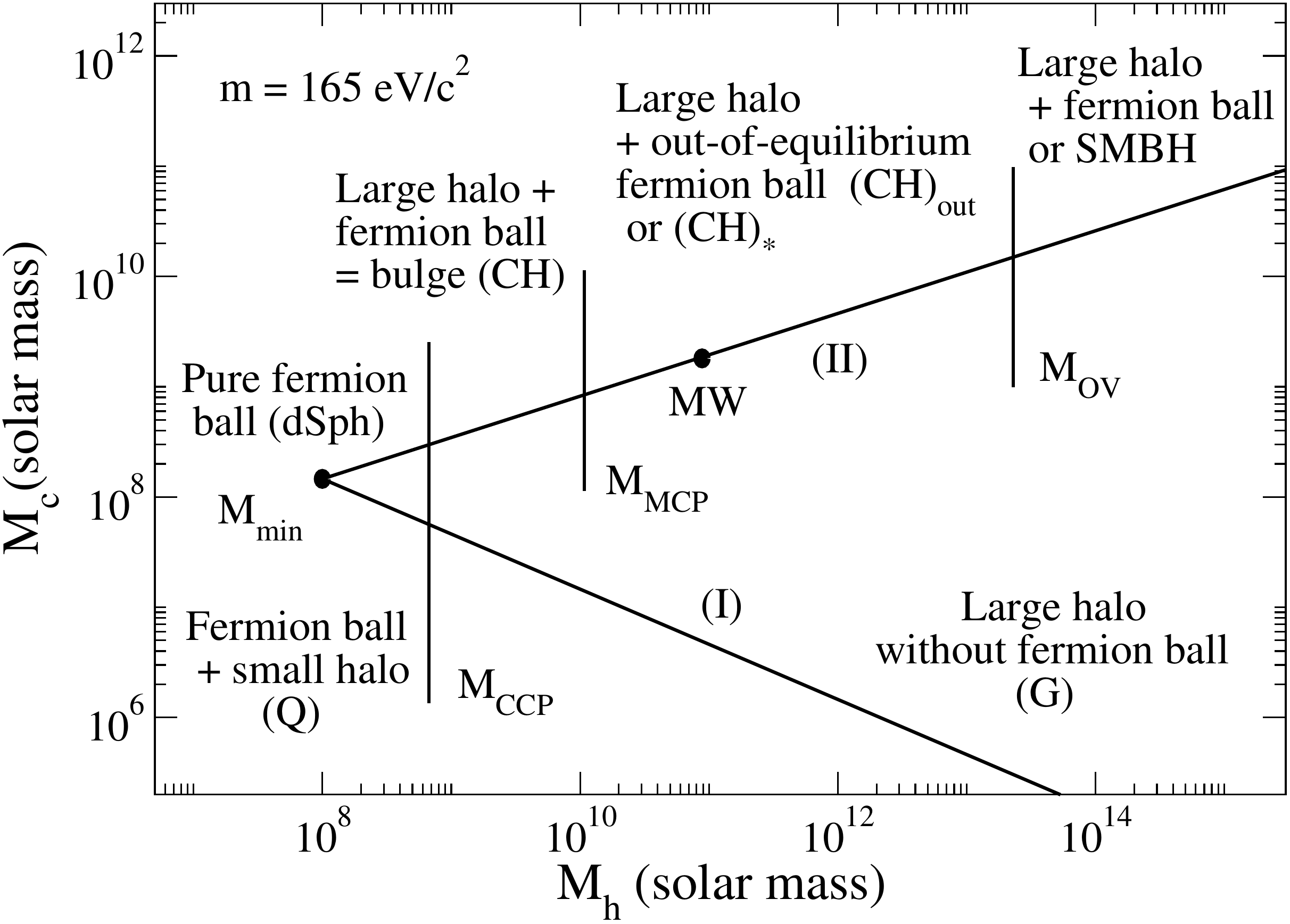}
\caption{ Phase
diagram for a fermion mass $m=165 \, {\rm eV}/c^2$ summarizing our main
results. It displays the minimum halo $(M_h)_{\rm min}=10^8\,
M_{\odot}$ (Fornax) where the DM halo is a pure fermion ball without isothermal
atmosphere (ground state). It also displays the canonical critical
point $(M_h)_{\rm CCP}=6.73\times
10^8\,
M_{\odot}$ at which a bifurcation occurs between the gaseous branch (G) where
the
DM halos are
purely
isothermal without quantum core and the core-halo branch (CH) where the DM
halos are made of a quantum core (bulge) surrounded by a
classical
isothermal
halo.
Finally, it displays the microcanonical
critical point $(M_h)_{\rm
MCP}=1.08\times 10^{10}\,
M_{\odot}$  above which the core-halo branch (CH) becomes
unstable [there may be, however, potentially relevant
solutions (CH)$_*$]. In that case, the DM
halos may undergo a gravothermal catastrophe leading to the formation of an
out-of-equilibrium fermion ball (CH)$_{\rm out}$ if $M_h<M_{\rm OV}$,  an
out-of-equilibrium fermion ball (CH)$_{\rm out}$ or a central SMBH if $M_{\rm
OV}<M_h<M'_*$, or
a SMBH if $M_h>M'_*$ (see Fig. \ref{conclusion48} for a better illustration of
these
different cases). We have located Fornax
(dSph) and the Milky Way (MW) for reference.}
\label{conclusion}
\end{center}
\end{figure}

The main results of our study for $m=165 \, {\rm eV}/c^2$ are summarized
in the phase diagram of Fig. \ref{conclusion}. The bullet corresponds to the
minimum 
halo of mass $(M_h)_{\rm min}=10^8\, M_{\odot}$. For $(M_h)_{\rm
min}=10^8\,
M_{\odot}<M_h<(M_h)_{\rm CCP}=6.73\times 10^8\, M_{\odot}$, there is only one
solution, the quantum solution (Q). The canonical critical point  $(M_h)_{\rm
CCP}=6.73\times 10^8\,
M_{\odot}$ determines a bifurcation between
the branch of purely gaseous solutions (G)  and the
branch of core-halo solutions (CH) where the core represents a large quantum
bulge. This bifurcation is associated with the occurence of a region of negative
specific heat in the caloric curve. 
The microcanonical critical point $(M_h)_{\rm MCP}= 1.08\times 10^{10}\,
M_{\odot}$ determines the moment at which the DM halo can experience a
gravothermal catastrophe. For $(M_h)_{\rm MCP}<M_h<M_{\rm OV}$ the  gravothermal
catastrophe is stopped by quantum mechanics and the DM halo harbors a possibly
out-of-equilibrium small quantum core. Therefore, the microcanonical critical
point $(M_h)_{\rm MCP}= 1.08\times
10^{10}\, M_{\odot}$ determines the
transition between DM halos possessing a large quantum bulge
and DM halos harboring a small quantum core (CH)$_{\rm out}$ [there are also
potentially relevant
solutions (CH)$_*$]. This transition is associated with
the instability of the large quantum bulge with respect to the gravothermal
catastrophe. The mass $(M_h)_{\rm MCP}= 1.08\times 10^{10}\,
M_{\odot}$ also determines the moment at which the behavior of the core mass
-- halo mass relation changes. On the other hand, the mass
$M_{\rm OV}=2.30\times
10^{13}\,
M_{\odot}$ determines the moment at which the core of the DM
halo may collapse towards a SMBH. For $M_{\rm OV}<M_h<M'_*$ the DM halo may
either
harbor a fermion ball or a SMBH. For $M_h>M'_*$  the DM halo can only harbor a
SMBH.

For $m=165 \, {\rm
eV}/c^2$, the value of $M_{\rm OV}=2.30\times
10^{13}\,
M_{\odot}$ is very large and may not be
astrophysical relevant. The value
of $M_{\rm OV}$ is reduced if the
fermion mass is larger. For $m\sim 1\, {\rm keV}/c^2$ we find 
 $M_{\rm OV}=6.26\times
10^{11}\, M_{\odot}$ but this value is still too large. It is comparable to
the mass the whole Milky Way instead of being comparable to the mass of Sgr
A$^*$. Therefore, the fermionic DM model with
a mass  $m=165 \, {\rm
eV}/c^2$ or $m\sim 1\, {\rm keV}/c^2$ cannot account for the presence of a 
supermassive compact object (either a SMBH or a fermion ball) of
mass $M_c=4.2\times
10^{6}\, M_{\odot}$ and radius $R_c<6\times 10^{-4}\,
{\rm pc}$ at the center of the Milky Way. It rather predicts the existence of a
large fermion
ball (bulge) of mass $M_c=9.45\times 10^{9}\, M_{\odot}$ and radius
$R_c=240\, {\rm pc}$ (Model II), or no fermion ball at all (Model I). In that
case, in order to account for the observation, we have to
generalize the fermionic
DM model by introducing a  primordial SMBH (Sgr A$^*$) at the center of the
Milky Way.

We note that the fermionic model developed in the present paper 
is based on the same ideas as those developed in Ref. \cite{modeldm} for bosonic
DM halos. The general scenario is the same (the fermion ball replacing the
soliton in the BEC model) but the $M_c(M_h)$ relation and the values of the
characteristic masses and radii are different. Therefore, a detailed comparison
between the two models may help determining whether DM is made of fermions or
bosons.

\subsection{Results for  $m=48\, {\rm keV/c^2}$}

We have also considered the possibility suggested by other authors
\cite{btv,rar} 
that the fermion ball may mimic a SMBH at the center of the galaxies. This
scenario requires a larger particle mass $m=48\, {\rm keV/c^2}$ (see footnote
42).

We have first considered the case of the usual Fermi-Dirac DF. 
We have shown that our simple semi-analytical box model, leading to the relation
from
Eq. (\ref{mcmh7}), reproduces the numerical results of Bilic {\it et
al.} \cite{btv} and Ruffini {\it et al.} \cite{rar}. However, the size
of the fermion ball is too large to satisfy the observational constraints
corresponding to Sgr A$^*$. On the other hand, in line with our previous
claims \cite{clm2,modeldm,acf}, we showed that the core-halo solution in these
models is thermodynamically
unstable. Therefore,  it cannot result from a process of violent relaxation.

We then mentioned the recent results of Arg\"uelles {\it et al.}
\cite{krut,rarnew} based 
on the fermionic King model. In a first work,  Arg\"uelles {\it et al.}
\cite{krut} obtained a density profile with a core-halo structure that
satisfies the observational constraints of Sgr A$^*$. In a second work, 
Arg\"uelles {\it et al.} \cite{rarnew} showed that this solution is
thermodynamically stable in the microcanonical ensemble so that it is likely to
result from a process of violent relaxation. However, it is not clear if the
process of violent relaxation can lead to a core-halo solution with such a high
value of the central density because of the problem of incomplete relaxation
\cite{lb,incomplete}.\footnote{We have mentioned that
this problem could be alleviated if the fermions are self-interacting.} The
purely gaseous solution (without quantum core) whether stable or metastable may
be
reached more easily.  This issue can be settled only
with direct numerical simulations. 

According to the work of Arg\"uelles {\it et al.}
\cite{rarnew}, medium size galaxies like the Milky Way may harbor a fermion
ball mimicking a SMBH of mass $M_c=4.2\times 10^{6}\, M_{\odot}$ and radius
radius $R_c=6\times 10^{-4}\, {\rm pc}$. This
corresponds to a stable configuration (CH)$_*$ located on the condensed branch
of the
caloric
curve of the self-gravitating Fermi gas close to the last turning point of
energy $E_*$  (see Fig.
\ref{mutresgrandprolongehenon}). Using
the results of Alberti and Chavanis \cite{acf},
we have argued that larger galaxies cannot harbor a fermion ball because, above
a
critical mass  $M_h>M'_*$, the condensed branch
disappears and the system forms a SMBH of mass $\sim M_{\rm OV}$ (presumably).
Therefore, if $m=48\, {\rm keV/c^2}$, very large galaxies are likely to contain
a SMBH of mass $M_{\rm OV}=2.71\times 10^8\, M_{\odot}$ possibly accounting for
AGNs. Medium size galaxies like the Milky Way may also follow the
branch of condensed states up to the turning point of energy $E''_c$ and
undergo core collapse
towards a SMBH. However the mass of the SMBH should  be much smaller than
$M_{\rm OV}=2.71\times 10^8\, M_{\odot}$ in order to account for the
characteristics of Sgr A$^*$. This may be achieved with a larger fermion mass.
For a 
fermion mass $m=386\, {\rm keV/c^2}$ the disappearance of the
condensed branch and the collapse of the core of the system
towards a SMBH already occur in galaxies like the Milky Way and lead to a SMBH
of mass $M_{\rm OV}=4.2\times 10^6\, {\rm keV/c^2}$ similar to Sgr A$^*$.

\begin{figure}
\begin{center}
\includegraphics[clip,scale=0.3]{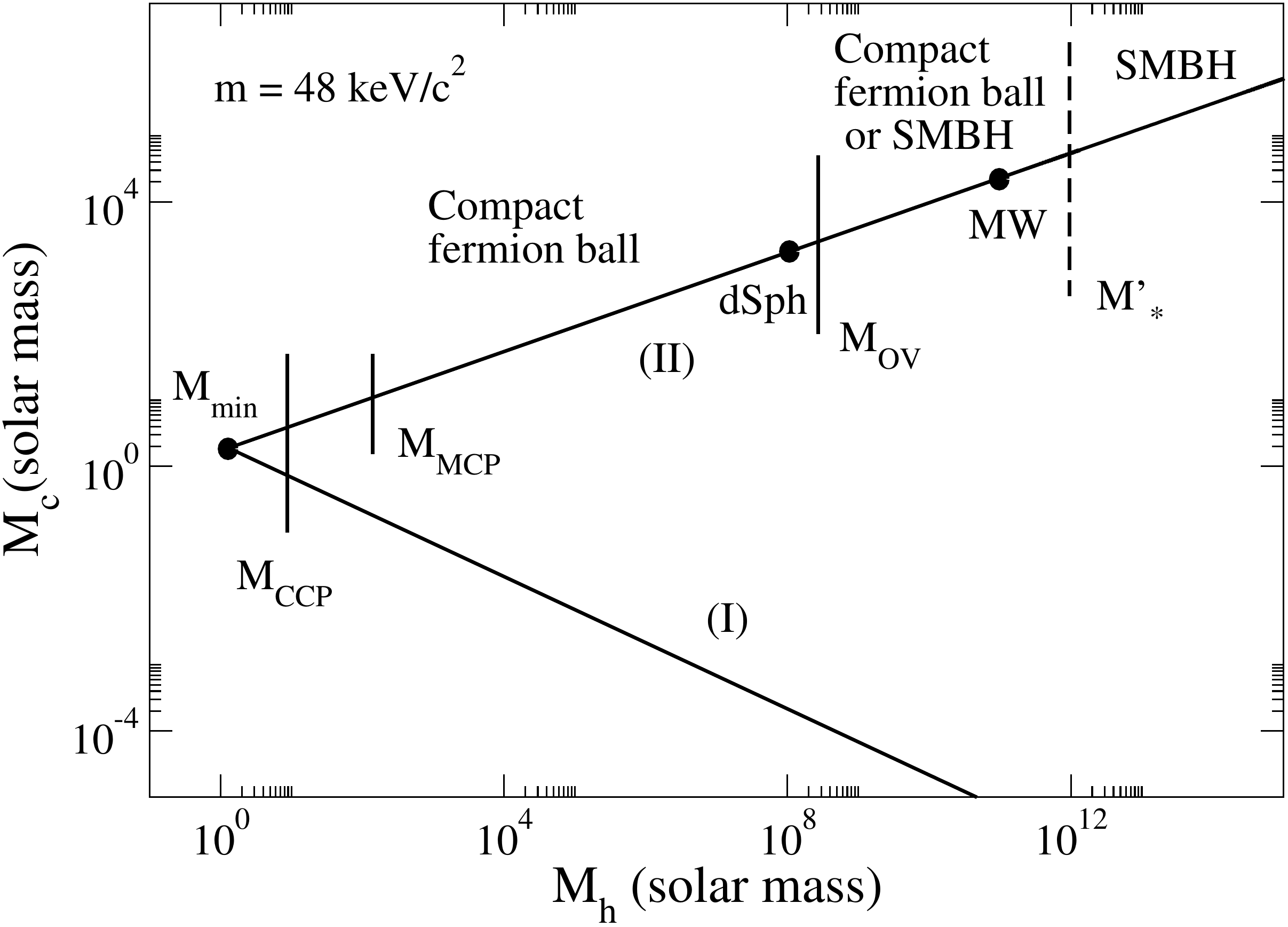}
\caption{Phase
diagram for a fermion mass $m=48 \, {\rm keV}/c^2$ summarizing our main
results. The mass of the minimum halo (ground state) is considerably reduced to
$(M_h)_{\rm min}=1.30\,
M_{\odot}$ leading to potential problems (see main text). For $M_h<M_{\rm
OV}=2.71\times 10^8\,
M_{\odot}$, the DM halos should harbor a small fermion ball mimicking
a SMBH.
For $M_{\rm OV}<M_h<M'_{*}$ they may harbor either a small fermion ball
mimicking a SMBH or a real SMBH (for $m=48 \, {\rm keV}/c^2$ an ultracompact
fermion ball in
the Milky Way is prefered over a SMBH which would have a too large mass,
$M_{\rm OV}=2.71\times 10^8\, M_{\odot}$, larger than the mass of Sgr A$^*$).
For $M_h>$ $M'_{*}$ they can harbor only a SMBH of mass $M_{\rm OV}=2.71\times
10^8\, M_{\odot}$ possibly accounting for AGNs. We have located Fornax
(dSph) and the Milky Way (MW) for reference.}
\label{conclusion48}
\end{center}
\end{figure}

We also mentioned potential difficulties with the model of 
Arg\"uelles {\it et
al.}
\cite{rarnew}. If the fermion mass
is $m=48\, {\rm keV/c^2}$, the mass of the minimum halo (ground state)  is
$(M_{h})_{\rm min}=1.30\, M_{\odot}$. Therefore,  there should exist DM
halos with a mass much below $10^8\, M_{\odot}$, up to $1\,
M_{\odot}$. On the
other hand, under the same conditions, DM halos of mass $M_{h}=10^8\,
M_{\odot}$ such as dSphs like Fornax should have a core-halo structure
(see Figs. \ref{fornaxBH} and \ref{fornaxBHvitesse}). As far
as we
know, these two features are not
observed: There are no DM
halos with a mass much smaller than $10^8\, M_{\odot}$ (Fornax) and the
density
profiles of dSphs have not a core-halo structure (they look like Figs.
\ref{profilepolytrope} and \ref{Vpoly} instead of  Figs. \ref{fornaxBH}
and \ref{fornaxBHvitesse}). More precisely, the fermionic DM model with a
fermion mass $m=48\, {\rm keV/c^2}$ predicts that dSphs of mass
$M_{h}=10^8\,
M_{\odot}$ should contain a fermion ball of
mass $M_c=1.57\times 10^4\, M_{\odot}$ and 
radius $R_c=5.42\, {\rm mpc}$ possibly mimicking an intermediate mass BH (see
Sec. \ref{sec_pot}). This
is either a very important prediction (if confirmed by observations) or the
evidence that this model is incorrect (if invalidated by observations).
The main results of our study for $m=48 \, {\rm keV}/c^2$ are summarized
in the phase diagram of Fig. \ref{conclusion48}.

\subsection{The importance of the DM particle mass}

There is a strong structural difference between the core-halo density profile 
corresponding to a large fermion mass $m=48\, {\rm keV/c^2}$ or a small
fermion mass $m=165 \, {\rm eV}/c^2$ when the fermionic model is applied
to a DM halo of mass $M_h=10^{11}\, M_{\odot}$ (Milky Way). In the first case,
the degeneracy parameter $\mu$  is
very
large ($\mu=3.09\times 10^{14}$) and the core-halo profile presents a strong
separation between the quantum core and the classical halo (see Fig.
\ref{profileBH}).
They are separated
by an extended plateau. Furthermore, the fermion ball has a mass 
$M_c=4.2\times 10^{6}\, M_{\odot}$ and a radius $R_c=6\times 10^{-4}\,
{\rm pc}$, mimicking a small SMBH (Sgr A$^*$). In the second case, $\mu$ is
relatively
small ($\mu=4.31\times 10^{4}$) and the separation between the core and the
halo is mild with no clear plateau between them (see Fig. \ref{profileMW}).
Furthermore, the fermion ball
has a mass $M_c=9.45\times 10^{9}\, M_{\odot}$ and a radius
$R_c=240\, {\rm pc}$ mimicking a large quantum bulge, not a small SMBH.
Therefore, depending on the DM particle mass, the fermionic model predicts very
different types of structures. Comparison with observations of the Milky
Way should
determine which type of
structure (a large quantum bulge or a small quantum core mimicking a SMBH) is
the most relevant in the fermionic model. In this respect, we note
that the 
BECDM model also leads to
core-halo configurations in which the fermion ball is replaced by a soliton. For
the commonly adopted boson mass $m\sim 10^{-22}\, {\rm eV/c^2}$ the core-halo
profiles obtained in direct numerical simulations
\cite{ch2,ch3,schwabe,mocz,moczSV,veltmaat,moczprl,moczmnras,veltmaat2}
 do  not show a very pronounced separation
between a core and a halo (there is no extended plateau) and look similar to
Fig. \ref{profileMW} rather than Fig. \ref{profileBH}. In addition, the
soliton mimics
a large quantum bulge rather than a SMBH. Such a large quantum
bulge seems to be necessary to account for the dispersion velocity peak
observed in the Milky Way \cite{martino}. This may be a strong observational
evidence for the presence of a large quantum bulge (bosonic or fermionic) at
the center of the galaxies. Therefore, the comparison of the
fermionic and bosonic models tends to favor a fermion mass of the order of
$m=165 \, {\rm eV}/c^2$ (or $1\, {\rm keV}/c^2$) instead of $m=48\, {\rm
keV/c^2}$. It would be interesting to consider BECDM models with a boson mass
much larger than $m\sim 10^{-22}\, {\rm eV/c^2}$ to see if they can lead to a
soliton mimicking a SMBH like in the model of Arg\"uelles {\it et al.}
\cite{krut,rarnew}. Considering a noninteracting boson for simplicity, we find
that its mass should be $m=1.84\times 10^{-18}\, {\rm eV/c^2}$ (see Sec. V.A of
\cite{mrjeans}). However, for such a large mass, BECDM is expected to behave
like CDM and present a central cusp instead of a core as demonstrated by Mocz
{\it et al.} \cite{moczSV} (see also the discussion in
Sec. \ref{sec_tqc}). Similarly, for a large fermion mass $m\sim 50\, {\rm
keV/c^2}$ such as the one considered in the models of Arg\"uelles
{\it et al.} \cite{krut,rarnew},  DM should behave like CDM and
may not be described by the Lynden-Bell DF as assumed by these authors. The
Lynden-Bell DF may be valid only for a smaller fermion mass $m\sim 1\, {\rm
keV/c^2}$ where the cusps are prevented by the Pauli exclusion principle.
However, these difficulties may disappear if DM is both quantum and
self-interacting (see the discussion in Sec. \ref{sec_tqc}). 
In that case, the Fermi-Dirac DF may be justified by the self-interaction
(collisions) of the fermions, not by a process of collisionless violent
relaxation. The same remarks apply to the bosonic model: One should consider a
repulsive self-interaction like in Ref. \cite{modeldm}.

The comparison between the bosonic and fermionic models that we have
initiated in this paper and in \cite{prd1,clm2,modeldm,mcmh,mcmhbh} may help
determining the DM particle mass and whether it is a fermion or a boson. We
would like to close this paper by suggesting that DM may be made of different
types of particles (fermions and bosons) with different characteristics (mass,
scattering length...). Some family of particles may be responsible for creating
a large quantum bulge (fermion ball or soliton) \cite{ch2,ch3,clm2,modeldm} at
the center of the galaxies which could explain the dispersion velocity peak
observed in the Milky Way \cite{martino} while other family of particles may be
responsible for creating a very compact object (fermion ball or soliton) at the
very center of the galaxies mimicking a SMBH \cite{btv,krut,rarnew}, or even
leading to the formation a real SMBH.\footnote{The
compact object at the center of the Galaxy (Sgr A$^*$ of mass $M=4.2\times
10^6\, M_{\odot}$) could be a mixed structure made of a SMBH surrounded by a
compact fermion or boson ball. In that case, the mass of the SMBH could be
smaller than commonly thought ($M_{\rm BH}\le 4.2\times
10^6\, M_{\odot}$) since part of the mass of the compact object (Sgr A$^*$)
would be in the
fermion or boson ball.} If this suggestion is correct, it would give interest
to all
kinds of research made on quantum (fermionic and bosonic) DM and SIDM. If not,
some physically interesting theoretical models may be ruled out by the
observations.

{\bf Acknowledgments:} I am grateful to Carlos
Arg\"uelles for a careful reading of the manuscript and
interesting comments.

\appendix

\section{Approximate equations of state}
\label{sec_aeos}

Instead of using the exact equation of state  for an ideal Fermi gas at
finite temperature, Eqs. (\ref{tf7}) and
(\ref{tf8}), we could consider the approximate equation
of state 
\begin{equation}
P=\frac{1}{20}\left (\frac{3}{\pi}\right
)^{2/3}\frac{h^2}{m^{8/3}}\rho^{5/3}+\rho\frac{k_B T}{m},
\label{aeos1}
\end{equation}
which is simply the sum of the polytropic equation of state (\ref{cdl5}) valid
at high densities and the isothermal equation of state (\ref{nd4}) valid at low
densities.

Similarly, in the case of self-interacting BECDM halos in the TF limit, we have
used in Ref. \cite{modeldm} an approximate equation of state of the form 
\begin{equation}
P=\frac{2\pi a_s \hbar^2}{m^3}\rho^{2}+\rho\frac{k_B
T}{m},
\label{aeos2}
\end{equation}
where $a_s$ is the scattering length of the bosons.

Finally, in the case of noninteracting BECDM halos, we have used in Refs.
\cite{chavtotal,nottalechaos} an approximate equation of state of the form
\begin{equation}
P=\left (\frac{2\pi G\hbar^2}{9m^2}\right )^{1/2}\rho^{3/2}+\rho\frac{k_B
T}{m},
\label{aeos3}
\end{equation}
where the first term mimics the quantum potential (see Appendix E of \cite{mcmh}
for the justification of this equation of state).

These equations of state are of the generic form
\begin{equation}
P=K\rho^{\gamma}+\rho\frac{k_B T}{m}\qquad (\gamma=1+1/n), 
\label{aeos4}
\end{equation}
involving a polytropic equation of state and an
isothermal (linear) equation of state. In the models discussed above, the
polytropic index is $n=3/2$, $n=1$ and $n=2$ respectively \cite{mcmh}. DM halos
described by the mixed equation of state (\ref{aeos4}) have been studied in
\cite{modeldm}. They are governed by a generalized Lane-Emden equation
introduced in Appendix E of \cite{modeldm}. The mixed equation of state
(\ref{aeos4}) has also been introduced and studied in a cosmological context (in
the framework of general relativity) in Refs.
\cite{cosmopoly1,cosmopoly2,cosmopoly3}.

\section{Entropy}
\label{sec_ent}

Using the Gibbs-Duhem formula (see Eqs. (40), (47) and (58) of \cite{gr1}), the
entropy of the
nonrelativistic self-gravitating Fermi gas is given by\footnote{It is shown in
\cite{gr1} that this expression is valid for an arbitrary form of
entropy.}
\begin{equation}
S=-\frac{\mu}{T}N+\frac{5E_{\rm
kin}}{3T}+\frac{2W}{T}.
\label{ent1}
\end{equation}
Using the virial theorem from Eq. (\ref{tf32}) and introducing the total energy
$E=E_{\rm kin}+W$, we obtain
\begin{equation}
\frac{S}{N k_B}=-\frac{\mu}{k_B T}+\frac{7E}{3Nk_B T}-\frac{P_{\rm
b}V}{Nk_B T}.
\label{ent2}
\end{equation}
Using
\begin{equation}
k=e^{-\beta\mu}e^{\beta m \Phi_{0}}
\label{ent3}
\end{equation}
and
\begin{eqnarray}
\psi(\alpha)&=&\beta m (\Phi(R)-\Phi_{0})\nonumber\\
&=&\beta
m\left (-\frac{GM}{R}-\Phi_{0}\right )=-\eta-\beta m\Phi_{0}
\label{ent4}
\end{eqnarray}
from Eq. (\ref{tf15}), we get
\begin{equation}
\beta\mu=-\ln k-\eta-\psi(\alpha).
\label{ent5}
\end{equation}
Substituting Eq. (\ref{ent5}) into Eq. (\ref{ent2}) and introducing the
dimensionless variables defined in Sec. \ref{sec_sgfgb} we finally obtain
\begin{equation}
\frac{S}{N k_B}=\ln
k+\eta+\psi_k(\alpha)-\frac{7}{3}\Lambda\eta-\frac{2\alpha^{6}}{9{\tilde\mu}^2}
I_{3/2}(ke^{\psi_k(\alpha)}),
\label{ent6}
\end{equation}
where $\tilde\mu$ denotes the degeneracy parameter from Eq. (\ref{tf29}) (we
use the notation $\tilde\mu$ here to distinguish it from the chemical potential
$\mu$).
This returns in a more direct manner the result obtained in \cite{pt}.

The entropy of the
nonrelativistic self-gravitating Boltzmann gas is also
given by Eq. (\ref{ent1}) (see footnote 55). Using Eq.
(\ref{a77b}) and introducing the total
energy $E=E_{\rm kin}+W$, we obtain
\begin{equation}
S=-\frac{\mu}{T} N+\frac{2E}{T}-\frac{1}{2}Nk_B.
\label{a85}
\end{equation}
On the other hand, applying Eq. (\ref{nd2}) at $r=R$ and using Eqs.
(\ref{a73})-(\ref{a75}) and $\Phi(R)=-GM/R$ we find that
\begin{equation}
\beta \mu=2\ln(\alpha)+\frac{1}{2}\ln\eta-\psi(\alpha)-\eta-\ln\tilde\mu+
\ln2-\frac{1}{2}\ln\pi.
\label{a87}
\end{equation}
Substituting Eq.
(\ref{a87}) into Eq.
(\ref{a85}), we finally obtain
\begin{eqnarray}
\frac{S}{Nk_B}=-\frac{1}{2}
\ln\eta-2\ln(\alpha)+\psi(\alpha)+\eta-2\Lambda\eta\nonumber\\
+\ln\tilde\mu+\frac{1}{2
}
\ln\pi-\ln2-\frac{1}{2}.
\label{a88}
\end{eqnarray}

\section{Basic equations and definitions}
\label{sec_hr}

For classical self-gravitating systems, or for quantum self-gravitating
systems in the TF approximation (where the quantum potential can be
neglected), the condition of hydrostatic equilibrium reads
\begin{eqnarray}
\label{gl1}
\nabla P+\rho\nabla\Phi={\bf 0}.
\end{eqnarray}
Combined with the Poisson equation
\begin{equation}
\label{gl2}
\Delta \Phi=4\pi G\rho,
\end{equation}
we obtain the fundamental differential equation
\begin{equation}
\label{gl3}
\nabla\cdot \left (\frac{\nabla P}{\rho}\right )=-4\pi G\rho.
\end{equation}
This equation determines the density profile $\rho(r)$ of a DM halo described
by a barotropic equation of state
$P(\rho)$.

The halo radius $r_h$ is defined as the distance at which the
central density  $\rho_0$  is divided by $4$:
\begin{eqnarray}
\label{hr1}
\frac{\rho(r_h)}{\rho_0}=\frac{1}{4}.
\end{eqnarray}
The mass $M(r)$ contained within a sphere of radius $r$ is given by
\begin{eqnarray}
\label{hr2}
M(r)=\int_0^r\rho(r')4\pi{r'}^2\, dr'.
\end{eqnarray}
The halo mass is
\begin{eqnarray}
\label{hr3}
M_h=M(r_h).
\end{eqnarray}
The circular velocity is defined by
\begin{eqnarray}
\label{hr4}
v^2(r)=\frac{GM(r)}{r}.
\end{eqnarray}
The circular velocity at the halo radius is
\begin{eqnarray}
\label{hr5}
v_h^2=v^2(r_h)=\frac{GM_h}{r_h}.
\end{eqnarray}
We note the identity
\begin{eqnarray}
\label{hr8}
\frac{v_h^2}{G\rho_0r_h^2}=\frac{M_h}{\rho_0 r_h^3}.
\end{eqnarray}

\section{Isothermal profile}
\label{sec_i}

\subsection{Emden equation}
\label{sec_ie}

We consider a DM halo with an isothermal equation
of state
\begin{eqnarray}
\label{i1}
P=\rho\frac{k_B T}{m},
\end{eqnarray}
where $T$ is the temperature \cite{chandrabook}. The
fundamental differential equation of
hydrostatic equilibrium (\ref{gl3}) takes the form
\begin{eqnarray}
\label{i2}
\frac{k_B T}{m}\Delta\ln\rho=-4\pi G\rho.
\end{eqnarray}
Writing
\begin{equation}
\label{i3}
\rho=\rho_0
e^{-\psi},
\end{equation}
where $\rho_0$ is the central density, introducing the normalized radial
distance
\begin{equation}
\label{i3b}
\xi=r/r_0, \qquad r_0=\left (\frac{k_B T}{4\pi G\rho_0
m}\right )^{1/2},
\end{equation}
where $r_0$ is the thermal core radius, and assuming that the DM
halo is spherically symmetric, we obtain the Emden equation \cite{chandrabook}
\begin{equation}
\label{i5}
\frac{1}{\xi^2}\frac{d}{d\xi}\left
(\xi^2\frac{d\psi}{d\xi}\right
)=e^{-\psi}
\end{equation}
with the boundary conditions
\begin{equation}
\label{i5b}
\psi(0)=\psi'(0)=0.
\end{equation}
The density profile has the self-similar (homology)
form $\rho(r)/\rho_0=e^{-\psi({r}/{r_0})}$.
Using Eqs. (\ref{hr2}), (\ref{i3}), (\ref{i3b})
and (\ref{i5}), the mass
contained within the
sphere of radius $r$ is given by
\begin{eqnarray}
\label{i7}
M(r)=4\pi \rho_0 r_0^3 {\xi}^2\psi'(\xi).
\end{eqnarray}
According to Eqs. (\ref{hr4}), (\ref{i3b}) and (\ref{i7}), the circular velocity
is
\begin{equation}
\label{i7q}
v^2(r)=4\pi G\rho_0r_0^2\xi \psi'(\xi).
\end{equation}
Using Eq. (\ref{i3b}), we find that the temperature satisfies the
relation
\begin{eqnarray}
\frac{k_B T}{m}=4\pi G\rho_0 r_0^2.
\end{eqnarray}
Therefore, we can rewrite  Eq. (\ref{i7q}) as
\begin{equation}
\frac{m v^2(r)}{k_B T}=\xi \psi'(\xi).
\end{equation}

\subsection{Halo mass and halo radius}
\label{sec_hmri}

The halo radius defined by Eq. (\ref{hr1}) is given by $r_h=\xi_h r_0$, where
$\xi_h$
is determined by the equation
\begin{eqnarray}
\label{i9}
e^{-\psi(\xi_h)}=\frac{1}{4}.
\end{eqnarray}
Solving the Emden equation (\ref{i5}) numerically, we find
\begin{eqnarray}
\label{i10}
\xi_h=3.63,\qquad \psi'(\xi_h)=0.507.
\end{eqnarray}
The normalized halo mass is
\begin{eqnarray}
\label{i11}
\frac{M_h}{\rho_0 r_h^3}=4\pi\frac{\psi'(\xi_h)}{{\xi_h}}=1.76.
\end{eqnarray}
The normalized circular velocity at the halo radius is
\begin{eqnarray}
\label{i12}
\frac{v_h^2}{4\pi G\rho_0r_h^2}=\frac{\psi'(\xi_h)}{\xi_h}=0.140.
\end{eqnarray}
The normalized temperature is
\begin{eqnarray}
\label{i13}
\frac{k_B T}{Gm\rho_0 r_h^2}=\frac{4\pi}{\xi_h^2}=0.954.
\end{eqnarray}
The normalized
inverse temperature of the halo is
\begin{eqnarray}
\label{app2}
\eta_v=\frac{\beta GM_hm}{r_h}=\xi_h\psi'_h=1.84.
\end{eqnarray}

\section{Polytropic profiles}
\label{sec_p}

\subsection{Lane-Emden equation}
\label{sec_ple}

We consider a DM halo with a polytropic equation
of state of the form
\begin{equation}
\label{gl4}
P=K\rho^{\gamma},
\end{equation}
where $K$ is the polytropic constant and $\gamma=1+1/n$ is the polytropic
index \cite{chandrabook}. The
fundamental differential equation of
hydrostatic equilibrium (\ref{gl3}) takes the form
\begin{equation}
\label{gl5}
K(n+1)\Delta\rho^{1/n}=-4\pi G\rho.
\end{equation}
In the following, we restrict ourselves to spherically symmetric distributions.
We also assume $K>0$ and $6/5<\gamma<+\infty$ (i.e. $0\le n<5$) in order to
have density profiles with a compact support (see below).

Writing
\begin{eqnarray}
\label{p3y}
\rho=\rho_0 \theta^{n},
\end{eqnarray}
where $\rho_0$ is the central density, introducing the normalized radial
distance
\begin{eqnarray}
\label{p4y}
\xi=r/r_0,\qquad r_0=\left \lbrack \frac{K(n+1)}{4\pi G\rho_0^{1-1/n}}\right
\rbrack^{1/2},
\end{eqnarray}
where $r_0$ is the polytropic core radius, and assuming that the DM
halo is spherically symmetric, we obtain the
Lane-Emden equation
\cite{chandrabook}
\begin{eqnarray}
\label{p5}
\frac{1}{\xi^2}\frac{d}{d\xi}\left (\xi^2\frac{d\theta}{d\xi}\right
)=-\theta^{n}
\end{eqnarray}
with the boundary conditions
\begin{eqnarray}
\label{p6bc}
\theta(0)=1,\qquad \theta'(0)=0.
\end{eqnarray}
The density profile has the self-similar (homology)
form $\rho(r)/\rho_0=\theta^n ({r}/{r_0})$.
Using Eqs. (\ref{hr2}), (\ref{p3y}), (\ref{p4y}) and (\ref{p5}), the mass
contained within the
sphere of radius $r$ is given by
\begin{eqnarray}
\label{p8}
M(r)=-4\pi \rho_0 r_0^3 {\xi}^2\theta'(\xi).
\end{eqnarray}
According to Eqs. (\ref{hr4}), (\ref{p4y})  and (\ref{p8}), the circular
velocity is
\begin{eqnarray}
\label{p9}
v^2(r)=-4\pi G\rho_0 r_0^2\xi\theta'(\xi).
\end{eqnarray}

\subsection{Mass and radius}
\label{sec_ptm}

When $n<5$, the polytropes are self-confined (their density has a compact
support) \cite{emden,chandrabook}. We denote by $\xi_1$ the normalized radius at
which the density
vanishes: $\theta_1=\theta(\xi_1)=0$. Their radius $R$ and their total
mass $M$ are
given by
\begin{eqnarray}
R=\xi_1 r_0,\qquad  M=-4\pi \rho_0 r_0^3 \xi^2_1\theta'_1
\end{eqnarray}
or, more explicitly, by
\begin{equation}
\label{rpol}
R=\xi_1\left\lbrack \frac{K(n+1)}{4\pi
G}\right\rbrack^{1/2}\frac{1}{\rho_0^{(n-1)/2n}},
\end{equation}
\begin{equation}
\label{mpol}
M=-4\pi \frac{\theta'_1}{\xi_1} \rho_0 R^3.
\end{equation}
Eliminating the central density between these two equations, we obtain the
mass-radius relation \cite{chandrabook}
\begin{eqnarray}
M^{(n-1)/n}R^{(3-n)/n}=\frac{K(n+1)}{G(4\pi)^{1/n}}\omega_n^{(n-1)/n},
\end{eqnarray}
where $\omega_n=-\xi_1^{(n+1)/(n-1)}\theta'_1$ is a constant determined by the
Lane-Emden equation (\ref{p5}). It can be shown that a
polytrope of index $n$ is dynamically stable with respect to
the Euler-Poisson equations if $n<3$ and linearly unstable if $n>3$
\cite{chandrabook}. On the other hand, the
gravitational energy of a polytrope of index $n$ is given by the
Betti-Ritter formula \cite{chandrabook}
\begin{equation}
W=-\frac{3}{5-n}\frac{GM^2}{R}.
\end{equation}

For the polytrope $n=3/2$, solving the Lane-Emden equation
(\ref{p5}) numerically, we find
\begin{equation}
\xi_1=3.65375,\qquad \theta'_1=-0.203302.
\label{nump}
\end{equation}
This polytrope represents a nonrelativistic fermion star at $T=0$. This leads to
Eqs. (\ref{cdl8})-(\ref{cdl10}) quoted in the main text. The
mass-radius relation may be written as
\begin{eqnarray}
\label{fdm3}
M R^3=\frac{9\omega_{3/2}}{8192\pi^4}\frac{h^6}{G^3m^{8}},
\end{eqnarray}
where $\omega_{3/2}=132.3843$.

\subsection{Halo mass and halo radius}
\label{sec_phmhr}

The halo radius  defined by Eq. (\ref{hr1}) is given by $r_h=\xi_h r_0$, where
$\xi_h$ is determined by the equation
\begin{eqnarray}
\label{p10}
\theta(\xi_h)^{n}=\frac{1}{4}.
\end{eqnarray}
The value of $\xi_h$ can be obtained by solving the Lane-Emden equation
(\ref{p5}) for a given value of $n$. The normalized  halo mass is
\begin{eqnarray}
\label{p11}
\frac{M_h}{\rho_0 r_h^3}=-4\pi\frac{\theta'(\xi_h)}{{\xi_h}}.
\end{eqnarray}
The normalized circular velocity at the halo radius is
\begin{eqnarray}
\label{p12}
\frac{v_h^2}{4\pi G\rho_0 r_h^2}=-\frac{\theta'(\xi_h)}{\xi_h}.
\end{eqnarray}

The halo radius $r_h$ and the halo mass may be written more explicitly as
\begin{eqnarray}
\label{p1}
r_h=\xi_h\left\lbrack \frac{K(n+1)}{4\pi
G}\right\rbrack^{1/2}\frac{1}{\rho_0^{(n-1)/2n}},
\end{eqnarray}
\begin{eqnarray}
\label{p3}
M_h=-4\pi\frac{\theta'(\xi_h)}{{\xi_h}}\rho_0 r_h^3.
\end{eqnarray}
Eliminating the central density between Eqs. (\ref{p1}) and (\ref{p3}), we
obtain the halo mass-radius relation
\begin{eqnarray}
\label{p4}
M_h
r_h^{(3-n)/(n-1)}=-4\pi\theta'(\xi_h)\xi_h^{(n+1)/(n-1)}\nonumber\\
\times\left\lbrack
\frac{K(n+1)}{4\pi
G}\right\rbrack^{n/(n-1)}.
\end{eqnarray}

Let us assume that the minimum halo corresponds to a polytrope of index $n$
(this includes the case of fermions corresponding to $n=3/2$, the case of 
noninteracting bosons corresponding to $n=2$ and the case of self-interacting
bosons corresponding to $n=1$ \cite{mcmh}).
Using Eqs. (\ref{p1}) and (\ref{p3}) and introducing the universal
surface
density of DM halos from Eq. (\ref{p5}) we find that the minimum halo radius,
the minimum halo mass, and the maximum
halo central density are given by
\begin{equation}
\label{p6}
(r_h)_{\rm min}=\xi_h^{2n/(n+1)}\left\lbrack
\frac{K(n+1)}{4\pi
G}\right\rbrack^{n/(n+1)}\frac{1}{\Sigma_0^{(n-1)/(n+1)}},
\end{equation}
\begin{eqnarray}
\label{p7}
(M_h)_{\rm min}=&-&4\pi \theta'(\xi_h) \xi_h^{(3n-1)/(n+1)}\nonumber\\
&\times&\left\lbrack
\frac{K(n+1)}{4\pi
G}\right\rbrack^{2n/(n+1)}\Sigma_0^{(3-n)/(n+1)},\nonumber\\
\end{eqnarray}
\begin{equation}
\label{sad2}
(\rho_0)_{\rm max}=\frac{1}{\xi_h^{2n/(n+1)}}\left\lbrack
\frac{4\pi
G}{K(n+1)}\right\rbrack^{n/(n+1)}\Sigma_0^{2n/(n+1)}.
\end{equation}

For the polytrope $n=3/2$, solving the Lane-Emden
equation (\ref{p5}) numerically, we find
\begin{equation}
\xi_h=2.27, \qquad \theta'_h=-0.360.
\label{p9num}
\end{equation}
If the minimum halo corresponds to a fermion ball at $T=0$ (equivalent to a
polytrope $n=3/2$) we obtain Eqs. (\ref{cdl11})-(\ref{cdl16}) quoted in the main
text.

\end{document}